\documentclass[longauth,subabstract]{aa} 

\usepackage{times}
\usepackage{graphicx}
\usepackage{longtable}
\usepackage{natbib}
\usepackage{lscape}
\usepackage{amssymb}

\usepackage{subfig} 
\usepackage{mathrsfs}
\DeclareMathAlphabet{\pazocal}{OMS}{zplm}{m}{n}
\usepackage{array}
\usepackage{txfonts}
\usepackage{color}

\usepackage{xspace}

\bibpunct{(}{)}{;}{a}{}{,}
\include{definitions}

\newcommand{\vsini} {$v$\,sin\,$i$}

\newcommand{\ov}{O\,V}
\newcommand{\ovz}{O\,Vz}
\newcommand{\Teff} {$T_{\rm eff}$}
\newcommand{\grav} {log\,{\em g}}
\newcommand{\gravc} {log\,{\em g$_{\rm c}$}}

\newcommand{\fastwind} {{\sc fastwind}}

\newcommand{\logq}{log\,{\em Q}}

\newcommand{\logl}{$\log{L/L_{\odot}}$}
\newcommand{\kms}{$\rm km\,s^{-1}$}
\newcommand{\bonnsai}{\mbox{\textsc{bonnsai}}\xspace}
\newcommand{\pp}{$\phantom{1}$}
\newcommand{\pl}{$\phantom{<}$}
\newcommand{\ps}{$\phantom{*}$}

\begin{document}

\title{The VLT-FLAMES Tarantula Survey\thanks{Based on observations at the 
European Southern Observatory Very Large Telescope in program 182.D-0222.}}
\subtitle{XXVI: Properties of the O-dwarf population in 30\,Doradus}
\author{C.~Sab\'in-Sanjuli\'an\inst{\ref{inst:dfuls},\ref{inst:uls}}, 
S.~Sim\'on-D\'iaz\inst{\ref{inst:iac},\ref{inst:ull}},
 A.~Herrero\inst{\ref{inst:iac},\ref{inst:ull}}, 
J.~Puls\inst{\ref{inst:munich}}, F.~R.~N.~Schneider\inst{\ref{inst:oxford}}, C.~J.~Evans\inst{\ref{inst:edinburgh}}, 
M. Garcia\inst{\ref{inst:csic-inta}}, F.~Najarro\inst{\ref{inst:csic-inta}}, 
I.~Brott\inst{\ref{inst:viena}}, N.~Castro\inst{\ref{inst:michigan}}, 
P.~A.~Crowther\inst{\ref{inst:sheffield}}, 
A.~de~Koter\inst{\ref{inst:pannekoek},\ref{inst:belgium}}, 
S.~E.~de~Mink\inst{\ref{inst:pannekoek}},
G.~Gr\"afener\inst{\ref{inst:bonn}}, 
N.~J.~Grin\inst{\ref{inst:bonn}}, 
G.~Holgado\inst{\ref{inst:iac},\ref{inst:ull}}, 
N.~Langer\inst{\ref{inst:bonn}}, D.~J.~Lennon\inst{\ref{inst:esac}}, 
J.~Ma\'iz~Apell\'aniz\inst{\ref{inst:cab-esac}}, 
O.~H.~Ram\'irez-Agudelo\inst{\ref{inst:edinburgh}}, 
H.~Sana\inst{\ref{inst:belgium}}, W.~D.~Taylor\inst{\ref{inst:edinburgh}}, \\
J.~S.~Vink\inst{\ref{inst:armagh}}, N.~R.~Walborn\inst{\ref{inst:baltimore}}}

\institute{Departamento de F\'isica y Astronom\'ia, Universidad de La Serena, Av. Cisternas 1200 Norte, La Serena, Chile\label{inst:dfuls}
\and Instituto de Investigaci\'on Multidisciplinar en Ciencia y 
Tecnolog\'ia, Universidad de La Serena, Ra\'ul Bitr\'an 1305, La Serena, 
Chile\label{inst:uls}
  \and Instituto de Astrof\'isica de Canarias, E-38200 La Laguna, Tenerife, 
Spain \label{inst:iac}  
  \and Departamento de Astrof\'isica, Universidad de La Laguna, E-38205 La 
Laguna, Tenerife, Spain \label{inst:ull}
  \and LMU Munich, Universit\"atssternwarte, Scheinerstrasse 1, 81679 Munchen, 
Germany \label{inst:munich} 
  \and Department of Physics, University of Oxford, Denys Wilkinson Building, 
Keble Road, Oxford OX1 3RH, United Kingdom\label{inst:oxford}
\and UK Astronomy Technology Centre, Royal Observatory, Blackford 
Hill, Edinburgh, EH9 3HJ, UK\label{inst:edinburgh}
  \and Centro de Astrobiolog\'ia (CSIC-INTA), Ctra. de Torrej\'on a Ajalvir 
km-4, E-28\,850 Torrej\'on de Ardoz, Madrid, Spain\label{inst:csic-inta}
    \and University of Vienna, Department of Astronomy, T\"{u}rkenschanzstr. 17, 
1180, Vienna, Austria\label{inst:viena}
  \and Department of Astronomy, University of Michigan, 1085 S. University 
Avenue, Ann Arbor, MI 48109-1107, USA\label{inst:michigan}
  \and Department of Physics \& Astronomy, Hounsfield Road, University of 
Sheffield, S3 7RH, UK\label{inst:sheffield}
  \and Astronomical Institute Anton Pannekoek, Amsterdam University, Science 
Park 904, 1098~XH, Amsterdam, The Netherlands\label{inst:pannekoek}
 \and Instituut voor Sterrenkunde, Universiteit Leuven, Celestijnenlaan 200D, 
3001, Leuven, Belgium\label{inst:belgium}
\and Argelander-Institut f\"{u}r Astronomie der Universit\"{a}t Bonn, Auf dem 
H\"{u}gel 71, 53121 Bonn, Germany\label{inst:bonn}
    \and European Space Astronomy Centre (ESAC), Camino bajo del castillo, s/n 
Urbanizaci\'on Villafranca del Castillo, Villanueva de la Ca\~{n}ada, E-28\,692 
Madrid, Spain\label{inst:esac}
  \and Centro de Astrobiolog\'{\i}a, CSIC-INTA, Campus ESAC, Camino bajo del 
castillo s/n, E-28\,692 Madrid, Spain\label{inst:cab-esac}
  \and Armagh Observatory, College Hill, Armagh, BT61 9DG, Northern Ireland, 
UK\label{inst:armagh}
  \and Space Telescope Science Institute, 3700 San Martin Drive, Baltimore, 
MD\,21218, USA\label{inst:baltimore}}

\offprints{C.~Sab\'in-Sanjuli\'an, email: \texttt{cssj@dfuls.cl}}

\date{Accepted 10th February 2017}

\titlerunning{Properties of the O-dwarf population in 30\,Doradus}
\authorrunning{Sab\'in-Sanjuli\'an et al.}

\abstract{The VLT-FLAMES Tarantula Survey has observed hundreds of
  O-type stars in the 30\,Doradus region of the Large Magellanic Cloud
  (LMC).}
{We study the properties of a statistically significant sample of
  O-type dwarfs in the same star-forming region and test the latest
  atmospheric and evolutionary models of the early main-sequence phase
  of massive stars.}
{We performed quantitative spectroscopic analysis of 105 apparently
  single O-type dwarfs. To determine stellar and wind parameters, we
  used the {\mbox{\textsc{iacob-gbat}}\xspace} package, an automatic
  procedure based on a large grid of atmospheric models that
are calculated
  with the \fastwind\ code. This package was developed for the analysis of optical
  spectra of O-type stars. In addition to classical techniques, we
  applied the Bayesian \bonnsai\ tool to estimate evolutionary masses.}
{We provide a new calibration of effective temperature vs. spectral
  type for O-type dwarfs in the LMC, based on our homogeneous analysis
  of the largest sample of such objects to date and including all
  spectral subtypes. Good agreement with previous results is found,
  although the sampling at the earliest subtypes could be improved.
  Rotation rates and helium abundances are studied in an evolutionary
  context. We find that most of the rapid rotators
  (\vsini\,$>$\,300\,\kms) in our sample have masses below
  $\sim$\,25~M$_{\odot}$ and intermediate rotation-corrected gravities
  (3.9\,$<$\,\gravc\,$<$4.1). Such rapid rotators are scarce at higher
  gravities (i.e.  younger ages) and absent at lower gravities
  (larger ages). This is not expected from theoretical evolutionary
  models, and does not appear to be due to a selection bias in our
  sample.  We compare the estimated evolutionary and spectroscopic
  masses, finding a trend that the former is higher for masses below
  $\sim$\,20~M$_{\odot}$.  This can be explained as a consequence of
  limiting our sample to the O-type stars, and we see no compelling
  evidence for a systematic mass discrepancy.
  For most of the stars in the sample we were unable to estimate the
  wind-strength parameter (hence mass-loss rates) reliably,
  particularly for objects with lower luminosity  (\logl\,$\lesssim$\,5.1). Only with ultraviolet spectroscopy
  will we be able to undertake a detailed investigation of the wind
  properties of these dwarfs.}
{}

  
   \maketitle
%

\section{Introduction}\label{intro}

Massive stars play a key role in many astrophysical areas because
of their
extreme physical properties (high masses, strong winds, and intense
radiation fields, see e.g. \citealt{mm2000,langer12}). They are
commonly used as tracers of young populations and to study
galactic physics. They have become powerful alternative tools to
\ion{H}{ii} regions and Cepheids to obtain information on present-day
chemical abundances in galaxies \citep[see e.g.][]{k08,k13} and
extragalactic distances \citep[][]{kp2000}. Their short lives end
dramatically as supernovae, leaving behind compact remnants such as neutron
stars and black holes \citep[][]{woosley02} and sometimes producing
long-duration gamma-ray bursts \citep[LGRB,][]{woosley06}. Lastly,
they may have contributed significantly to the reionisation of the
Universe and its early chemical evolution \citep[][]{bromm09}.
Characterisation of their physical properties 
over a range of different environments (metallicities) is therefore
an essential task in contemporary astrophysics.

Despite the progress in the field over the past decades, many
unanswered questions remain regarding the formation, evolution, and
final fate of massive stars. For example, their formation processes
are still poorly understood, mainly because of the very short pre-main-sequence phase and observational limitations
of studying the earliest stages of their lives in heavily obscured
regions \citep[e.g.][]{hanson,zinnecker07}.  Submillimetre and radio
observations are starting to probe the dense cores of massive
protostars \citep[e.g.][]{ilee16}, but a better understanding of their
main-sequence phase will also provide important constraints on their
origins.

When it has formed, the initial mass of a star is the dominant factor on its
subsequent evolution, but there are other important effects that also
need to be taken into account. Firstly, massive stars can lose a
significant fraction of their outer envelopes through their intense winds,
which modifies their path in the Hertzsprung--Russell (H--R) diagram. This
is particularly important for luminous stars near the Eddington
limit, or for objects in an advanced evolutionary phase, such as red
supergiants, Wolf--Rayet stars, or luminous blue variables
\citep[e.g.][]{vink2012}. Unfortunately, our understanding of these
outflows is limited by different phenomena such as micro-clumping
\citep[][]{hillier91,fullerton06,puls06_clumping}, macro-clumping,
or porosity \citep{muijres,surlan13,sundqvist14}.

A second factor that affects their internal structure and evolution is
rotation. Stellar rotation reduces the effective gravity through the
associated (latitude-dependent) centrifugal acceleration, leading to
an oblate shape with a larger equatorial than polar radius 
and to gravity darkening \citep[e.g.][]{gray}. It also
contributes to the transport of chemical elements and angular
momentum. At rapid rotation rates, the stellar wind is predicted to
become aspherical (in this case, prolate, as a result of gravity darkening),
and the integrated mass-loss rates might be affected
\citep[][]{muellervink}.  These processes significantly influence the
evolution, lifetime, and final fate of the star
\cite[][]{mm2000,brott}.

The fact that the majority of massive stars are found in binary
systems \citep[see e.g.][]{sana12,sota14} means that binarity must
also be an important factor in their formation and evolution. Binarity
affects the rotation and chemical composition of the stellar
atmosphere by means of mass transfer and/or mergers, and may explain
chemical enrichment in slow rotators \citep{langer12}.  Other factors
such as magnetic fields \citep[see,
e.g.][]{morel2015,wade2015,wade2016} may also have an effect on their
evolution \citep{petit2016}.

The VLT-FLAMES Tarantula Survey \citep[VFTS,][]{evans11} is an ESO
Large Programme to study the properties of an unprecedented number of
massive stars in the 30~Doradus star-forming region in the Large
Magellanic Cloud (LMC). The primary objective of the VFTS was to
obtain multi-epoch intermediate-resolution optical spectroscopy of a
large sample of O-type stars to investigate rotation,
binarity or multiplicity, and wind properties, particularly during the
main-sequence phase where they spend the majority of their lives.

To date, the VFTS has reported serendipitous findings of outstanding
objects, such as VFTS\,682 (WN5h), which, with a current mass of
$\sim$150~M$_{\odot}$, is one of the most massive isolated stars known
to date \citep{besten11}, and the discovery of two stars with
extremely rapid ($\sim$\,600\,\kms) rotational velocities, namely
VFTS 102 (O:\,Vnnne) and VFTS\,285 (O7.5\,Vnnn) \citep{dufton11, w11,
  walborn13, oscar}. More comprehensive studies of the O-type stars in
the survey include detailed spectral classifications
\citep{walborn13}, analysis of their multiplicity through multi-epoch
observations \citep{sana13}, and investigation of their
rotational-velocity distributions
\citep{oscar,oscar15} --\,similar studies have also been
  presented for the B-type stars \citep[see e.g.][]{dufton13,evans15,dunstall15}. These efforts are an
important step towards estimates for physical parameters and chemical
abundances for the whole O-type sample, which will ultimately be used
to address fundamental questions in both stellar and cluster
evolution.

In a previous study, we analysed 48 O\,Vz and 36 O\,V stars from the
VFTS to test the hypothesis that O~Vz stars\footnote{O\,Vz stars are
  characterised by \ion{He}{ii}\,$\lambda$\,4686 absorption stronger
  than any other helium line in their optical spectra \citep[][]{w09}.}
are at a different (younger) evolutionary stage compared to normal
O-type dwarfs \citep[][hereafter Paper~XIII]{cssj}. Here we extend
that work by characterising the physical properties and evolutionary
status of the complete sample of (apparently) single O dwarfs and
subgiants identified by \cite{sana13} and \cite{walborn13}. In
parallel, \citet{oscar17} have investigated the O-type
giants and supergiants from the VFTS, with analysis of the nitrogen
abundances for the dwarfs and giants and supergiants presented by
Sim\'on-D\'iaz et al. (in prep.) and \citet{grin16}, respectively.

This paper is structured as follows. The sample is introduced in
Sect.~2, with the relevant data introduced in Sect.~3.  The methods to
determine the stellar and wind parameters are described in Sect.~4,
together with discussion of some of the limitations of the analysis.
The general properties of the sample are discussed in Sect.~5,
including a new calibration of effective temperature vs. spectral
type.  Discussion of our results in an evolutionary context is given
in Sect.~6, with our main conclusions summarised in Sect.~7.

\begin{figure*}
 \centering
\includegraphics[scale=0.50,angle=90]{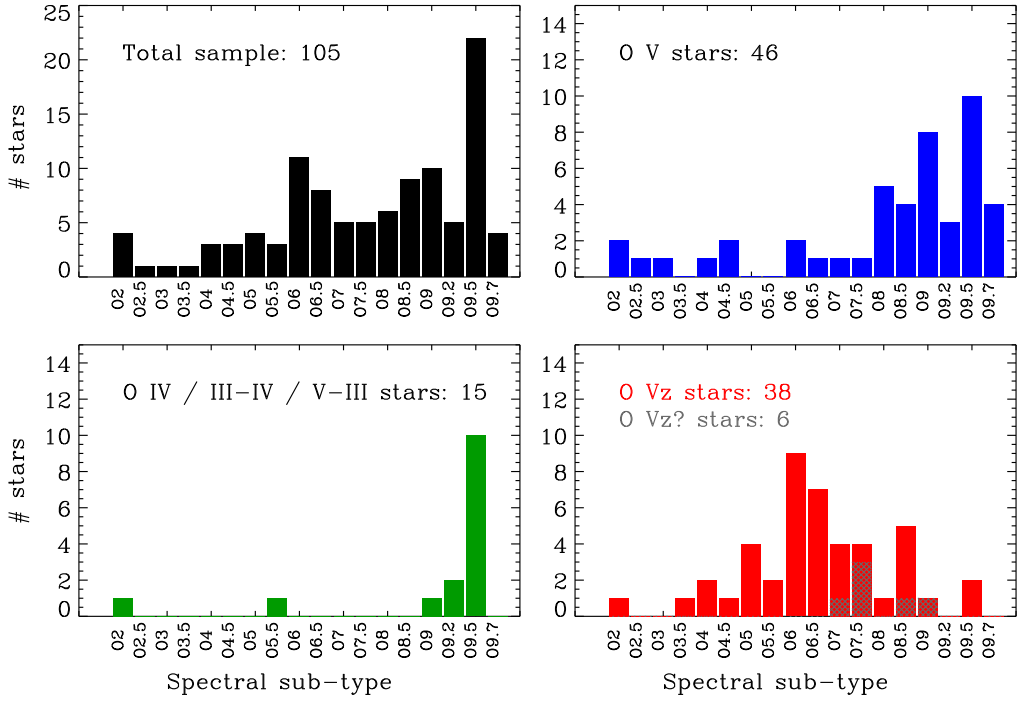}
 \caption{Spectral type distributions for the O-type stars. The whole sample is shown in the upper left panel in
   black.  The remaining panels show the spectral type distribution
   for the luminosity subclasses within the sample: \ov\ stars in
   blue, O\,IV, O\,III-IV, and O\,V-III in green, \ovz\ in red (and \ov
   z? by the grey hatched areas).\label{distrib_spt_dwarfs}}
\end{figure*}

\section{Sample selection}\label{sample}

For this study we selected the apparently single unevolved O-type
stars identified by \citet{sana13} and \citet{walborn13} that have
sufficiently good spectra for reliable quantitative analysis.
\cite{walborn13} divided the 340 O-type stars in the VFTS into two
groups, `AAA' and `BBB', comprising 213 and 127 stars, respectively.
The BBB group were spectra rated as lower quality as a result
of problems such
as low signal-to-noise ratio, strong nebular contamination, double-lined
binaries, and difficulties in precise classification. The BBB stars
were excluded from our sample, as were AAA stars identified by
\citet{sana13} as single-lined binaries (i.e. those for which
radial-velocity variations of $>$\,20\,\kms\ were detected).  This led
to a sample of 105 stars with luminosity classes V and IV (including
the Vz subclass, the uncertain classification V-III, and III-IV, which indicates a precise interpolation between III and IV), as listed in Tables~\ref{tab1_HHe} and
\ref{tab1_HHeN} (with a complete description of the information in the
tables given in Sect.~\ref{iacob-gbat}). Although we describe these
stars as apparently single objects, we note that some spectra
probably include the contribution of more than one star (which may
not necessarily be physically bound, see Sect.~\ref{obs}).

\begin{figure}
 \centering
\includegraphics[scale=1]{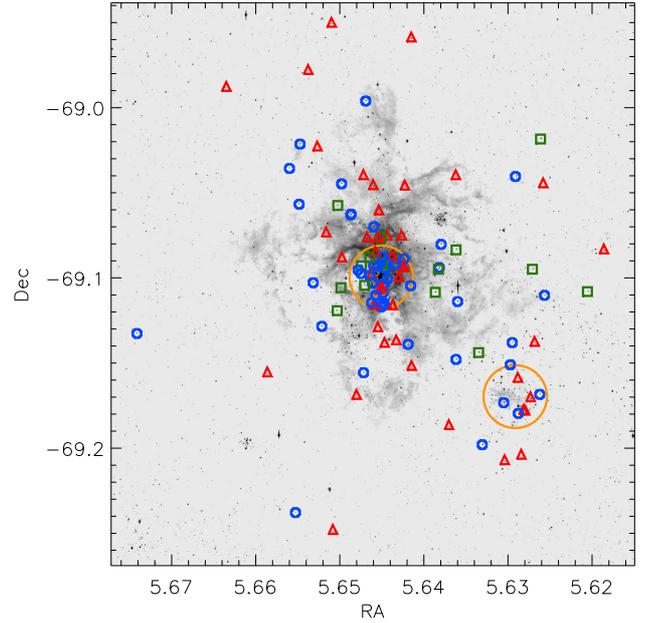}
 \caption{Spatial distribution of our sample stars in 30~Dor.
   Luminosity subclasses are indicated as follows: blue circles for
   \ov\ stars, red triangles for \ovz\ stars (including those with
   doubtful V/Vz classifications), and green squares for O\,IV stars
   (including V-III and III-IV classes).  The central and southwestern
   orange circles (with adopted radii of 2\farcm4, see e.g.
   \citealt{sana13}) indicate the approximate extent of NGC\,2070 and
   NGC\,2060, respectively.\label{30dor_clases}}
\end{figure}

The distribution of spectral types of our sample is shown in
Fig.~\ref{distrib_spt_dwarfs}. The sample is concentrated at medium and
late types, with peaks at O6 and O9.5, with only seven stars earlier
than O4. Interestingly, different distributions are found when the
stars are separated by luminosity class. Contrary to the distribution
of the `normal' \ov\ stars, which are concentrated mainly at spectral types
later than O8, the distribution of \ovz\ stars dominates at medium
subtypes; we refer to Paper XIII for a detailed
discussion of the origin of this bimodal distribution of \ov\ and
\ovz\ stars. Only a few O\,IV stars are in the sample, and these are
mostly O9.5 objects. For completeness, the location of the various
subgroups of our sample in the 30~Dor region is shown in
Fig.~\ref{30dor_clases}.

\section{Observations}\label{obs}

The VFTS observations were obtained at the Very Large Telescope (VLT)
at Paranal in Chile, using the Fibre Large Array Multi-Element
Spectrograph instrument \citep[FLAMES;][]{pasquini}. Details of the
VFTS observations and data were given by \cite{evans11}.

We employed the same data as those used by \cite{oscar}, \cite{walborn13},
and \cite{cssj}. These comprise spectra obtained using the fibre-fed
Medusa mode of FLAMES and three of the standard settings of the
Giraffe
spectrograph\footnote{http://www.eso.org/sci/facilities/paranal/instruments/flames/inst/
  Giraffe.html}: LR02, LR03, and HR15N (see Table~\ref{flames_table}).
As described by \cite{evans11}, the VFTS gathered multi-epoch
observations of all stars in the survey. Details regarding the
combination of the multi-epoch LR02 and LR03 data to produce
the final spectra discussed were given by \cite{walborn13}.

\begin{table}
\caption[]{Spectral coverage and resolving power ($R$) of each setting 
in the FLAMES--Medusa mode used by the VFTS.\label{flames_table}}
\begin{center}
\begin{tabular}{ccc}
\hline
\hline
\multicolumn{3}{c}{} \\ [-2 ex]
Setting & Range (\AA) & $R$ \\
\hline
\multicolumn{3}{c}{} \\ [-2 ex]
LR02  & 3960-4564 & 7000\\
LR03  & 4499-5071 & 8500\\
HR15N & 6442-6817 & 16000\\ 
\hline\hline
\end{tabular}
\end{center}
\end{table}

\begin{figure*}[t!]
 \centering
\includegraphics[scale=0.65, angle=90,trim=0mm 0mm 0mm 0mm,clip]{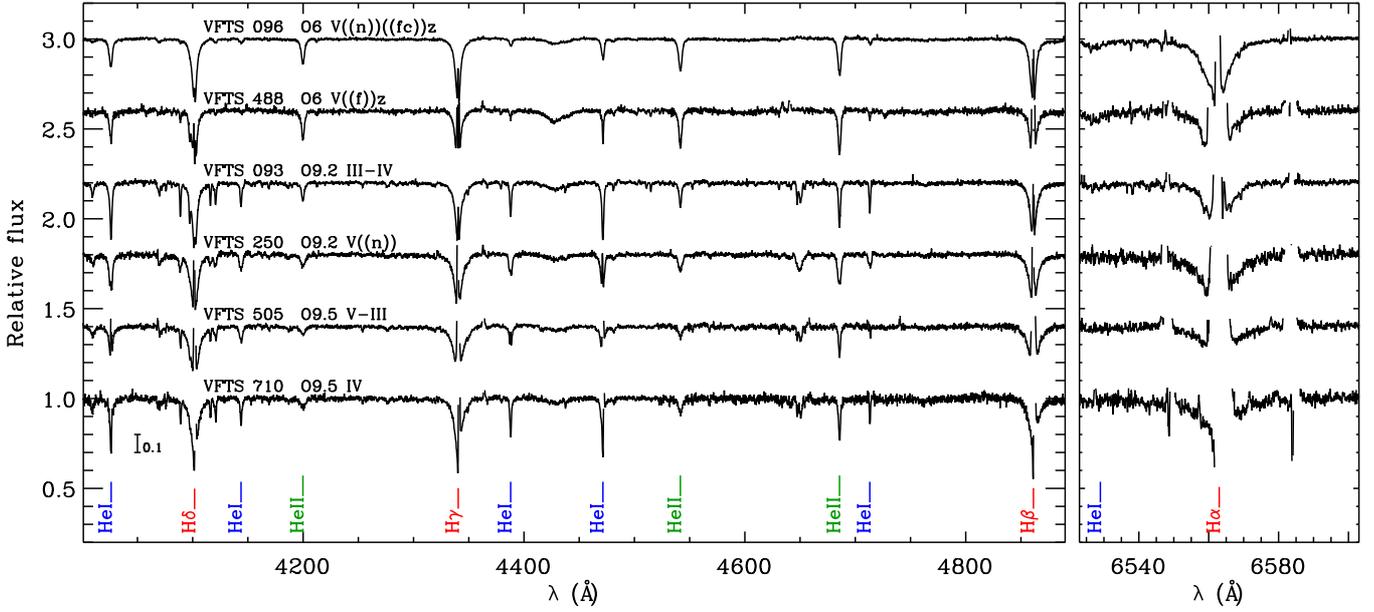}
 \caption{Example O-type dwarf spectra 
for the range of luminosity subclasses in our sample (IV, III-IV, V-III, V,
   and Vz); the diagnostic H and He lines are indicated. The uppermost
   spectrum (VFTS\,096) suffers from
   stellar contamination in the Medusa fibre (see
   Fig.~\ref{hst_images_muestra} and Sect.~\ref{caution:contamination}).
   \label{spectra_examples}}
\end{figure*}

Figure~\ref{spectra_examples} shows six representative examples of
Medusa spectra from our sample, covering the different luminosity
classes (IV, III-IV, V-III, V, and Vz). As described in Paper XIII,
the observed spectral range includes \ion{H}{i} and \ion{He}{i/ii}
lines (plus several \ion{N}{iii/iv/v} lines, see
Sect.~\ref{iacob-gbat}) suited to determine the important physical
parameters of O-type stars. We also note the presence of nebular
emission lines in most of the spectra.

Absolute magnitudes (col.~5 in Tables \ref{tab2_HHe} and
\ref{tab2_HHeN}) were calculated using the ground-based $B$- and
$V$-band photometry from \citet{evans11}, adopting a distance modulus
of 18.5\,mag \citep[e.g.][]{gibson}, and using the Bayesian code
\textsc{chorizos} \citep{jma2004} to take the
line-of-sight extinction into account; a complete description of the method
and model spectral energy distributions used (calculated for the
metallicity of the LMC) was given by \citet{jma2014}.

\begin{figure}
 \centering
\subfloat[VFTS\,021 ]{\includegraphics[scale=0.3,trim=10mm 10mm 10mm 0mm, 
clip]{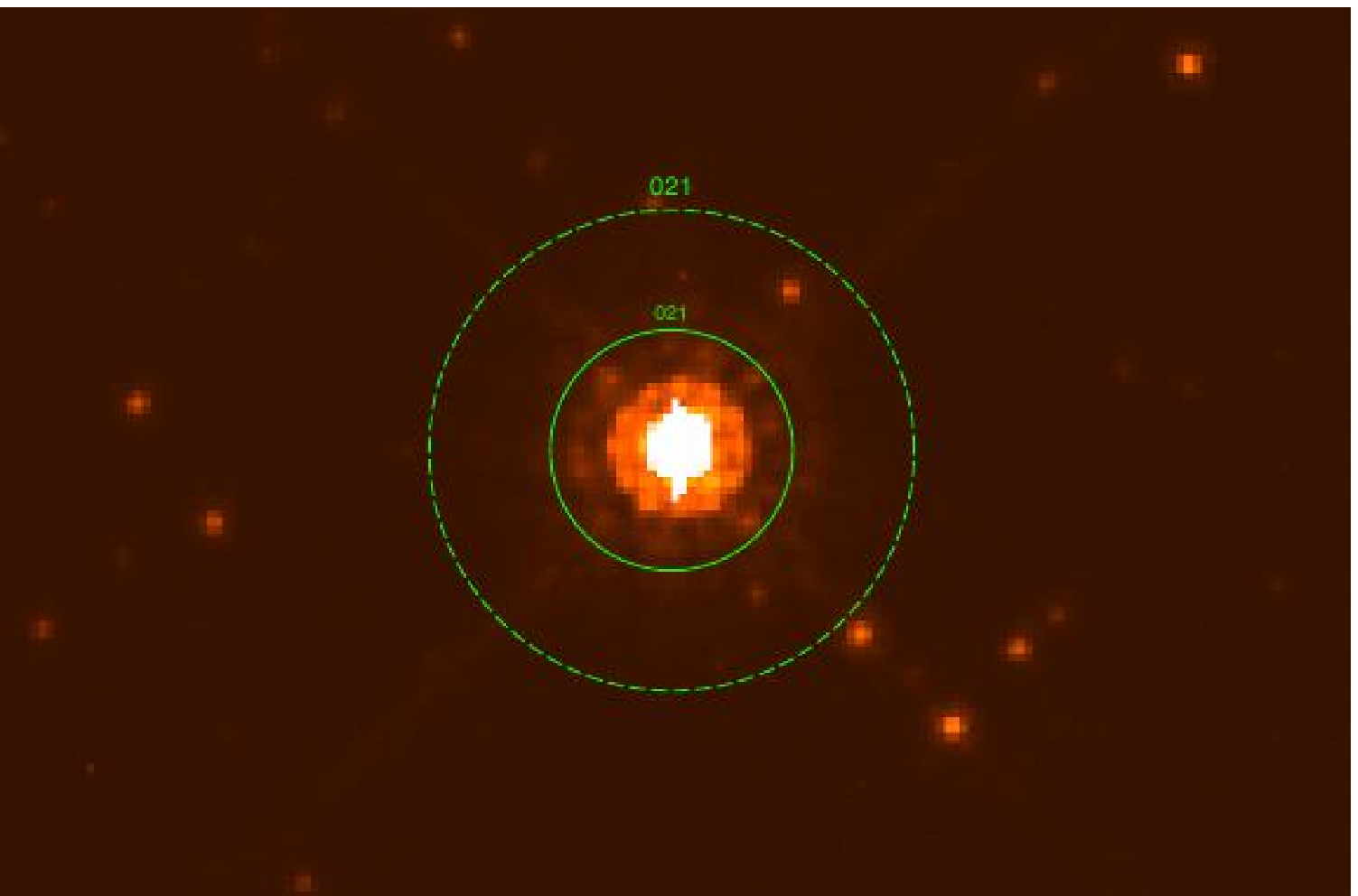}}~~
\subfloat[VFTS\,677 ]{\includegraphics[scale=0.3,trim=10mm 10mm 10mm 0mm, 
clip]{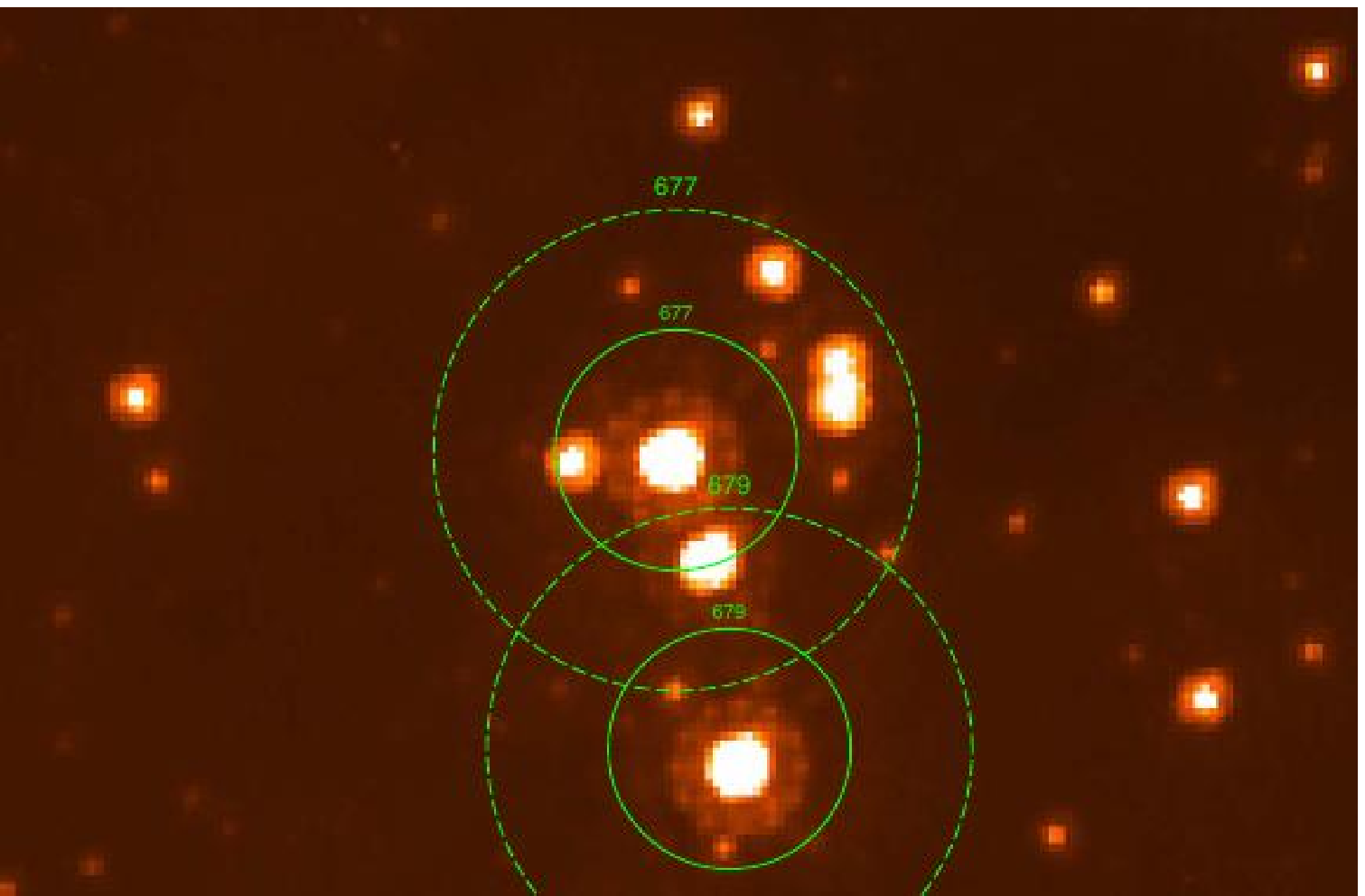}}\\
\subfloat[VFTS\,096 ]{\includegraphics[scale=0.3,trim=10mm 10mm 10mm 0mm, 
clip]{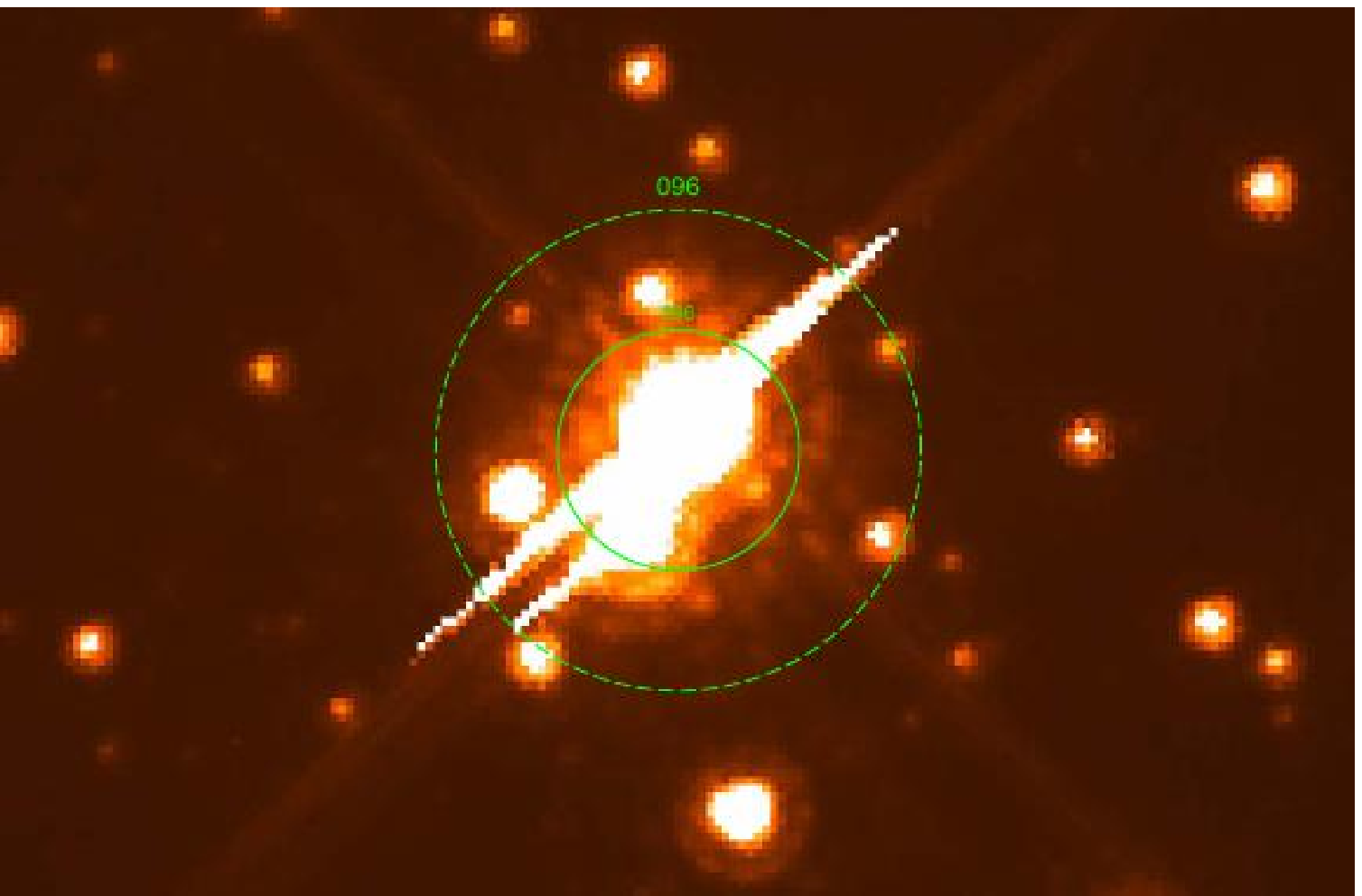}}~~
\subfloat[VFTS\,468 ]{\includegraphics[scale=0.3,trim=10mm 10mm 10mm 0mm, 
clip]{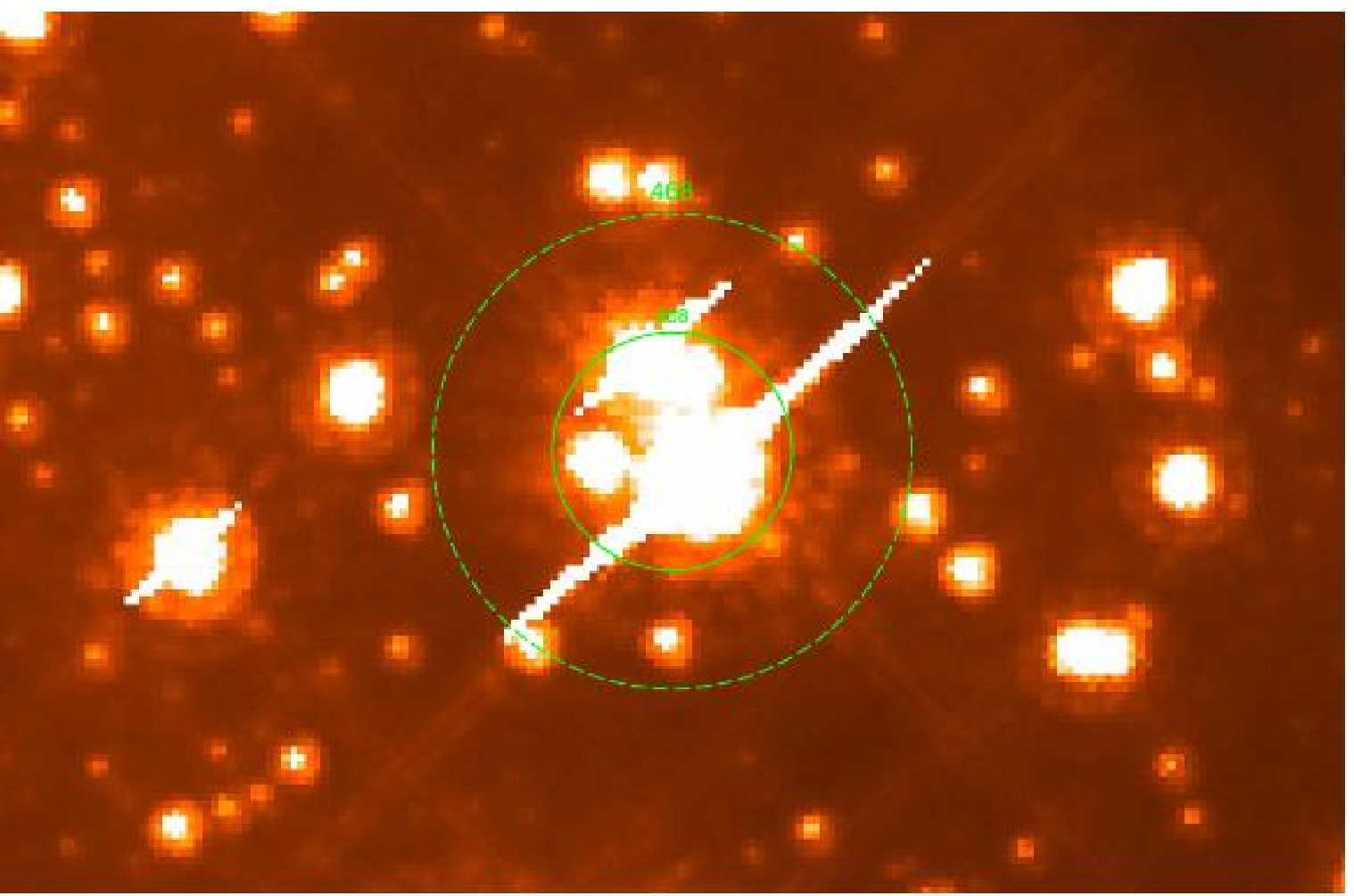}}\\
\subfloat[VFTS\,506 ]{\includegraphics[scale=0.3,trim=10mm 10mm 10mm 0mm, 
clip]{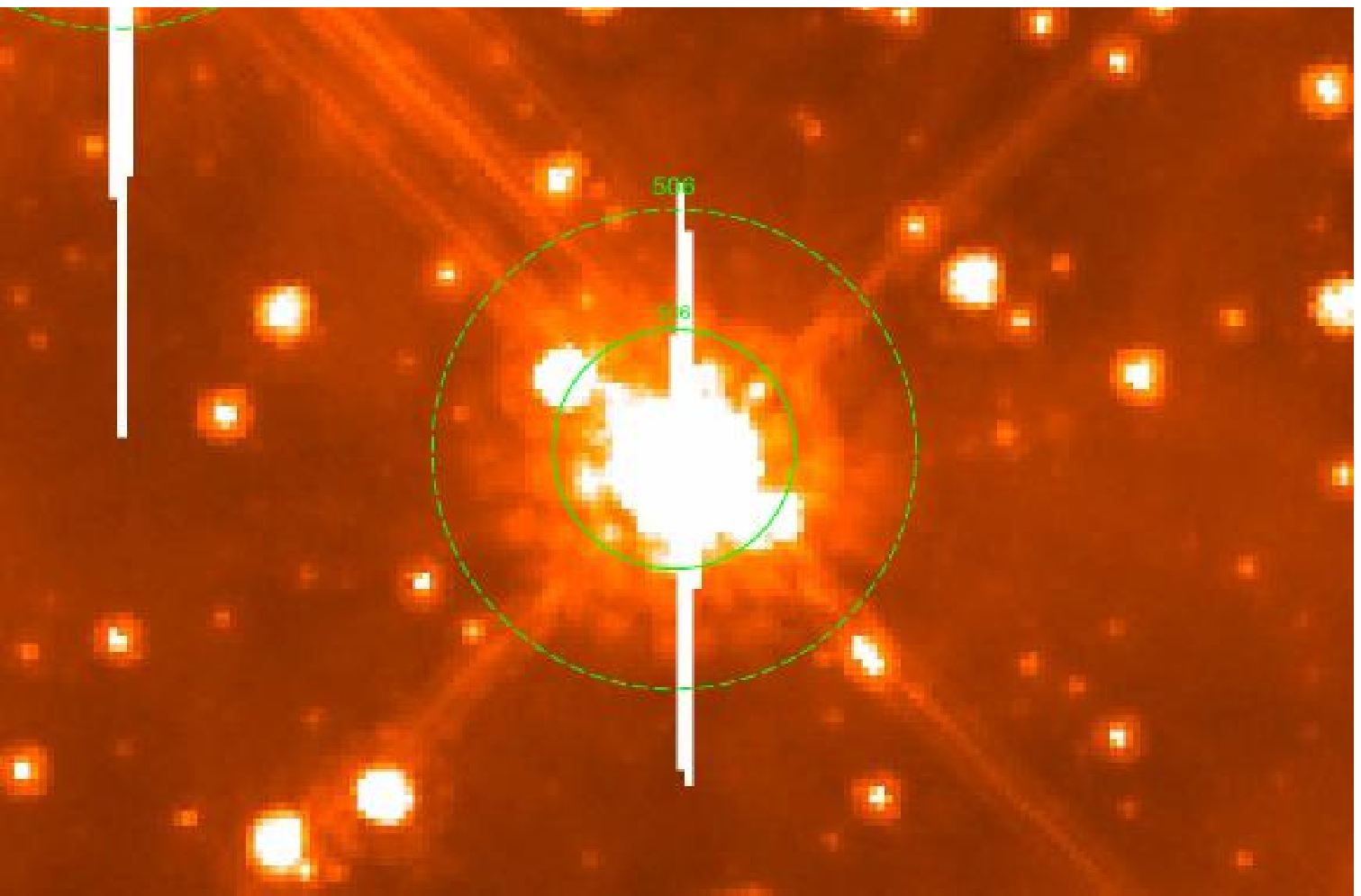}}~~
\subfloat[VFTS\,621 ]{\includegraphics[scale=0.3,trim=10mm 10mm 10mm 0mm, 
clip]{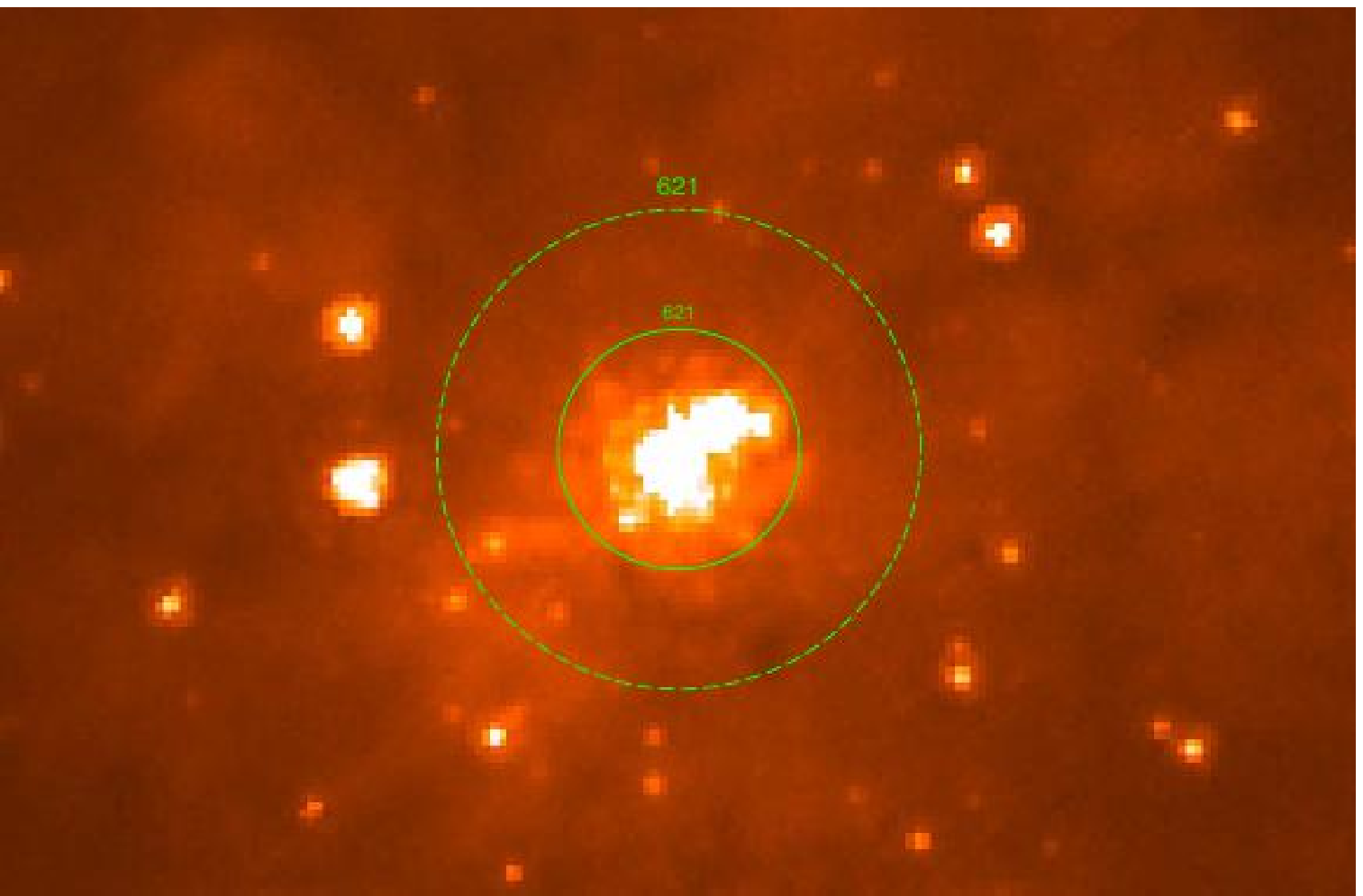}}
 \caption{Example {\em HST}/WFC3 images of O dwarf stars in the VFTS.
   The aperture of the Medusa fibres (1\farcs2) is indicated by the
   inner green circles; the outer circles have twice this diameter to
   help guide the eye.  VFTS\,021 is shown as an example of a
   uncontaminated Medusa fibre, while the spectra of the others
   shown here are expected to have some degree of contamination from
   close-by companions.}
\label{hst_images_muestra}
\end{figure}

Lastly, we had access to a series of $I$-band images obtained with the
{\em Hubble Space Telescope (HST)}\footnote{From the `Proper motions
  of massive stars in 30 Doradus' program (GO\,12499, P.~I.:
  D.~J.~Lennon).}, which includes more than 80\,\% of the stars in the
VFTS. These images were helpful to investigate possible
multiple sources within the Medusa fibres \citep[see
Fig.~\ref{hst_images_muestra} for examples, and
also][]{sabbi2012,sabbi2016}, allowing detection of composite spectra
that are otherwise undetected from radial-velocity measurements, and/or stars
with contaminated photometry in the ground-based imaging.  For
completeness, details of possible or confirmed contamination
\cite[provided by][]{walborn13} are included in the final
column of Tables~\ref{tab1_HHe} and \ref{tab1_HHeN}.

\label{sample}

\section{Stellar and wind parameter determination}\label{iacob-gbat}

\subsection{Spectroscopic parameters}
\subsubsection{HHe analysis}\label{sec:HHe}

Stellar and wind parameters were determined using the IACOB grid-based
automatic tool \citep[{\sc iacob-gbat}, see][]{iacob11}. This tool is
based on standard techniques for the quantitative spectroscopic
analysis of O stars \citep[see e.g.][]{h92,h02,rph04} and has been
automated by applying a $\chi^2$ algorithm to a large
grid\footnote{The grid was computed using the Condor workload
  management system (\textit{http://www.cs.wisc.edu/condor/}) implemented at
  the Instituto de Astrof\'isica de Canarias.} of synthetic spectra
computed using {\sc fastwind} \citep{santolaya,puls05,rivero12iii}.
The grid comprises $\sim$\,180\,000 atmosphere models, covering a wide
range of stellar and wind parameters, and is optimised for the analysis of
O-type stars \citep[see also][for similar approaches using {\sc
  fastwind}]{lefever2007,castro2012}.

We followed the same strategy and criteria as were used in analysis of the \ov
z subsample presented in Paper XIII. In brief, we used the grid of
\fastwind\ models computed for the metallicity of the LMC
\citep[Z\,=\,0.5\,Z$_{\odot}$, see e.g.][]{mokiem2007LMC}. The
parameter ranges of the \fastwind\ grid and the hydrogen and helium
lines used in the analysis are shown in Tables 2 and 3 of Paper XIII.
We adopted the projected rotational velocities (\vsini) from
\cite{oscar}, except for 17 stars where the global broadening of the
synthetic and the observed lines did not agree; for these stars the
\vsini\ values were iterated on
until achieving a good fit. Although macroturbulent broadening of the
line profiles appears to be a relatively
ubiquitous feature of O-type stars \citep[see e.g.][]{ssd16}, we do not expect
this to be a significant factor in the context of the physical
parameters studied here with \ion{H}{i} and \ion{He}{i/ii}, and do not include its effects separately in our
line-broadening convolution. Lastly, we fixed the microturbulence and
the $\beta$-parameter (i.e., the exponent of the wind velocity-law) to
5 \kms\ and 0.8, respectively, as these two parameters could not be
properly constrained with the current sample (see Sects.~3.4.1 and
3.4.2 of Paper XIII).

Results from the {\sc iacob-gbat} analysis of our sample (except for
the targets discussed in Sect.~\ref{analisis_nitro}) are presented in
Table~\ref{tab1_HHe}. The column entries are as follows: (1) VFTS
identifier; (2) spectral classification from \cite{walborn13}; (3)
\vsini\ considered in the analysis; (4-11) derived effective
temperature (\Teff), gravity (\grav), rotation-corrected gravity\footnote{\gravc\ is defined as the
  logarithmic gravity derived from the spectral analysis corrected for
  centrifugal effects \citep[see e.g.][]{h92,rph04}.}
(\gravc), helium abundance\footnote{Y(He) is the helium-to-hydrogen number fraction, Y(He)\,=\,$N$(He)/$N$(H).} (Y(He)), and wind-strength $Q$-parameter (see Sect.~\ref{sect:winds}),
and their formal uncertainties; (12) estimated values for
the wind momentum $D_{\rm mom}$ (also Sect.~\ref{sect:winds}); (13)
comments from \citeauthor{walborn13} regarding possible
binarity or  multiplicity.
 
We adopted a minimum value of 0.1~dex for the formal errors on
\grav\ as we consider that uncertainties below this value are not
realistic. We note that several sources of uncertainty need to
be taken into account besides those arising from the spectral fitting.
These include the continuum renormalisation and the sampling of the
\fastwind\ grid. For the same reasons, the uncertainties in Y(He) were set to 0.02 since the errors estimated by the \textsc{iacob-gbat}
were unrealistically low \citep[see e.g.][]{h92,h02,rph04}. Other
points to be taken into account when interpreting results from the
analysis are discussed in Sect.~\ref{caution}.

\subsubsection{HHeN analysis} \label{analisis_nitro}

In 20 cases the HHe analysis using the \textsc{iacob-gbat} failed to
provide reliable results, as indicated by the broad $\chi^2$\,--\,\Teff\
distributions (with widths in excess of 5000\,K) and/or unexpectedly
low temperatures for the spectral type. After careful inspection of
all the cases, we found that this was mainly due to a lack of reliable
\ion{He}{i} diagnostic lines (because of strong nebular contamination
or very high temperatures).  In these cases, we turned to the N
ionisation balance, following the guidelines from
\cite{rivero12i,rivero12iii,rivero12ii}.

The grids of \fastwind\ models for different metallicities
incorporated in the \textsc{iacob-gbat} have recently been
extended to
include nitrogen as an explicit model atom; however, the nitrogen
lines are not yet included in the \textsc{iacob-gbat}
computations.  A detailed HHeN analysis of the complete O-dwarf sample
(also including N abundances) is now in progress and will be presented
in a forthcoming paper (Sim\'on-D\'iaz et al., in prep.). For the
purposes of the current study we performed a traditional by-eye HHeN
analysis of the 20 stars for which the automated HHe analysis did not
provide reliable results. In these analyses we fixed the helium
abundance to Y(He)\,$=$\,0.10, considered the parameters associated
with the best-fitting model, and adopted the following formal errors
for \Teff, \grav, Y(He), and \logq: 1500~K, 0.10~dex, 0.02, and 0.20,
respectively. The results for these 20 stars are summarised in
Table~\ref{tab1_HHeN} (which includes the same information as
Table~\ref{tab1_HHe}, but for quantities fixed in the
analysis).

\subsection{Radii, luminosities, and masses} \label{determination:phys}

Stellar radii, luminosities, and spectroscopic masses were calculated following
the relations described by \cite{kudri80}, which connect the absolute
magnitude in the $V$ band and the stellar radius \citep[see
also][]{h92,rph04}.

Evolutionary masses ($M_{\rm ev}$) were estimated in the classical
way, that is,  by interpolating between evolutionary tracks by \cite{brott} in the H--R
(\logl\,vs.\,log\,\Teff) and Kiel
(\gravc\,vs.\,log\,\Teff) diagrams. 
  
We also used the Bayesian
\bonnsai tool\footnote{The \bonnsai web-service is available at
\textit{http://www.astro.uni-bonn.de/stars/bonnsai}.} \citep{bonnsai}.
In contrast to more traditional approaches, \bonnsai simultaneously
accounts for all
available observables (i.e. \logl, \gravc, \Teff,
\vsini) and, assuming prior knowledge of the initial mass
function, for example, it computes full posterior probability distributions of the
various stellar parameters for a given set of evolutionary models.
For our sample we used \bonnsai to match the derived luminosities,
effective temperatures, surface gravities, and projected rotational
velocities to the evolutionary models of \citet{brott} and
\citet{kohler2015}. We assumed a Salpeter initial mass function
\citep{salpeter55} as the initial-mass prior, adopted the distribution
of rotational velocities of O-type stars in 30~Dor from \citet{oscar}
as the initial rotational velocity prior (and assuming that their
rotational axes are randomly orientated in space), and adopted a uniform prior for stellar ages.
Tables~\ref{tab2_HHe} and \ref{tab2_HHeN} summarise this second set of
stellar parameters, in which the column entries are (1) VFTS identifier;
(2) spectral classification; (3) effective temperature; (4)
rotation-corrected gravity; (5) absolute visual magnitude $M_V$;
(6-11) radii, luminosities, and spectroscopic masses and their
corresponding formal errors; (12, 13) evolutionary masses calculated
from the Kiel diagram, using evolutionary tracks with an initial
rotational velocity of 171~\kms\ (see Sect.~\ref{sect:vsini}) and
their corresponding errors; (14, 15) the same as (12, 13), but using the
H--R diagram; and (16) evolutionary masses derived using
\textsc{bonnsai} and their errors.

\subsection{Cautionary remarks}\label{caution}

During the analysis of our sample we found various cases for which we
could only provide upper or lower limits. We also detected a few stars
for which a quantitative spectroscopic analysis based exclusively on the
H and He lines could not provide reliable estimates of the effective
temperature (see Sect.~\ref{analisis_nitro}). A summary of the number of stars with problematic
analyses is shown in Table~\ref{tab:caution}.

\begin{table*}[t]
\centering
\caption{Number of stars in our analysis with problematic  $\chi^2$ 
distributions of wind strength (\logq), effective temperature (\Teff), helium abundance (Y(He)), 
and surface gravity (\grav) caused by degeneracies or because
they reach the boundaries of the grid. \label{tab:caution}}
\begin{tabular}{lcl}
\hline\hline
\multicolumn{3}{c}{} \\ [-2 ex]
Parameter & \# stars & Comments \\
\hline
\multicolumn{3}{c}{} \\ [-2 ex]
\logq & 69 & Limitations from the H$\alpha$ and \ion{He}{ii}\,$\lambda$4686 
diagnostics \\
  \Teff & 20 & Strong nebular contamination and/or weak or nonexisting 
\ion{He}{i} lines, improved by HHeN diagnostics \\
  Y(He) & 14 & He abundance too low -- possible undetected binarity \\
  \grav & 10 & Gravity too high -- possible undetected binarity \\
\hline
\end{tabular}
\end{table*}

\subsubsection{Stellar contamination in the fibre aperture} \label{caution:contamination}

Known spectroscopic binaries were omitted from our sample, but as
illustrated in Fig.~\ref{hst_images_muestra}, some of the VFTS spectra
are contaminated by light from nearby companions on the sky (which are
not necessarily in bound binary or multiple systems). We therefore need
to consider the effect of this on the parameters estimated by our
analysis (\Teff, \grav, Y(He), etc.) and/or the parameters inferred from
potentially erroneous absolute magnitudes ($R, L, M$).

When a certain shift between both components is present, Balmer lines may appear too broadened and therefore a higher surface gravity will be derived, which also affects \Teff. Additionally, 
the dilution effect in a composite spectrum could weaken the He lines, thus leading to lower helium abundances.  

The spectroscopic mass ($M_{\rm sp}$) depends on the stellar radius. 
When the photometry of the target star is contaminated by a nearby
companion, the derived radii can therefore be overestimated by up to
$\sim$\,35\,\% (in the worst-case scenario of two equally bright
components), leading to an overestimate of $M_{\rm sp}$ of
$\sim$\,70\%. Similarly, such contamination in the fibre would also
lead to overestimated evolutionary masses through a higher inferred
luminosity, which moves the star to higher masses in the H--R diagram.

\subsubsection{Nebular contamination in the fibre aperture}

Extreme cases of nebular contamination were omitted from our sample
by excluding the BBB stars from \citet{walborn13}.  Still,
$\sim$70\% of the spectra in our sample display relatively strong
nebular contamination in H$\alpha$ and some degree of contamination
in the \ion{He}{i} lines. This may affect the determination of stellar
parameters and must be handled with care. The most critical parameters
that can be affected are \Teff\ and \logq\, because the He ionisation
balance and H$\alpha$ line are the main diagnostics for
these parameters, respectively; \grav\ and Y(He) may also be affected to a lesser
extent.

We checked each case individually and found that the situation is not
critical for most stars and only results in larger uncertainties.
However, in 11 objects (with spectral types later than O4) the
\ion{He}{i} lines should be sufficiently strong to be used in the HHe
analysis, but a strong nebular contamination rendered
them unusable. These cases were therefore analysed using HHeN
diagnostics (see Sect.~\ref{analisis_nitro}, with results given in
Table~\ref{tab1_HHeN}).

\subsubsection{Earliest spectral types}\label{early_types}

\begin{figure}[t]
\centering
\includegraphics[scale=0.46,trim=5mm 0mm 0mm 0mm, clip]{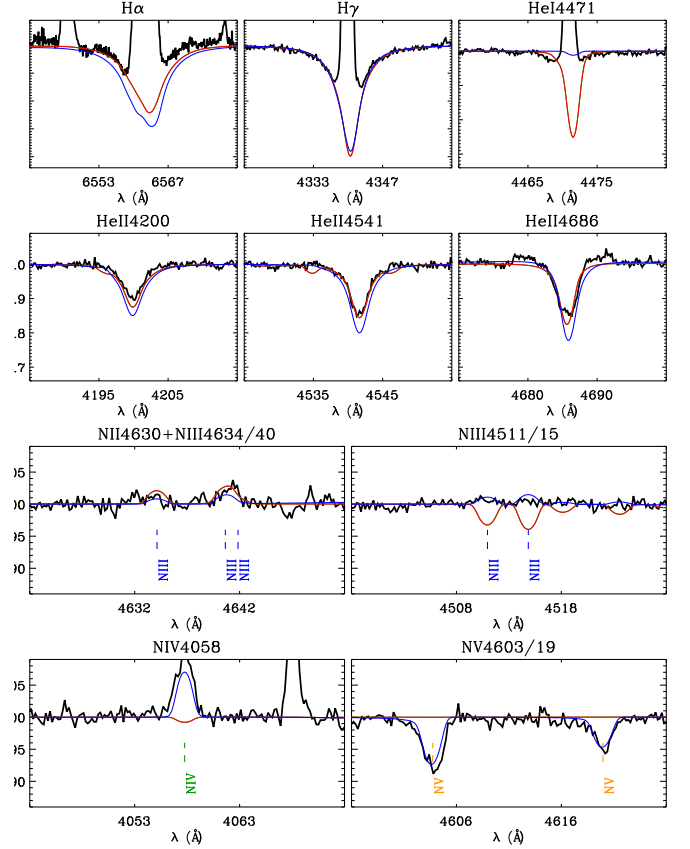}  
\caption{\label{621n} Model fits to the spectrum of VFTS\,621, with
  the best fit from the HHe (\Teff\,$=$\,36\,kK, \grav\,$=$\,3.8)
  and HHeN (\Teff\,$=$\,54\,kK, \grav\,$=$\,4.2) analyses shown
  in red and blue, respectively.  The \logq\ as well as the\  helium and nitrogen
  abundances are the same in both models.  These single-star models
  both suffer deficiencies that are suggestive of a composite observed
  spectrum, as seen in {\em HST} images (Fig.~\ref{hst_images_muestra}).}
\end{figure}
For nine stars with spectral types earlier than O4, the \ion{He}{i}
lines were too weak or even nonexistent (e.g. VFTS\,072).  The
$\chi^2$ distribution for \Teff\ for these stars showed degeneracies,
therefore we resorted to analysing them using the nitrogen lines described in
Sect.~\ref{analisis_nitro}. Results for these stars are presented in
Table~\ref{tab1_HHeN}.

In the course of this analysis we noted that \ion{He}{i} lines are
present in the spectra of VFTS\,468, 506 and 621, which is unexpected
given their classification as O2 type. This qualitative
(morphological) argument is reinforced by the results of our HHeN
analysis. Inclusion of the diagnostic \ion{N}{iii/iv/v} lines
confirmed our suspicion that the HHe analysis returned estimates
of \Teff\ (and hence \grav) that were too low, but the
best-fitting HHeN models do not predict \ion{He}{i} intensities as
strong as in the data (particularly for \ion{He}{i} $\lambda$4471).

In these three cases we also noted that the broadening required to
fit the observed \ion{He}{i} lines is much larger than the broadening needed to
fit the nitrogen lines. An example of this is shown Fig.~\ref{621n},
where the best-fitting HHe and HHeN models for VFTS\,621 are shown in
red and blue, respectively. The HHe fit predicts \ion{N}{iv/v} lines
that are too weak (or nonexistent) and \ion{N}{iii} absorption that
is too strong, while the HHeN model predicts \ion{He}{i}$\lambda$4471
absorption that is too weak, together with \ion{He}{ii} lines that are
too strong. 

These inconsistencies could be related to multiplicity and/or nitrogen
peculiarities: VFTS\,621 was noted by \citet{walborn13} as a visual
multiple with three components \citep[VM3, see also][]{walborn2002},
VFTS\,468 is classified as O2\,V((f*))\,$+$\,OB and noted as a visual
multiple with four components \citep{walborn13}, and VFTS\,506 is
classified as an ON2 star \citep[and classified as a (small-amplitude)
single-lined binary by][]{sana13}.  These examples of probably composite spectra
appear similar to the case of Sk~183 in the Small Magellanic Cloud
\citep{evans12}. Sk~183 was initially classified as an O3 dwarf
according to its nitrogen lines, but He absorption in its spectrum
suggested a later type.  Spectral analysis by \citet{evans12} found
that the \ion{He}{i} and \ion{He}{ii} lines could not be fitted
simultaneously using a single model, and that better fits could be
achieved using a composite (O\,$+$\,B) model.  Similar inconsistencies
were also noted by \cite{rivero12ii,rivero12iii} in their HHeN
analysis of O-type stars in the Magellanic Clouds. Particularly for their ON2-type giants, they failed to
simultaneously fit the observed \ion{He}{i}\,$\lambda$\,4471 and
nitrogen lines using both the \fastwind\ and \textsc{cmfgen}
\citep{hm98} codes.

These three O2-type stars show the importance of including information
on possible close-by companions (that are not necessarily physically
bound) to elucidate the difficulties found in the analysis of some stars,
and the apparent bimodality found in the literature for the derived
\Teff\ at the earliest spectral types (see
Sect.~\ref{sect:calibrations}).  In this context of interpreting
ground-based spectroscopy, an increasingly important input is results
from surveys of massive stars at high spatial resolution
\citep[e.g.][]{mason1998,mason2009,sabbi2012,sabbi2016,sana14,aldoretta}.
Also promising is the use of spatially resolved spectroscopy to separate different components \citep[][]{w99,w02,sota}.
The fourth O2 star in our sample, VFTS~072, appears to be a simpler
case as no \ion{He}{i}$\lambda$4471 is present. The derived
temperature is higher when including the nitrogen lines, but no
signatures of a companion are present in the spectrum. Unfortunately,
we do not have the corresponding {\em HST} image to investigate if
this star appears genuinely isolated.

\subsubsection{Thin winds}

As we discussed in Paper XIII, for some stars we encountered
degeneracies in the $\chi^2$ distributions of \logq\ due to weak
winds, which render the H$\alpha$ and \ion{He}{ii}\,$\lambda$4686
lines insensitive to changes in this wind parameter. For
$\sim$\,66\% of the sample we were therefore only able to provide upper limits for
\logq\ (see col.~10 in Table~\ref{tab1_HHe}).

\subsubsection{Helium and gravity}\label{caution:heliumgrav}

We found 14 cases (4 O~Vz, 9 O~V, and 1 O~IV) for which we
could provide only upper limits for the helium abundance (see col.~9
in Table~\ref{tab1_HHe}), and 10 cases (5 with low He
abundance as well) with \grav\,$>$\,4.2 (marked with asterisks in col.~6
of Table~\ref{tab1_HHe}) that are relatively high compared to
predictions by theoretical evolutionary models.  These features could
indicate binarity (see Sect.~\ref{caution:contamination}), which would affect the position of these stars in
the Kiel or H--R diagrams, thus modifying the estimated spectroscopic
and evolutionary masses.

\section{General properties of the sample}

\subsection{Stars in the Kiel and H--R diagrams}\label{sect:diagrams}

\begin{figure*}
\centering
\subfloat{\includegraphics[scale=0.5,angle=0]{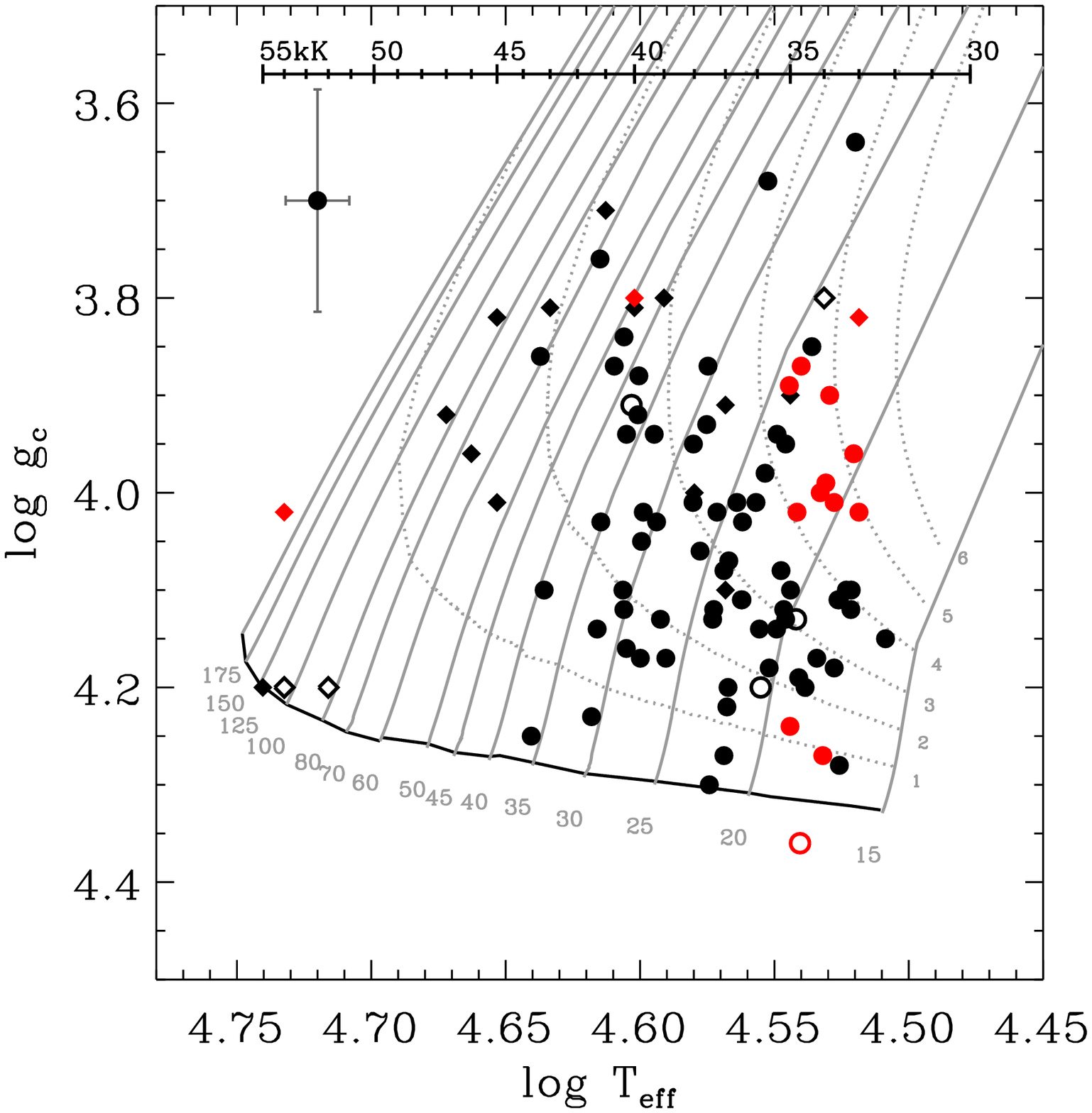}\label{gtd_all}} ~~ 
\subfloat{\includegraphics[scale=0.5,angle=0]{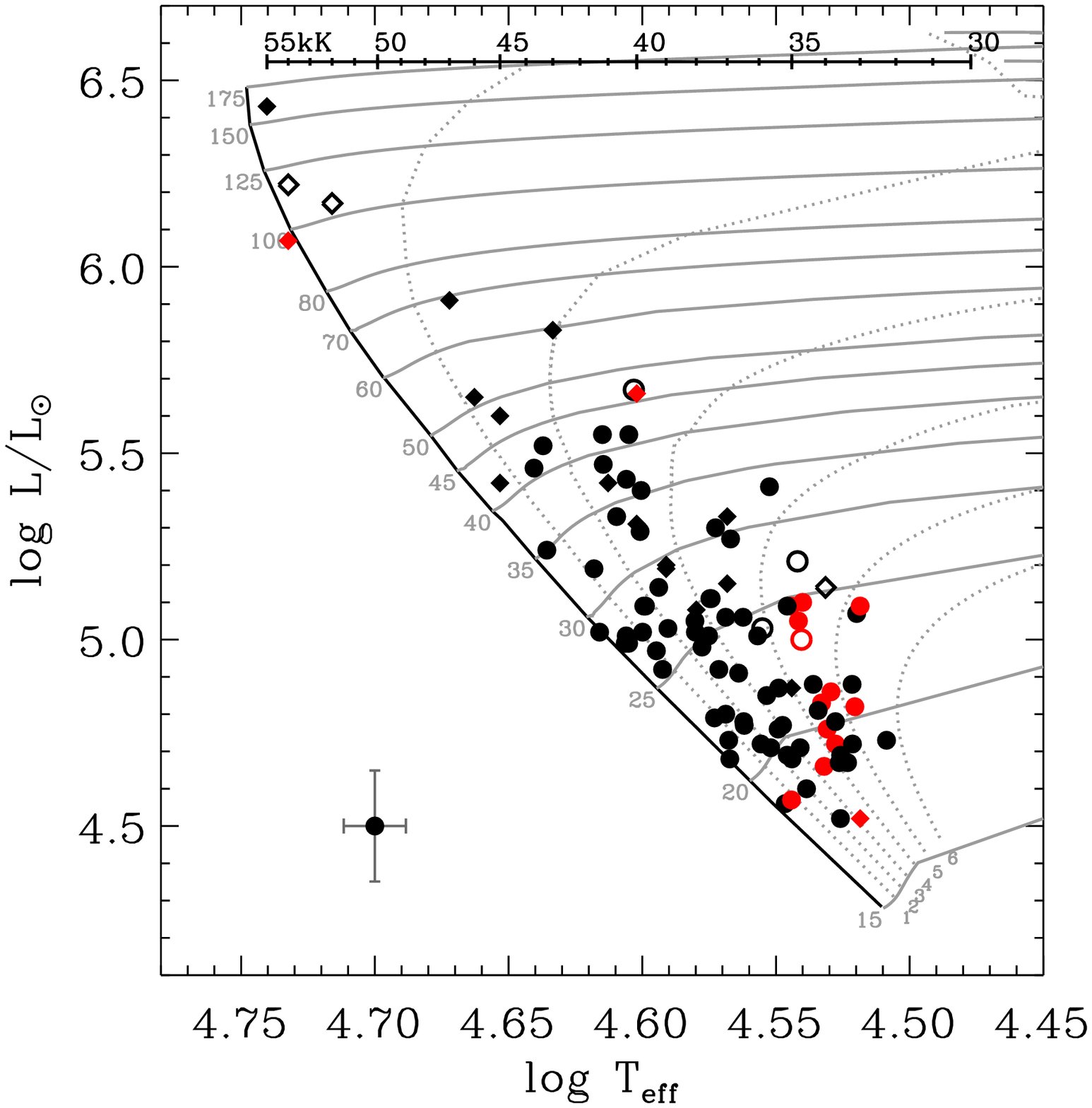}\label{hrd_all}}  
\caption{Kiel (left) and H--R (right) diagrams for our sample.  Class V
  objects are plotted in black, those with other luminosity classes
  (IV, III-IV, and V-III) in red. Stars analysed using HHeN
  lines are plotted as diamonds, while those with (stellar)
  contamination in the fibre aperture (see
  Sect.~\ref{caution:contamination}) are plotted as open symbols.
  Evolutionary tracks and isochrones for the LMC are from \cite{brott}
  and \cite{kohler2015}, with an initial rotational velocity of 171
  \kms\ (see Sect.~\ref{sect:vsini}). The zero-age main-sequence (ZAMS) is indicated in the two plots by the bold
  black line, and typical uncertainties are indicated in the upper and lower corner, respectively.
  \label{diagramas_glt}}
\end{figure*}

To investigate the evolutionary state of our sample of stars, we plot
them in the Kiel (\gravc\,--\,log\,\Teff)  and H--R diagrams, as shown in
Fig.~\ref{diagramas_glt} (in which we have highlighted the different
luminosity classes)\footnote{Five stars for which photometry was
unavailable are not included.}.
Evolutionary tracks and isochrones by \cite{kohler2015} and \cite{brott} for an LMC-like metallicity are shown.

We remark that a reliable determination of effective temperatures is
critical when studying the position of the stars with the earliest
spectral types in the H--R and Kiel diagrams, as this has a direct
effect on the determination of their evolutionary masses and
luminosities. For example, the inferred evolutionary masses can range
from 70~M$_{\odot}$ to 150~M$_{\odot}$ for effective temperatures
spanning 50\,000~K to 55\,000~K.  Because of this and the apparently
composite nature of three of our O2-type stars (see
Sect.~\ref{early_types}), we treat them with particular care in the
following sections.

\subsubsection{Kiel diagram}

The rotation-corrected gravities of our sample mostly span
3.8\,$<$\,\gravc\,$<$\,4.2. We note that relatively few stars
lie below the 1\,Myr isochrone, which might be a consequence of the
\fastwind/\textsc{cmfgen} discrepancy noted by \cite{massey13}. These
authors found that gravities derived with {\sc cmfgen} were typically
0.1\,dex higher than those obtained using \fastwind. We are therefore
careful when considering masses inferred from the Kiel diagram, denoting the evolutionary state `in terms of gravity'. That said, we note
that if the difference between the codes were the sole cause, then
results from {\sc cmfgen} would be expected to place at least six stars
below the zero-age main-sequence (ZAMS).

Except for the O2 stars, our Kiel diagram has a dearth of objects
close to the theoretical ZAMS for masses above
$\sim$\,35\,M$_{\odot}$. A similar result was found by
\cite{castro2014} for a sample of 575 Galactic OB stars. 
The authors pointed out that such stars could
still be embedded in their birth clouds, which would hamper their
detection at optical wavelengths \citep{yorke86}. 
Alternatively, if this
absence of stars with masses above $\sim$\,35\,M$_{\odot}$ were real,
this empirical result may present an important challenge to theories
of massive-star formation.

The class IV stars are mostly concentrated at
log\,\Teff\,$\lesssim$\,4.55, with 3.8\,$<$\,\gravc\,$<$\,4.0. As the
intermediate class between dwarfs and giants, the O\,IV stars are
expected to be more distant from the ZAMS (in terms of gravity) than
the class V objects. Most of the class IV stars define an upper
envelope in gravity for stars with 4.52\,$<$\,log\,\Teff\,$<$\,4.55,
confirming the expectation of a slightly more evolved state (in terms
of their gravities).  However, there are three stars (VFTS\,303, 505,
and 710) with 4.53\,$<$\,log\,\Teff\,$<$\,4.55 with gravities that
appear too high (\gravc\,$>$\,4.2) for their luminosity class.
Possible binarity or composite spectra could explain these results:
either by confirmed (VFTS\,303) or possible (VFTS\,505) contamination
in the fibre aperture, or suspected spectroscopic binarity (VFTS\,710). We highlight that \break \citet{oscar17} have found
similar cases in their analysis of the giants and supergiants from the
VFTS, where several late-type class II and III stars have estimated
gravities of \grav\,$=$\,4.0 to 4.5 (more in line with those
expected of dwarfs).  Intricacies in the luminosity classification may
have played a role in these cases, where the \ion{Si}{iv} 
absorption is weaker than expected for giants, pointing to a dwarf or
subgiant classification rather than a giant classification \citep[see
also][]{walborn13}.

\subsubsection{H--R diagram}

As shown in Fig.~\ref{diagramas_glt}, most of our sample have
luminosities ranging between \logl\,$=$\,4.5 and 5.7.  Most of the
O\,IV stars have luminosities in the range
$\sim$\,4.5\,$<$\,\logl\,$<$\,5.1, with inferred ages of
$\sim$3-5\,Myr, somewhat more evolved than the dwarf subsample, as
expected. As in the Kiel diagram, VFTS\,505 and 710 appear to
be too young
(on or below the ZAMS), although VFTS\,303 is located within the main
group of class IV stars. Interestingly, a larger number of stars are
located between the ZAMS and the 1~Myr isochrone in the H--R diagram,
which is less affected by the possible underestimation of the
gravities discussed above.

\subsection{Spectral calibrations}\label{sect:calibrations}

Spectral calibrations are useful tools to characterise the physical
properties of massive stars, with applications in several
astrophysical fields, such as studies of \ion{H}{ii} regions and
population synthesis. As the largest sample to date (and with complete
coverage of spectral types), the VFTS results offer a unique
opportunity to characterise the \Teff\ scale of O dwarfs in the LMC.

\subsubsection{\Teff\ and \grav\ calibrations}

\begin{figure}[t]
\centering
\subfloat{\includegraphics[scale=0.3,angle=90,trim=22mm 0mm 0mm 0mm,clip]{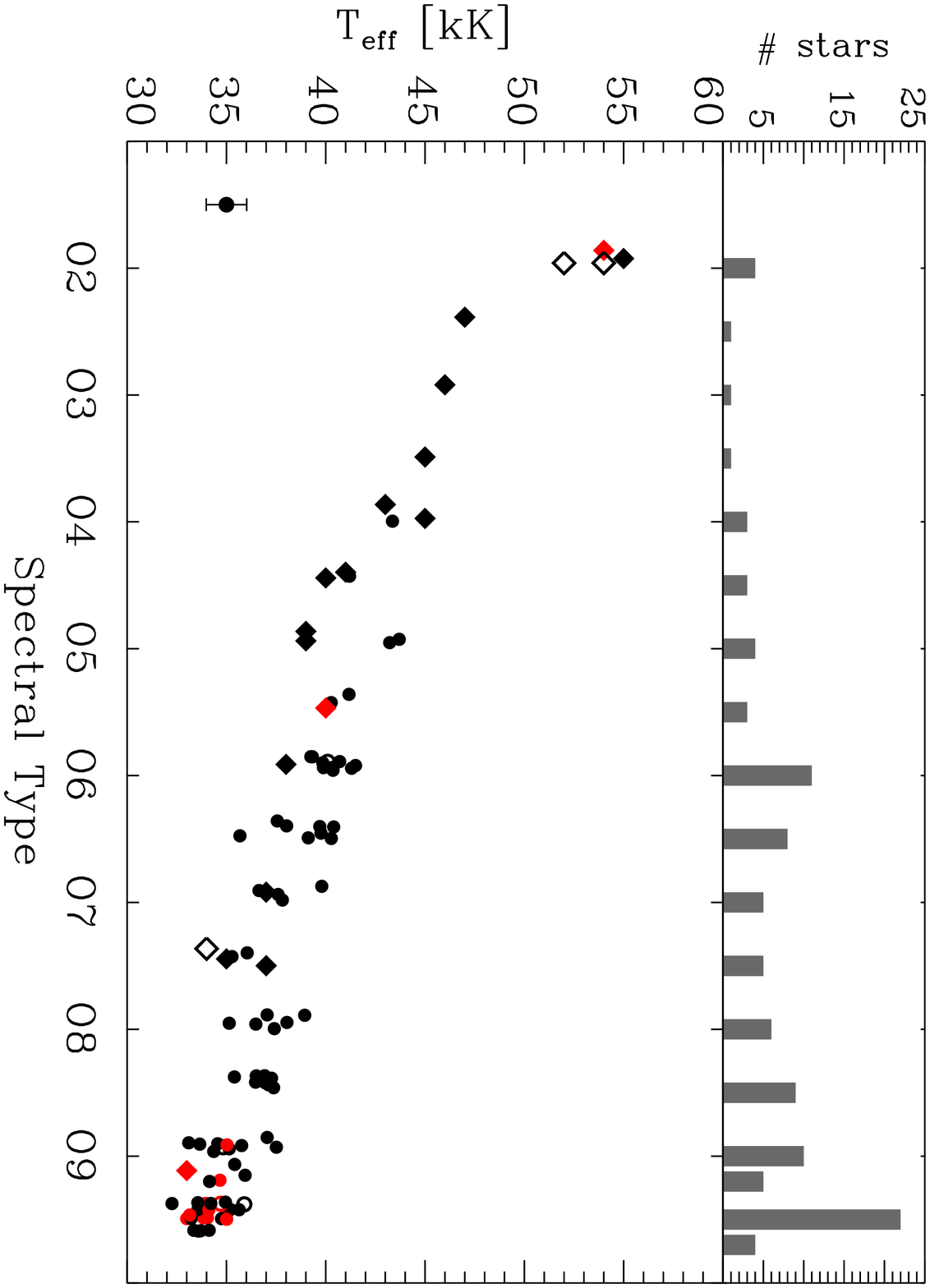}}\vspace{-0.2cm}\\
\subfloat{\includegraphics[scale=0.3,angle=90,trim=0mm 0mm 0mm 0mm,clip]{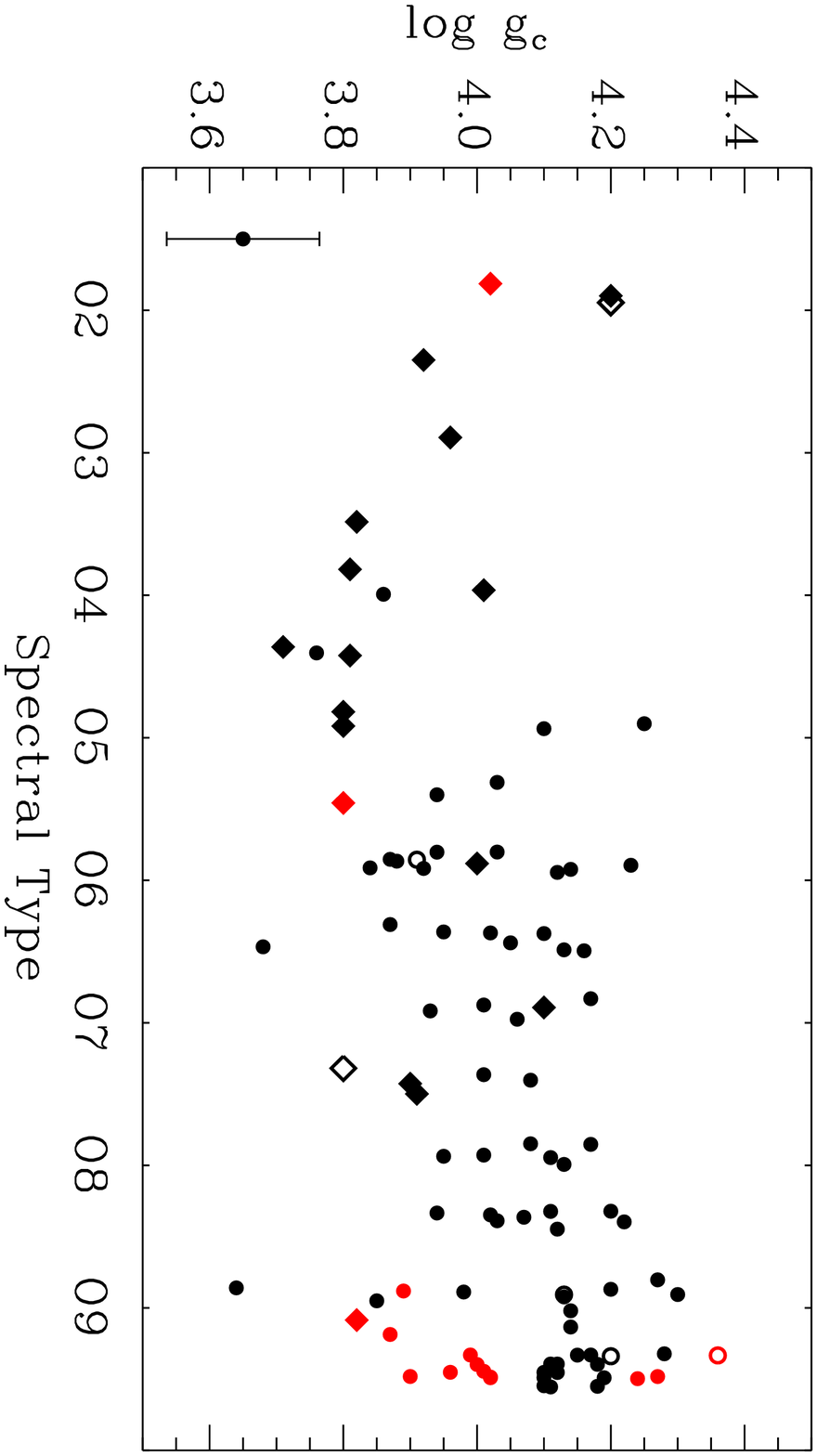}} 
\caption{\label{calib_spt_teff_y_g} Estimated effective temperatures
  (\Teff) and rotation-corrected gravities (\gravc) for our sample as
  a function of spectral subtype (using the same symbols as in
  Fig.~\ref{diagramas_glt}); small shifts have been added to the
  abscissae of each star to avoid overlap. The upper panel shows the
  number of stars per subtype.}
\end{figure}

To construct calibrations from our sample, we plot the \Teff\
and \gravc\ estimates in Fig.~\ref{calib_spt_teff_y_g} as a function of
spectral subtype, including results from both the HHe and HHeN
analyses (plotted as circles and diamonds, respectively, and red
symbols for stars with luminosity classes IV, III-IV, and
V-III).  Our sample covers the whole range of O subtypes, from O2 to
O9.7, but has two peaks (at O6 and O9.5) and is dominated by stars
with types of O6 or later.  Moreover, we only have one
object for each subtype from O2.5 to O3.5.

The effective temperature decreases from $\sim$\,55\,000~K for O2
stars to $\sim$\,34\,000~K by O9.7. There is a linear trend
between O2.5 and O9.7, even taking into account the stars analysed
with HHeN lines (diamonds in the figure), but the four O2-type stars
seem to break this trend.

There is a notable dispersion in \Teff\ and \grav\ for almost every
spectral subtype. This effect was discussed by \citet{ssd_letter14},
who compared the \Teff\ and \grav\ scales for populations of O dwarfs
in the Galaxy \citep[analysed within the IACOB project,
see][]{ssd13iacob} and the results for the LMC stars presented in
Paper~XIII (which are also included in the current study).
\citeauthor{ssd_letter14} warned against the use of spectral
calibrations based on small samples of O-type dwarfs, as a
relatively evolved O dwarf population is expected to show a wide range
of gravities because of the different evolution of early- and late-type O
dwarfs.

\subsubsection{Comparison with  \citeauthor{rivero12iii}}

\begin{figure}[t]
\centering
 
\includegraphics[scale=0.3,angle=90]{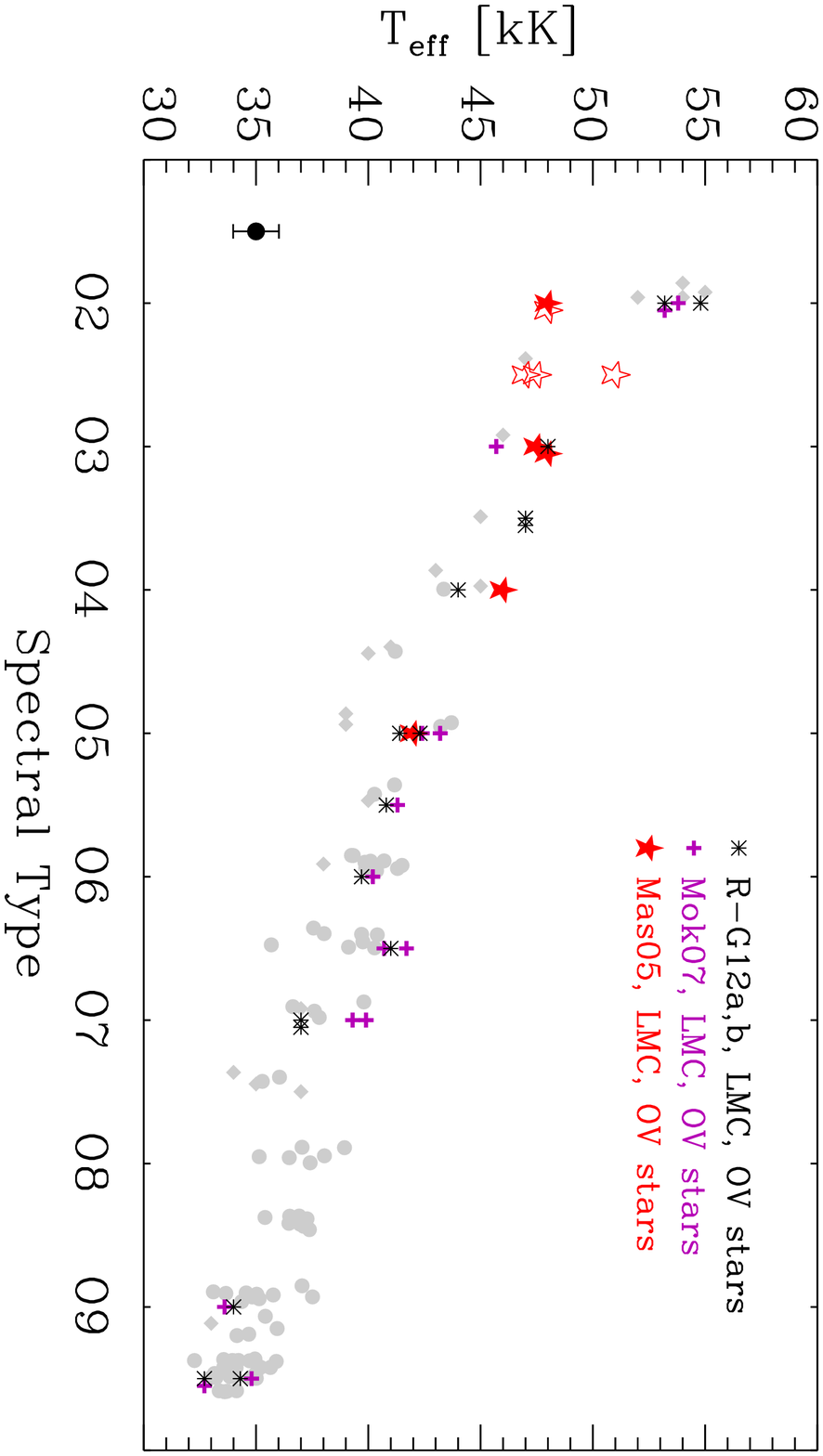}\\
\caption{\label{calib_spt_teff_y_g_datosLMC} Effective-temperature
  scale for our sample of stars (in grey) compared with results from
  \cite{rivero12i,rivero12ii,rivero12iii}, \cite{mokiem2007LMC}, and
  \cite{massey05}.}
\end{figure}

The most complete \Teff\ calibration for LMC O-type stars before
this study is the work of \citet{rivero12iii,rivero12ii}, who used
HHeN diagnostics to analyse optical spectroscopy of 25 stars
(including 16 dwarfs); our temperatures are compared to their results
in Fig.~\ref{calib_spt_teff_y_g_datosLMC}, with generally good
agreement between the two distributions. There is one star in
  common between the two studies, namely VFTS\,072 (\,$=$\,BI\,253).
  We obtain a similar temperature (\Teff\,$=$\,54\,000\,K vs. their
  estimate of 54\,800\,K), while our gravity is slightly lower
  (\grav\,$=$\,4.00 vs. 4.18). When we compare the spectra, a small difference in the wings of the Balmer lines in the
  4000-5000\,\AA\ range is visible (likely an artefact of the continuum
  normalisation, although we note that the H$\alpha$ profiles are very
  similar), which most likely relates to differences in spectral resolution
  (the ``older'' data were obtained at $R$\,$=$\,40\,000) and different
  levels of nebular contamination in the line cores (fibres vs. slit
  spectroscopy). 
  
  For a more quantitative comparison, a linear fit to
our data (excluding the O2 stars) is given by

\begin{figure}[t]
 \centering
\includegraphics[scale=0.3,angle=90]{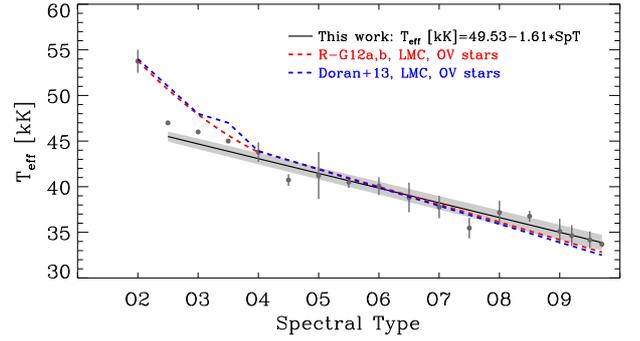}
 \caption{Comparison of our \Teff\ calibration for O dwarfs in
   the LMC with those derived by \cite{rivero12iii,rivero12ii} and
   adopted by \cite{doran13}. Grey dots and
   bars represent the mean value and standard deviation of our \Teff\ estimates for
   each spectral type, with our linear fit shown by the black line
   (and the grey zones indicating the associated uncertainties).
   \label{spt_teff_nuestroajuste}}
\end{figure}

\begin{center}
\begin{equation}
T_{\rm{eff}}\,[{\rm kK}] = 49.75(\pm 0.54) - 1.64(\pm 0.07) \times {\rm SpT},
\end{equation}
\end{center}

where SpT is the number corresponding to the spectral subtype. This
calibration is shown in Fig.~\ref{spt_teff_nuestroajuste}, compared to
the fit from \citeauthor{rivero12iii}, who divided their calibration
into quadratic and linear components for spectral types earlier and
later than O4, respectively. We find excellent agreement between both scales for types
later than O4, while \Teff\ values for O2.5-O4 stars
are slightly hotter than our fit (albeit lower than their quadratic
fit), and our results for the O2 stars agree with their calibration.

As commented above, the small number of stars observed at the earliest
types, in combination with the intrinsic scatter of a \Teff\,--\,SpT
calibration and difficulties of both the observations and analysis, do
not allow us to reach a firm conclusion on the need for a change in
slope of the calibration in the O2.5-O4 range. A larger sample of
early O-type stars, including information on the possible
binary or composite nature of their spectra, is needed to shed more light
on the temperature scale. This should be possible using data from a
recent {\em HST} program by \citet{crowther2016} to obtain ultraviolet
and optical spectroscopy of the massive stars in R136, the central
cluster of 30~Dor. There are a large number of O2-3~V stars in R136,
and analysis of their optical spectroscopy from {\em HST} should provide a firmer
understanding of the temperatures at the earliest spectral types.

\subsubsection{Comparison with other calibrations}

There are other results in the literature for O dwarfs in the LMC,
although they do not have such thorough coverage in terms of spectral
type. \cite{massey04,massey05,massey09} performed HHe \fastwind\
analyses of optical spectra from 19 stars (9 \ov) in the LMC,
together with 35 stars in the Milky Way and Small Magellanic Cloud
(SMC). Results from \citeauthor{massey09} for LMC dwarfs are shown in
red in Fig.~\ref{calib_spt_teff_y_g_datosLMC}. These were limited to
spectral types earlier than O5, and include three stars with uncertain
classifications of O2-3.5~V and one O2 star with only a lower limit on
its temperature (due to the absence of \ion{He}{i} lines); these four
stars are plotted with open symbols in the figure. The results from
\citeauthor{massey04} are generally consistent with our distribution
(given the uncertainties in classification for some of their stars).
The authors found lower temperatures and gravities for the two O2 stars in
their sample, but this could be related to the effects discussed above
in Sect.~\ref{early_types}.

In Fig.~\ref{calib_spt_teff_y_g_datosLMC} we also include the results
from a HHe analysis with \fastwind\ of 13 dwarfs in the LMC (spanning
O2 to O9.5 types) by \cite{mokiem2007LMC}. Again, subject to problems
for the O2-type stars with only a HHe analysis, their results agree
reasonably well with ours.

\citet{doran13} adopted a \Teff\ scale for the O dwarfs in 30~Dor
based on the calibration from \citet{martins05gal}, revised upwards by
1000\,K to adapt it for the metallicity of the LMC. For spectral types
earlier than O3.5, they used individual estimates from
\cite{rivero12iii} and \cite{doran11}. The scale from
\citeauthor{doran13} is practically identical to that from
\citet{rivero12iii}, with the exception of the O3.5 stars (with the
former $\sim$1400\,K hotter), as shown in Fig.~\ref{spt_teff_nuestroajuste}.

Lastly, we note that eight of our O-type dwarfs\footnote{VFTS\,072,
  169, 216, 468, 506, 621, 755, and 797.} from the HHeN analysis
overlap with the VFTS sample analysed using {\sc cmfgen} by
\citet{besten14} to understand the wind properties of the luminous O
and Wolf--Rayet stars. The estimated temperatures and luminosities
from \citeauthor{besten14} are in reasonable agreement for the four
stars later than O2 (with differences of $\Delta$\Teff\,$\sim$1000\,K
and $\Delta$\logl\,$\sim$\,0.01, i.e. compatible within the estimated
uncertainties). For the four O2-type stars (VFTS\,072, 468, 506, and
621) we obtain effective temperatures up to $\sim$14\,\% higher, with
luminosities of 1 to 3\,\% higher. However, as their main objective
was to investigate the wind properties, \citeauthor{besten14} adopted
\grav\,=\,4.0 in their analyses, which affects the ionisation
equilibrium and might explain the differences in \Teff\ for VFTS\,468,
506 and 621, for which we derive \grav\,=\,4.2. The composite nature
of the spectra discussed in Sect.~\ref{early_types} and the different
analysis methods probably also contribute to the different \Teff\
estimates.

\section{Discussion}

\subsection{Rotation and helium abundance in an evolutionary 
context}\label{sect:vsini}

\subsubsection{\vsini\ distribution}

\begin{figure}[]
 \centering
\includegraphics[angle=90,scale=0.7,trim=92mm 120mm 0mm 0mm, clip]{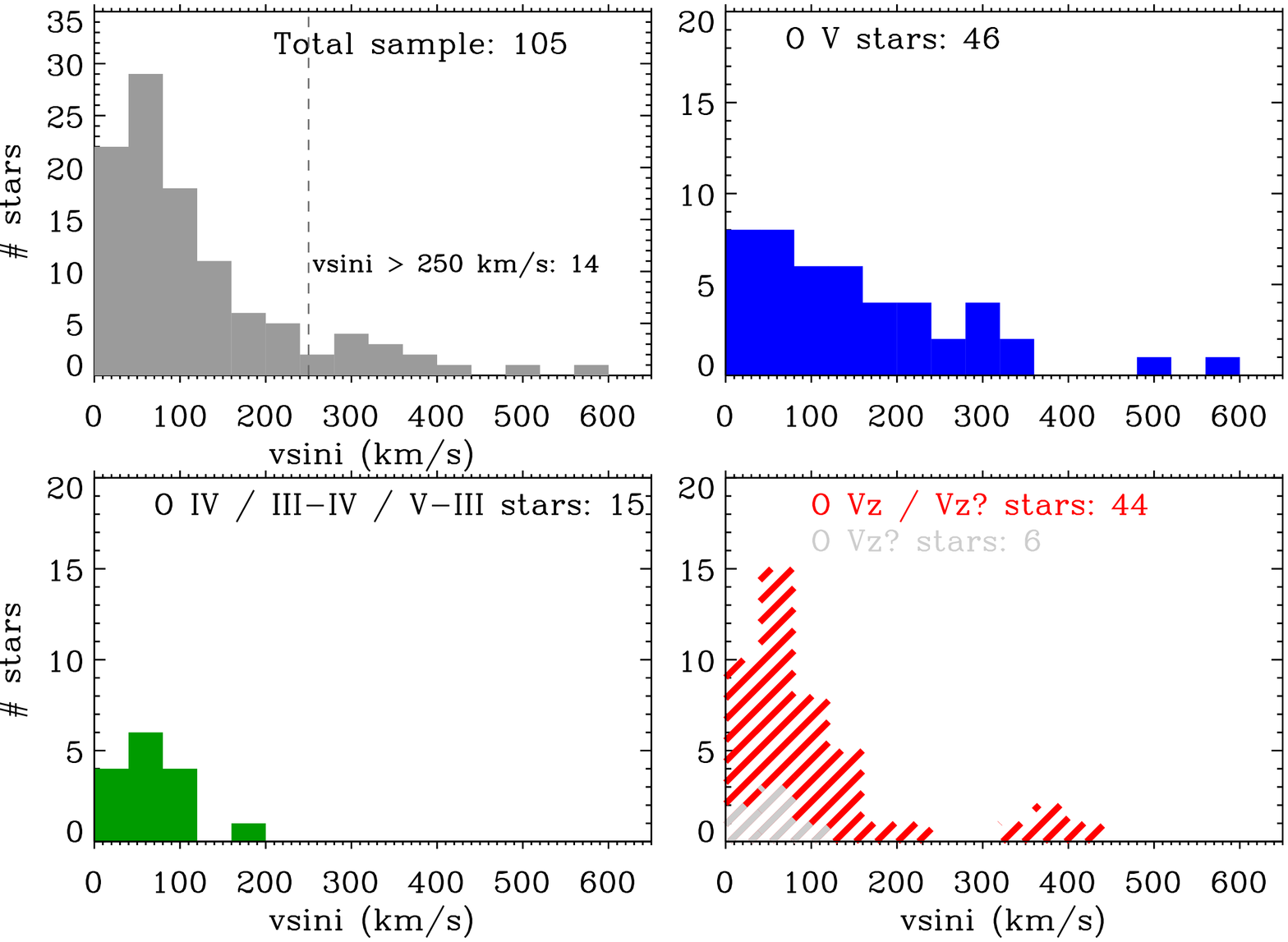}
 \caption{Projected rotational velocities for our sample of O-type
   dwarfs.  Fourteen stars have \vsini\,$>$\,250\,\kms\ (with the
   velocity threshold indicated by the vertical dashed
   line).\label{distrib_vsini_dwarfs}}
\end{figure}

The \vsini\ distribution for our O dwarfs is presented in
Fig.~\ref{distrib_vsini_dwarfs}. It peaks in the 40\,--\,80~\kms\ bin,
with most stars found at \vsini\,$\lesssim$\,150\,\kms.  Given the
limitations of the strategy applied to determine \vsini\ in the O-type
sample \citep[see][]{oscar}, the first two bins in
Fig.~\ref{distrib_vsini_dwarfs} must be considered with care.  Given
the lack of useful metal lines in the observed spectral range, using \ion{He}{i} lines (or even
\ion{He}{ii} in some critical cases) to estimate \vsini, combined with
the presence of nebular contamination and/or low signal-to-noise
ratio,
leads to large relative uncertainties in estimates below
$\sim$\,100\,\kms.

The tail of the distribution extends to high rotational velocities.
The star at $\sim$600\,\kms\ is VFTS\,285, one of the fastest rotators known
to date (see \citealt{w11,walborn13}).  We find a low but noteworthy
peak over the 240\,--\,440~\kms\ range \citep[see also][]{oscar},
which could originate from the effects of mergers and mass transfer
(via Roche-lobe overflow) in binary evolution \citep{demink13,demink14}.

\begin{figure*}[t]
 \centering
\includegraphics[trim=0mm 0mm 0mm 0mm,clip,scale=0.65,angle=90]{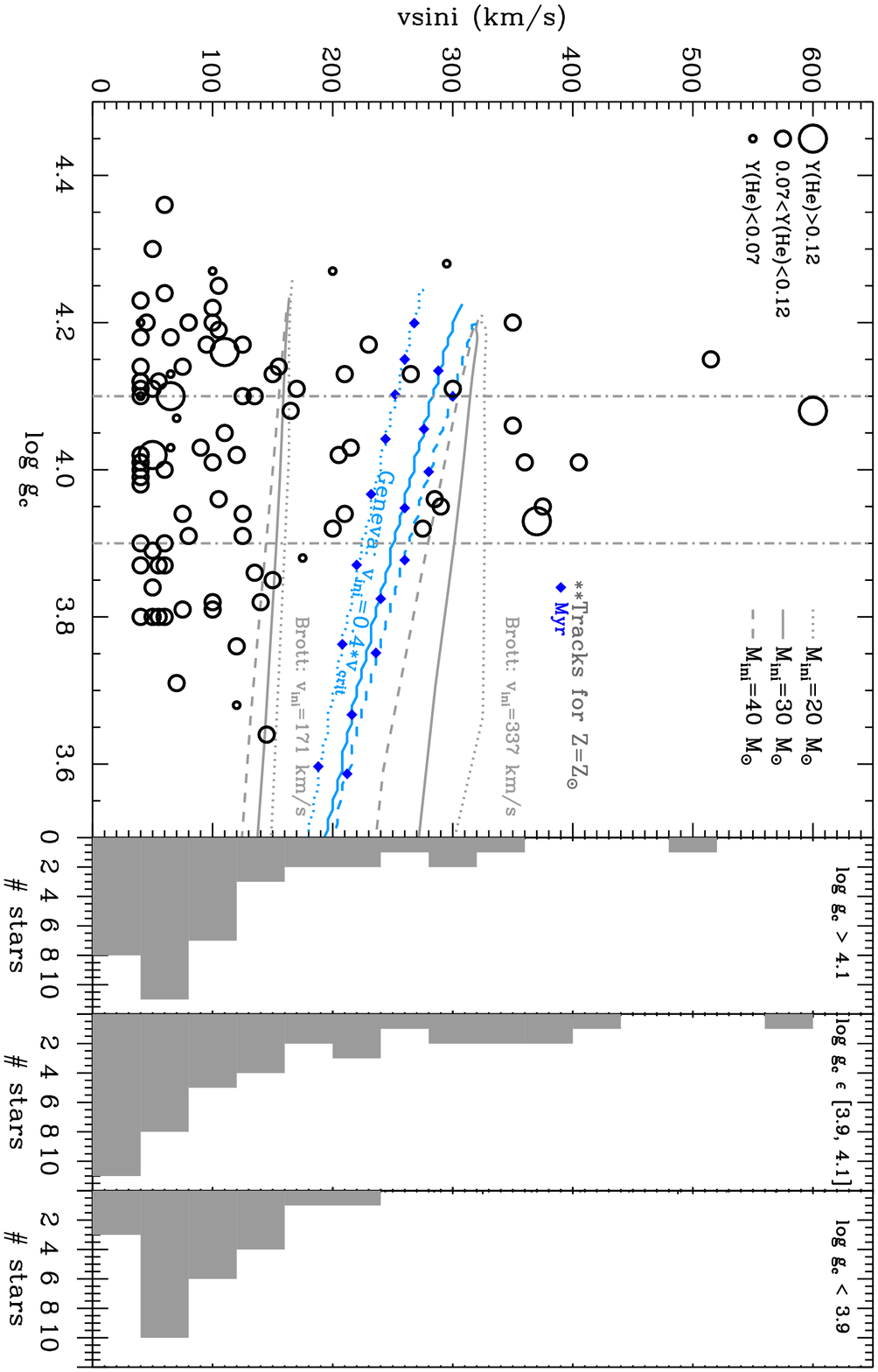} 
 \caption{\textit{Left:} Projected rotational velocities (\vsini) vs.
   rotation-corrected gravities (\gravc) for our sample, grouped
   according to helium abundance. Theoretical predictions for the LMC
   from \citet{brott} are shown in grey (for three initial masses and
   two initial rotational velocities as indicated).  Predictions for
   solar metallicity and an initial $v_{\rm ini}$ of 40\% the critical
   velocity from \citet{ekstrom2012} are shown in blue.
   \textit{Right:} \vsini\ distributions for three ranges of
   gravity: (1) low (\gravc\,$<$\,3.9), (2) intermediate
   (3.9\,$\leq$\,\gravc\,$\leq$\,4.1), and (3) high
   (\gravc\,$>$\,4.1).
   \label{logg_vs_vsini}}
\end{figure*}

\subsubsection{Rotation vs. gravity}\label{sect:rot_vs_grav}

Our projected rotational velocities were investigated as a function of
rotation-corrected gravity, as shown in Fig.~\ref{logg_vs_vsini}. Overlaid on the figure
are theoretical predictions from \citet{brott} for three initial
stellar masses (20, 30 and 40\,$M_\odot$) and two initial rotational
velocities (171 and 337\,\kms).

The distribution of stars in Fig.~\ref{logg_vs_vsini} can be roughly
separated into three groups when compared with the evolutionary
tracks. Firstly, most of the stars of the sample are found at lower
velocities than the $v_{\rm ini}$\,$=$\,171\,\kms\ tracks. Taking into
account that the \vsini\ values are lower limits on $v_{rot}$ due to projection
effects, the use of models with $v_{\rm ini}$\,$=$\,171\,\kms\ to
study our sample in an evolutionary context seems a reasonable
approach. Secondly, evolutionary tracks with $v_{\rm
  ini}$\,=\,337\,\kms\ can explain the stars above the 171\,\kms\
tracks, which includes most of the fast rotators. Finally, there is a
third group of extremely rapid rotators, above the $v_{\rm
  ini}$\,=\,337\,\kms\ track -- it is unclear if these were born with
such rapid rotational velocities or if they are the products of binary
evolution. Estimates of nitrogen abundances and proper motions may
help clarify their nature.

In the right-hand panels of Fig.~\ref{logg_vs_vsini} we show the
\vsini\ distribution for three ranges in \gravc\ ($>$\,4.1, [4.1,
3.9] and $<$\,3.9). The rapid rotators seem to be more frequent at
intermediate gravities, less common at higher values (i.e. younger),
and absent at \gravc\,$<$\,3.9 (older). We performed a Kuiper test
\citep{kuiper} to see if this possible evolution in the \vsini\
distributions was statistically significant. In general terms, the
Kuiper test compares the cumulative distributions of two data samples
to estimate if they can be drawn from the same parent distribution.
Importantly, the test is sensitive to the extremes of the
distribution, where we find our fast rotators.  For the test, we
denoted the three distributions in the right-hand panel of
Fig.~\ref{logg_vs_vsini} as `1', `2', and `3' (from left to right).
From applying the Kuiper test, distributions 1 and 2 have a
probability of 94\% of being derived from the same parent
distribution, whereas distributions 2 and 3 only have a 6\%
probability to be drawn from the same parent sample.

The evolutionary models (for LMC metallicity) from \citet{brott}
predict that rotation remains constant down to \gravc\,$\sim$\,3.5,
which appears somewhat incompatible with the trend above.  In
Fig.~\ref{logg_vs_vsini} we also include evolutionary tracks from the
Geneva group \citep[][for Solar metallicity]{ekstrom2012}. Their
models predict a decrease in rotational velocity from $\sim$300 to
$\sim$200\,\kms, which is still not sufficient to explain the observed
behaviour. As a word of caution we recall that the evolutionary models 
we compare with here solve for the stellar structure in one dimension. 
They do not show the possible variations in effective surface gravity with latitude, 
which may be important in stars that rotate at a significant fraction of their break-up rate. 
These results may imply that the angular momentum transport is
not properly taken into account in current evolutionary models and
that braking should be more efficient at earlier ages. Whether this
braking is related to mass loss at the surface, magnetic fields,
or other physical processes cannot be assessed with the current data.


\subsubsection{Rotation in the Kiel and H--R diagrams}

\begin{figure*}[t!]
 \centering
\subfloat{\includegraphics[scale=0.45,angle=0]{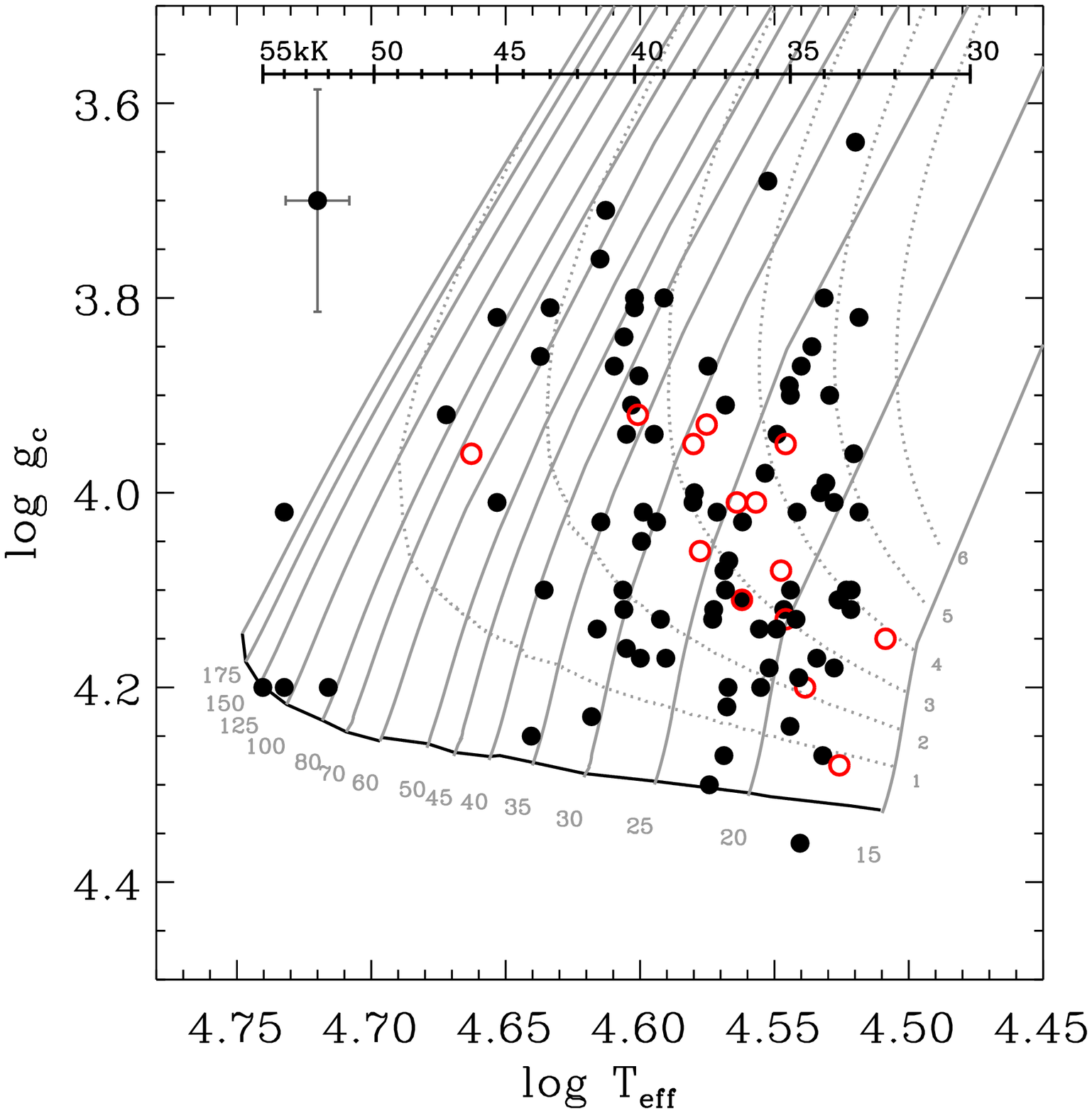} } ~
\subfloat{\includegraphics[scale=0.45,angle=0]{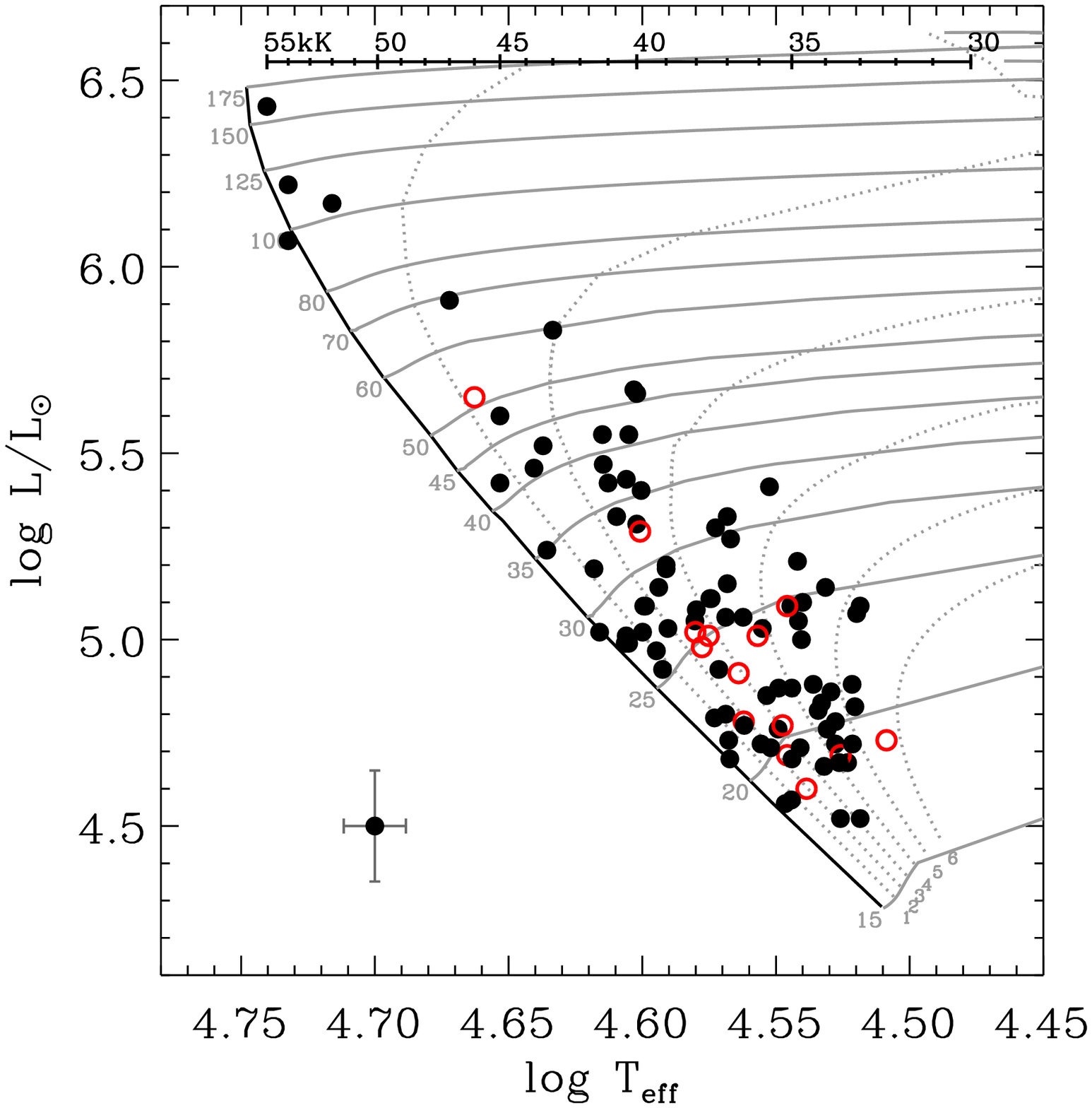} }
 \caption{Kiel (left) and H--R (right) diagrams for our sample grouped
   according to \vsini. Black filled circles: \vsini\,$<$\,250\,\kms; red open circles:
\vsini\,$\ge$\,250\,\kms. Evolutionary tracks and isochrones for an initial 
\vsini\,$=$\,171\,\kms\ from \citet{brott} are also shown.
\label{diagramas_glt_runaway}}
\end{figure*}

Further Kiel and H--R diagrams for our sample stars are shown in
Fig.~\ref{diagramas_glt_runaway}, where our sample is now grouped by
\vsini\ (lower and higher than 250\,\kms, plotted as black and red circles, 
respectively).
Stars with \vsini\,$<$\,250\,\kms\ do not appear to have a particular
pattern in the H--R diagram, but those with more rapid rotation appear to
be mostly concentrated at $M$\,$<$\,30\,M$_\odot$ and closer to the ZAMS
compared to the rest of the sample. As indicated by \citet{walborn13}, 
most of the rapid rotators are located outside the main clusters in 30~Dor
(NGC\,2060 and NGC\,2070), which
may indicate a runaway nature, a binary formation scenario, or both
\citep[][]{packet81,langer08,walborn13,platais2015}. Future analyses with {\em
  HST} imaging and nitrogen abundances \citep[Sim\'on-D\'iaz et al. in
prep.,][]{grin16} will help constrain the origin of these extremely
rapid rotators, providing ideal targets to refine our understanding of
rotational-mixing processes, chemical evolution, and binary interaction \citep[see,
e.g.][]{hunter2008}.

The fastest rotators (above the \vsini\,$=$\,337\,\kms\ tracks in
Fig.~\ref{logg_vs_vsini}) have $M_{\rm ev}$\,$<$\,25~M$_{\odot}$, and
some even below 20~M$_{\odot}$ (see
Table~\ref{tab2_HHe}). 
\ From the Kiel diagram in Fig.~\ref{diagramas_glt_runaway}, a star
with an initial mass of 20~M$_{\odot}$, will have
\Teff\,$\sim$\,30\,000\,K when reaching \gravc\,=\,3.9. With these
properties the star would be classified as an O-giant or even a B
star, and thus outside our O-dwarf sample. Indeed, the stars with
\gravc\,$<$\,3.9 in Fig.~\ref{logg_vs_vsini}, are mostly class IV,
III-IV, and V-III objects.

In view of this result, we were concerned that the drop of \vsini\ at relatively low gravities (see Sect.~\ref{sect:rot_vs_grav})
might arise from stars with initial masses below 25\,$M_\odot$ but
that will no longer be classified as O-type dwarfs once they reach
\gravc\,$<$\,3.9, particularly at large \vsini.  To check this we
included provisional results for stars with initial masses higher than
15~M$_{\odot}$ that have evolved into O and B giants
\citep[][Schneider et al. in prep.]{oscar17}. A Kuiper test of the
updated distributions for groups 2 and 3 found that the probability to
be drawn from the same parent distribution is still
  $\sim$\,10\%. This preliminary test indicates that this possible
bias can probably not explain the lack of fast rotators below
\gravc\,$\sim$\,3.9, although we remark that the parameters for some
stars still have to be confirmed, and we will revisit this in a future
study.

\subsubsection{Helium abundances}

Helium abundances for our sample were also indicated in
Fig.~\ref{logg_vs_vsini}. Most of our stars have Y(He)\,$<$\,0.12,
i.e. normal abundances within the estimated uncertainties. There were
14 stars for which we could not obtain good determinations of Y(He),
and where only upper limits of $\sim$0.08 could be estimated (with the
best-fitting models being 0.06 or less). As explained in Paper~XIII,
these low abundances could be a consequence of undetected binarity,
the effects of low signal-to-noise ratio, and/or nebular contamination in the
spectra.

We found five stars with enhanced helium abundances (Y(He)\,$>$\,0.12)
and \gravc\,$>$\,3.9. Two, VFTS\,724 and VFTS\,285, are rapid rotators
with \vsini\,$=$\,370 and 600\,\kms, respectively. The rapid rotation
of these two stars could explain their high helium abundances
\citep[see e.g.][]{maeder87,langer92,denissenkov94,h99}. The other
three stars (VFTS\,089, 123, and 761) have intermediate and low rotation
rates. Rapid rotation might still explain the He excess of these stars
given that their true rotational velocities might be higher because
of projection effects. Equally, their He enhancements might indicate past
mass transfer from an undetected binary \citep[e.g.][]{hunter2008},
and they could be magnetic stars \citep[see e.g.][]{fabian2016}.
We note that stellar winds are unlikely to be the cause for the He
enrichment, as such strong winds should generate a larger surface
enhancement, with wind features also seen in the spectra.

\subsection{Mass-discrepancy problem}\label{sec:mass_discrep}

As noted earlier, evolutionary masses ($M_{\rm ev}$) are typically
estimated by placing a star in the H--R diagram (using results from
spectral analysis) and interpolating between evolutionary tracks.
Ideally, $M_{\rm ev}$ should agree with the spectroscopic mass
($M_{\rm sp}$), calculated from \Teff, \grav, and $R$ from the
spectral analysis, but this is not always the case and is known as the
`mass-discrepancy problem'.

This discrepancy has been a challenge in the determination of masses
in O-type stars since it was identified by \cite{h92}. Possible
explanations have included uncertain distance estimates to Galactic
stars, problems with stellar atmosphere models (e.g.  underestimated
gravities or the treatment of mass-loss), and problems in the
evolutionary models (treatment of overshooting, rotation, and/or
binary evolution). However, given a diversity of results across a
range of metallicities, this issue remains unresolved \citep[e.g.][]
{h02,massey05,mokiem2007WLR,weidner10,morrell2014,mahy2015}, therefore we used
our mass estimates to investigate if this effect is present in our
sample of O dwarfs in the LMC.

\subsubsection{Best approach to estimate evolutionary  mass?}
Although H--R diagrams are generally used to estimate $M_{\rm ev}$,
some authors prefer using the Kiel diagram, which allows a comparison
of parameters obtained directly from spectroscopic analysis (thus
uncontaminated by insecure distances, although less relevant in the
case of the LMC).  An alternative approach is provided by the Bayesian
method of \cite{bonnsai}, which makes use of all the available
spectroscopic parameters simultaneously (see
Sect.~\ref{determination:phys}). A further alternative is the
  so-called spectroscopic H--R diagram
  \citep{langerkudri2014,castro2014}, which also only uses parameters
  from spectral analysis by constructing the \Teff$^4$/$g$ ratio
  (which is proportional to luminosity at constant stellar mass). It was not be used here since the distance to the LMC is well constrained.

\begin{figure}[h!]
  \centering
\includegraphics[clip,trim=0mm 0mm 0mm 00mm,angle=90,scale=0.35]{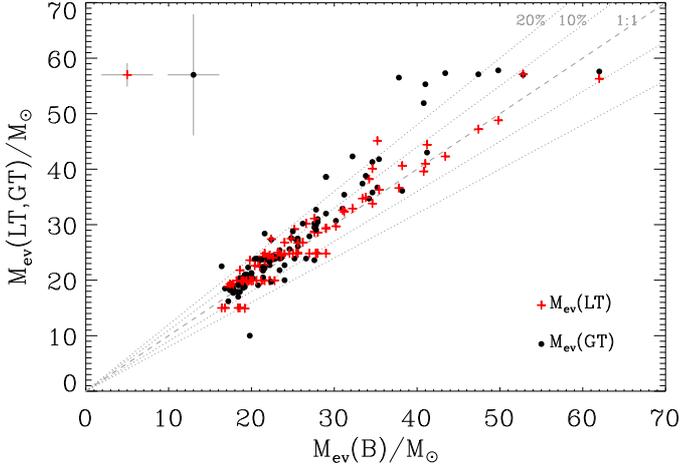}
  \caption{Evolutionary masses ($M_{\rm ev}$) estimated
    using the H--R ($M_{\rm ev}(\rm LT)$, red crosses) and 
    Kiel diagrams ($M_{\rm ev}(\rm GT)$, black dots) compared to those from 
    \textsc{bonnsai} ($M_{\rm ev}(\rm B)$). Error bars are shown in the
    upper left corner. The dashed line traces the one-to-one ratio,
    with differences of $\pm$\,10\,\% and 20\,\% shown by the dotted lines. O2 stars are not included here.
    \label{mass_LT_GT_B}}
 \end{figure}

Before investigating the masses of our sample, we examined how the
choice of method affected on the results. We note that O2 stars have been excluded from this study because their  photometry  is strongly affected (see Sect.~\ref{caution}).
Figure~\ref{mass_LT_GT_B}
compares our $M_{\rm ev}$ estimates from the H--R diagram
($M_{\rm ev}(\rm LT)$), the Kiel diagram ($M_{\rm ev}(\rm GT)$), and
\bonnsai\ ($M_{\rm ev}(\rm B)$). We find reasonable agreement between
$M_{\rm ev}(\rm LT)$ and $M_{\rm ev}(\rm B)$: the majority agree to
within $\pm$\,20\% (with 65\% agreeing to $<$10\%, and with
differences of $>$20\% for only 15 stars).  In contrast, there is a
non-negligible number of stars for which $M_{\rm ev}(\rm GT)$ is
significantly higher ($>$20\%) than $M_{\rm ev}(\rm LT)$ and $M_{\rm
  ev}(\rm B)$.

The mean and standard deviation for $M_{\rm ev}(\rm B)$\,$-$\,$M_{\rm
  ev}(\rm LT)$ and $M_{\rm ev}(\rm B)$\,$-$\,$M_{\rm ev}(\rm GT)$ are
$-$0.6\,$\pm$\,2.3~M$_{\odot}$ and 2.2\,$\pm$\,4.0~M$_{\odot}$,
respectively. When not all parameters are available, our
findings therefore indicate that $M_{\rm ev}(\rm LT)$ is a better proxy for
$M_{\rm ev}(\rm B)$ than $M_{\rm ev}(\rm GT)$ (both smaller dispersion
and error bars). However, we conclude that all three methods give
globally consistent results, in the sense that the differences will not
affect investigation of the mass discrepancy. In the following, we
adopt mass estimates from \bonnsai\ because the tool makes use of all
the available parameters (see Sect.~\ref{determination:phys}), and
provides valuable information on the significance of the results.

\subsubsection{Evolutionary vs. spectroscopic masses}\label{Comp_mev_msp}

\begin{figure}[th!!!!]
\centering
\includegraphics[scale=0.5,angle=90,trim=0mm 140mm 0mm 0mm,clip]{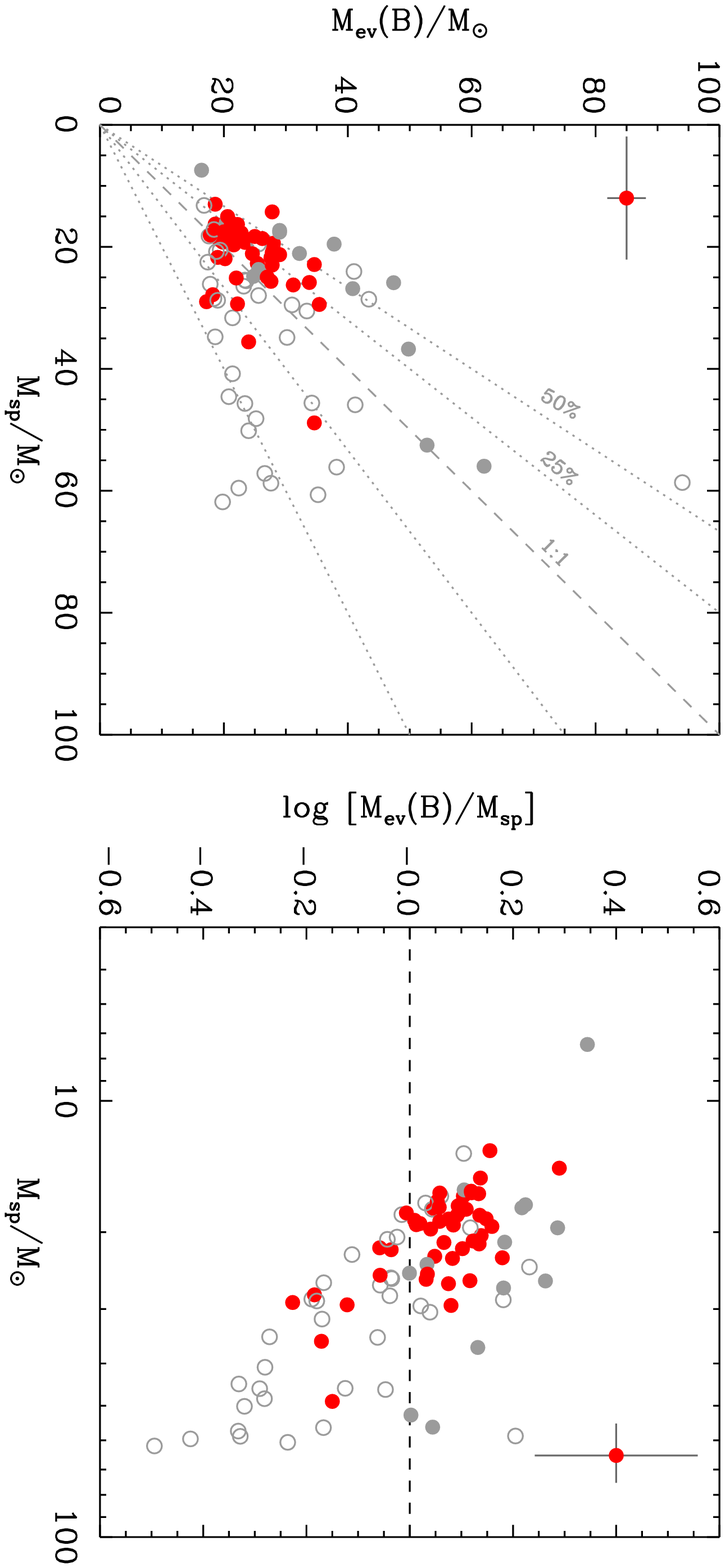}
\caption{Evolutionary-mass estimates from \textsc{bonnsai} ($M_{\rm
    ev}(\rm B)$) vs. spectroscopic masses ($M_{\rm sp}$). Grey open
  symbols: possible single-lined spectroscopic binaries
  \citep{sana13,walborn13}, stars with poor fits to He lines, high
  gravities (\grav\,$>$\,4.2) or low helium abundance, and objects
  with possible or confirmed fibre contamination. Filled grey symbols:
  objects from HHeN analysis and no indications of binarity or fibre
  contamination. Red symbols: rest of the sample. The dashed grey line
  indicates the one-to-one relation, with $\pm$25 and $\pm$50\,\%
  shown by the dotted lines. Typical error bars are given in the
  upper left corner. O2 stars were excluded.
  \label{mass_discrep}}
\end{figure}

\begin{figure*}[t]
\centering
\includegraphics[clip,trim=0mm 0mm 0mm 90mm,angle=90,scale=0.6]{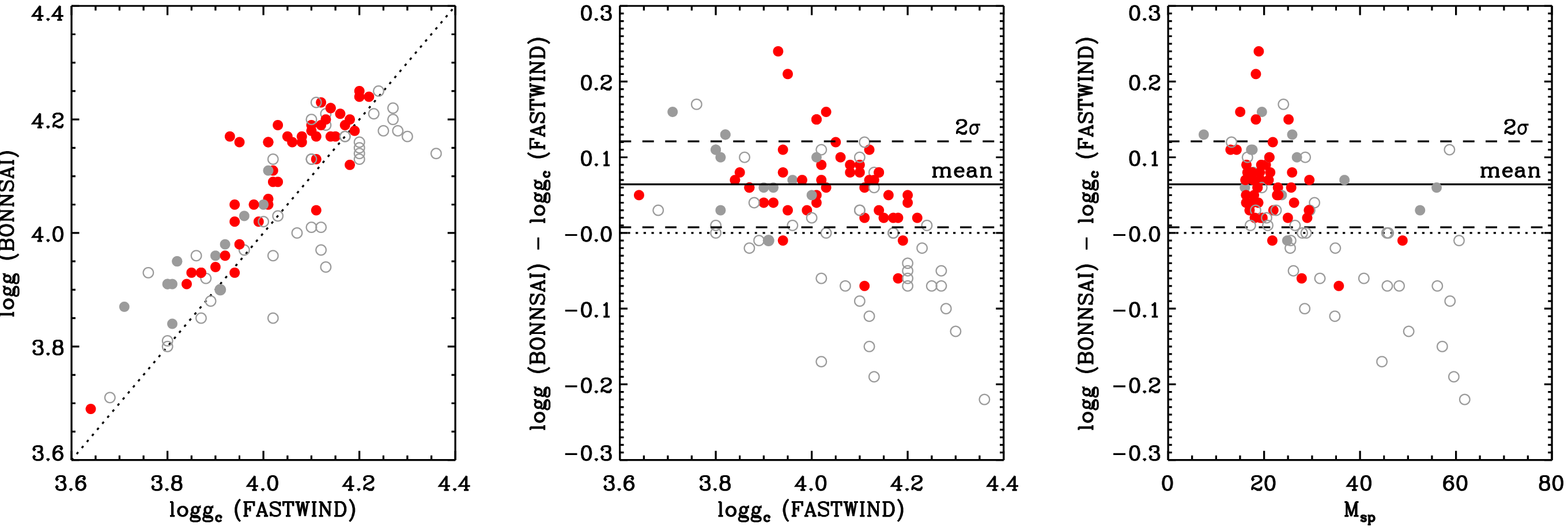}
\caption{Differences between gravities derived from \bonnsai and those
  obtained from the spectroscopic analysis (\gravc) as a function of
  the latter (left-hand panel) and spectroscopic mass ($M_{\rm sp}$,
  right-hand panel). The dashed lines mark the $\pm$\,2$\sigma$ region
  used to calculate the mean differences; symbols have the
  same meaning as in Fig.~\ref{mass_discrep}. O2 stars were excluded.\label{logg_B_F}}
 \end{figure*}
 
 The evolutionary masses obtained with \bonnsai\ are compared with
 those obtained from the spectroscopic analysis in
 Fig.~\ref{mass_discrep}.  The parameters for all stars except
 VFTS\,303 (O9.5~IV) were reproduced by the evolutionary models in the
 \bonnsai\ analysis, that is to say,  the resolution and goodness-of-fit
 tests\footnote{For a complete description see \cite{bonnsai}.} were
 passed at a 5\% significance level. In addition, the most extreme
 rotators (e.g.  VFTS\,285) are not covered by the models from
 \citet{brott} and \citet{kohler2015} used by \bonnsai.

Most of the results in Fig.~\ref{mass_discrep} are consistent with the
one-to-one relation within the uncertainties, although there is a large
dispersion and there are some objects with large discrepancies. This
remains the case even when ignoring the results for stars with
possible or confirmed contamination in the fibre and/or binary nature
(see Sect.~\ref{caution}), plotted as grey open circles in
Fig.~\ref{mass_discrep}, which dominate the $M_{\rm
  sp}$\,$>$\,$M_{\rm ev}$ region of the plot. 

A trend evident in Fig.~\ref{mass_discrep} is that except for a few
points close to unity, $M_{\rm ev}$\,$>$\,$M_{\rm sp}$ for stars with
$M_{\rm sp}$\,$\le$\,20~M$_{\odot}$, similar to results from
\citet{groen89} and \citet{h92} for Galactic stars. We refer to this
as a `positive' discrepancy and explore below if it constitutes a
real mass discrepancy.

As mentioned in Sect.~\ref{sect:diagrams}, \cite{massey13} noted a
difference between {\sc fastwind} and {\sc cmfgen} results that
could affect our gravity estimates. This systematic difference would
then also directly underestimate $M_{\rm sp}$ by an average of
$\sim$\,25\%. As our spectroscopic analyses were performed with
\fastwind, we investigated if the derived gravities are indeed too
low.

We calculated the mean difference between the inferred gravities from
\bonnsai\ and those determined with \fastwind\ (\gravc), as shown in
the left-hand panel of Fig.~\ref{logg_B_F} as a function of the
latter. We see no evidence for a trend with \gravc\ and, after
discarding stars with possible fibre contamination and/or binarity
(open grey circles in the figure), and clipping out points with
differences larger than 2$\sigma$, the mean difference is
0.06\,$\pm$\,0.03\,dex. That is, the gravities from \bonnsai\ are, on
average, 0.06\,dex higher than the spectroscopic results. This is well
within the quoted uncertainties, although it does not rule out a small
effect of underestimated gravities using \fastwind.

An underestimate of spectroscopic gravities might arise from
underestimating the radiative acceleration in the photosphere in
\fastwind, resulting in too little gravitational acceleration.
However, we find no correlation of the differences in estimated
gravities with effective temperature or luminosity, as we would expect
if the treatment of the radiative acceleration were incorrect.
Alternatively, the difference might arise from how \bonnsai\ obtains
the gravities. \bonnsai\ takes into account that stars spend more time
near the ZAMS than far away from it (see e.g. the isochrones in
Fig.~\ref{diagramas_glt}), so that the density of evolutionary models used by
\bonnsai\ is more concentrated towards higher gravities and is therefore
expected to provide higher \grav\ values.

A general shift of all \fastwind\ gravities would not solve the
situation in Fig.~\ref{mass_discrep}. The estimates of $M_{\rm sp}$
would increase by about 11\%, which would not help much at $M_{\rm
  sp}$\,$\ge$\,20~M$_{\odot}$. This is more clearly seen in the
right-hand panel of Fig~\ref{logg_B_F}, showing the difference in the
estimated gravities vs. $M_{\rm sp}$. Below 20~M$_{\odot}$ $M_{\rm
  sp}$\,$<$\,$M_{\rm ev}$, therefore if the mass discrepancy were to be
solved by correcting the \fastwind\ estimates, these would only need
to be increased at low masses.

\begin{figure}[]
\centering
\includegraphics[scale=0.5,angle=90,trim=0mm 0mm 0mm 140mm,clip]{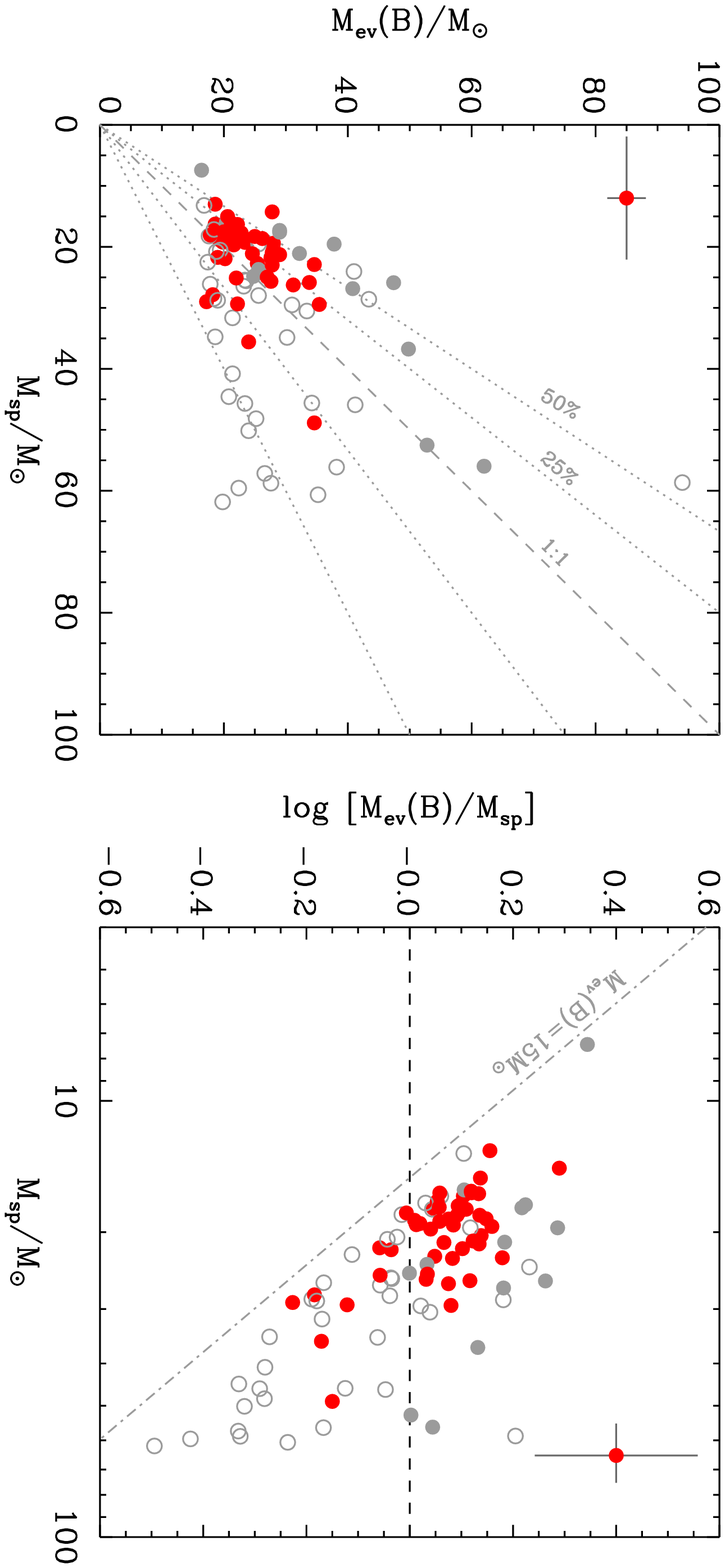}
\caption{Logarithmic ratio of evolutionary masses from
  \textsc{bonnsai} ($M_{\rm ev}(B)$) to spectroscopic masses ($M_{\rm
    sp}$), with the same symbols as in Fig.~\ref{mass_discrep}.
  Typical error bars are given in the upper right corner. The dashed
  line indicates points where $M_{\rm ev}(\rm B)$\,=\,15\,M$_{\odot}$. O2 stars were excluded.
  \label{mass_discrep_bias}}
\end{figure}

Figure~\ref{mass_discrep_bias} provides a different view of our results,
showing log($M_{\rm ev}(B)$/$M_{\rm sp}$) vs. $M_{\rm sp}$ (with the
latter plotted logarithmically). The figure shows a clear trend of a
positive ($M_{\rm ev}/M_{\rm sp}>1$) mass discrepancy for most of the
sample. Only for stars with $M_{\rm sp}$\,$\ge$\,20~M$_{\odot}$
do we find both positive and negative ($M_{\rm ev}/M_{\rm sp}<1$)
discrepancies, similar to the trend seen in Figs.~\ref{mass_discrep}
and ~\ref{logg_B_F}.

However, we think this trend arises from an observational bias.  The
O-type dwarfs in this study systematically have M$_{\rm
  ev}$\,$>$\,15\,$M_{\odot}$ (e.g. Fig.~\ref{diagramas_glt}).  This
implies that any stars having a negative mass discrepancy below
$M_{\rm ev}\sim15~$M$_{\odot}$ are missed by our sample, hence are not
present in Fig.~\ref{mass_discrep_bias}. The boundary associated with
this effect is shown by the dash-dotted line in
Fig.~\ref{mass_discrep_bias}. As expected, no stars are below that
line in our sample, and to investigate these effects further will
require quantitative analysis of the early B-type dwarfs from the
VFTS. We therefore conclude that we find no compelling evidence for a
systematic mass discrepancy in our sample of LMC O-type dwarfs.

\subsection{Wind properties} \label{sect:winds}

\begin{figure*}[t]
 \centering
\includegraphics[trim=0mm 0mm 0mm 0mm,clip,scale=0.6,angle=90]{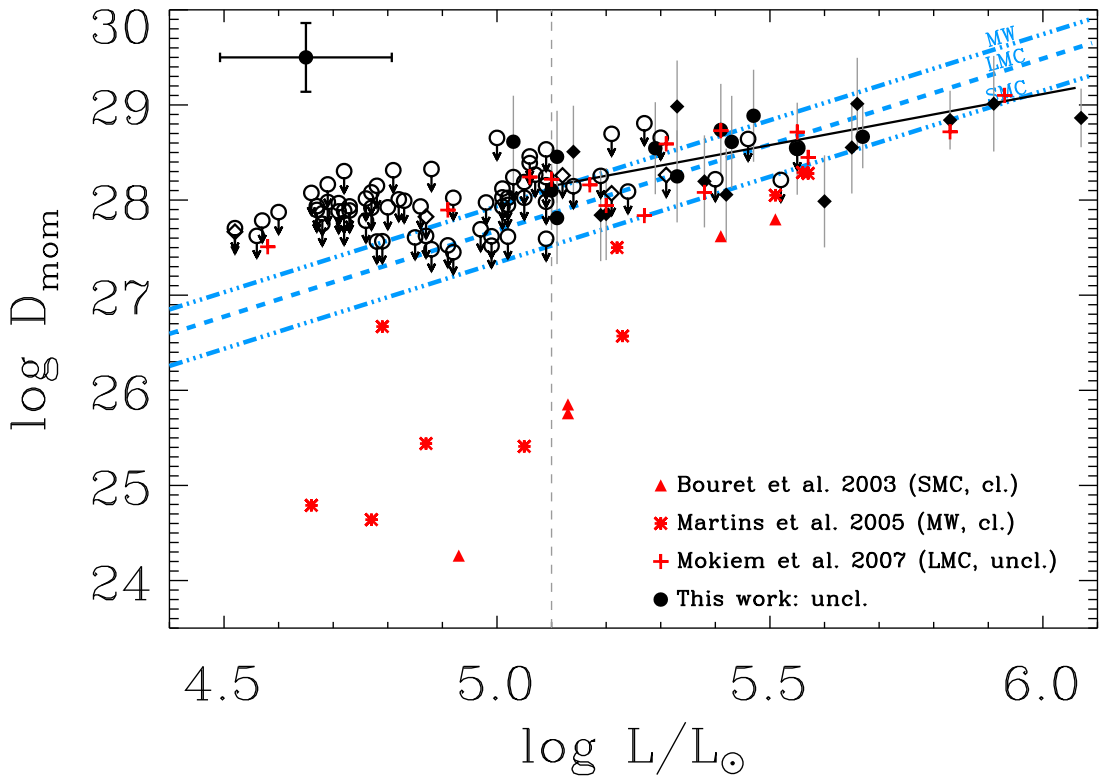}
 \caption{Wind-momentum--luminosity relationship (WLR) for our sample
   compared with theoretical predictions from \cite{vink2001} for the
   metallicities of the Milky Way and the Magellanic Clouds. Stars
   with upper limits on \logq\ (hence $D_{\rm mom}$) are plotted with
   open symbols, and those analysed using the nitrogen lines are
   plotted with diamonds.  Typical uncertainties are shown in the
   upper left corner, and the vertical dashed line indicates
   \logl\,=\,5.1. Results from \cite{b03}, \cite{martins05gal}, and
   \cite{mokiem2007LMC} for their samples of O dwarfs in the SMC, MW,
   and LMC O, respectively, are plotted in red; the use of clumped
   (`cl.') or unclumped (`uncl.') wind models in each work is also
   indicated in the legend.\label{wlr_theo_vink} }
\end{figure*}

We also investigated the wind properties of our sample in
the context of a wind-momentum\,--\,luminosity relationship (WLR, see \citealt{k92,k95}),
compared to past results and theoretical predictions.

The winds in the \fastwind\ models used by the \textsc{iacob-gbat} are
characterised by the wind-strength parameter, \logq\ (where
$Q$\,$=$\,$\dot{M}(Rv_{\infty})^{-3/2}$). If estimates for the
terminal velocities ($v_{\infty}$) and stellar radii ($R$) are
available, then the \logq\ values from the spectral fits can be used
to estimate mass-loss rates ($\dot{M}$). However, we lack the UV
spectroscopy required to estimate $v_{\infty}$ directly for our
targets.

We therefore followed the same approach as
\citet{mokiem2007LMC,mokiem2007WLR}, using our estimates of $g$ and
$R$ to estimate the escape velocity ($v_{\rm esc}$):
\begin{equation}
v_{\rm esc}\,=\,\left[ 2gR(1-\Gamma) \right]^{1/2}, 
\end{equation}
where $\Gamma$ is the Eddington factor\footnote{Defined here as the
  ratio of radiative acceleration due to Thomson scattering compared
  to gravity, $\Gamma$\,=\,5.765$\times$10$^{-16}$ $\frac{T_{\rm
      eff}^4}{g}$ $\frac{1+2Y}{1+4Y}$, where $Y$ is the helium-to-  hydrogen number fraction.}. From this we obtained estimates of
$v_{\infty}$, employing $v_{\infty}/v_{\rm esc}$\,=\,2.65
\citep{kp2000}, scaled to a metallicity of 0.5\,Z$_{\odot}$ by a
factor of Z$^{0.13}$ \citep{leitherer92}, leading to
estimates of $\dot{M}$ via the $Q$-parameter. We then calculated the
modified wind-momentum rates for our sample, $D_{\rm
  mom}$\,=\,$\dot{M}v_{\infty}R^{1/2}$, which are shown as a function of
stellar luminosity in Fig.~\ref{wlr_theo_vink}. Our results follow the
expected linear trend from \cite{vink2001}, with log\,$D_{\rm mom}$
increasing with luminosity. We now discuss our results in two groups,
separated at \logl\,$\sim$\,5.1, as indicated in
Fig.~\ref{wlr_theo_vink}.

\subsubsection{\logl\,$\ge$\,5.1}

For the more luminous stars in our sample we were generally able to
arrive at reliable estimates of \logq\, and therefore estimate
$\dot{M}$ and $D_{\rm mom}$. For a WLR of the form log\,$D_{\rm
  mom}$\,$=$\,$x$\,\logl\,$+$\,$D_0$, we obtain
$x$\,$=$\,1.07\,$\pm$\,0.18 and $D_0$\,$=$\,22.67\,$\pm$\,0.99 from a
linear fit to our results for \logl\,$\ge$\,5.1. These values differ
from those from \cite{mokiem2007WLR}, who obtained
$x$\,$=$\,1.87\,$\pm$\,0.19 and $D_0$\,$=$\,18.30\,$\pm$\,1.04, in
good agreement with predictions from \cite{vink2001}.

These differences may have two origins. A first factor might be the
inhomogeneity of the \citeauthor{mokiem2007WLR} sample, which included
more luminous stars as well as dwarfs. In the standard theory of
radiatively driven winds, luminosity class should not play a role
\citep{kp2000}, but clumping may introduce a differential effect.
Some of the luminous stars studied by \citeauthor{mokiem2007WLR} had
large wind momenta, possibly reflecting the impact of significant
clumping. When corrected for clumping in selected objects
\citep[following][]{rph04}, they found $x$\,$=$\,1.49\,$\pm$\,0.18 and
$D_0$\,$=$\,20.40\,$\pm$\,1.00, in much better agreement with our
values. This effect is reinforced by the strong dependence of
$\dot{M}$ on $\Gamma$. \cite{vink2011} and \cite{grafener2011} have
argued that there is a kink in the relationship between $\dot{M}$ and
$\Gamma_e$ at log\,$\Gamma_e$\,$\sim$\,--\,0.58, when the effect of
$\Gamma$ sets in.  The result is increased $\dot{M}$ for supergiants
as compared to dwarfs, as for instance shown by the low $\Gamma_e$ supergiants
compared to their dwarf counterparts in Fig.~8 from \citet{besten14}. Thus,
the WLR slope obtained by \citeauthor{mokiem2007WLR} will be steeper
because of the influence of supergiants close to the Eddington limit.

Secondly, at slightly lower luminosities, when the
\citeauthor{mokiem2007WLR} sample was also dominated by dwarfs, they
obtained estimates for $\beta$ (the exponent of the wind velocity-law)
that were larger than the 0.8 adopted in our study. Lower $\beta$
means faster wind acceleration and higher velocities in the region
where H$\alpha$ and \ion{He}{ii}\,$\lambda$\,4686 are formed, leading
to higher mass-loss rates \citep[for a discussion of the dependence of
the derived mass-loss rate on $\beta$, see][]{markova04}. An example
of how this can reconcile the derived WLRs is given by
\citet{oscar17}.

In short, both effects push our estimated WLR to a smaller slope and a
larger vertical offset than that from Mokiem et al. We thus find a
flatter slope for the WLR (also flatter than theoretical expectations)
that should be investigated further.
 
\subsubsection{\logl\,$<$\,5.1}

The winds of the less luminous stars in the LMC become thin, and the
optical diagnostics (mainly H$\alpha$ and
\ion{He}{ii}\,$\lambda$\,4686) are insensitive to changes in density.
We could therefore only derive upper limits on $D_{\rm mom}$ for
stars with \logl\,$<$\,5.1 (except for one object fairly close to the
\logl\,$=$\,5.1 boundary); similar problems were also encountered by
\citet{mokiem2007WLR}.

We also include results in Fig.~\ref{wlr_theo_vink} for the SMC from
\cite{b03} and Milky Way from \cite{martins05gal}, who both found from UV diagnostics weaker winds than predicted by theory in this lower-luminosity regime.
These stars remain a challenge for the theory of radiatively
driven
winds \citep[see][]{puls08}. Several explanations for these weak winds
have been proposed, such as X-rays \citep{h09,huenem12} or coronal winds
\citep{lucy12}, although the question remains open.  A similar problem
might be present even for some higher-luminosity stars, for example around
\logl\,$\sim$\,5.5\,--\,5.6 in our results, although we note that both
\citeauthor{b03} and \citeauthor{martins05gal} included
clumped winds in their analyses, which led to lower mass-loss rates.

To obtain better constraints on the wind properties of the less
luminous stars in our sample (and investigate if similarly weak winds
are found as in other studies), we require UV spectroscopy spanning
1100\,--\,1800\,\AA, which contains diagnostic lines far more
sensitive to wind effects than those available in the optical
(particularly in the case of weak winds).

\section{Summary and conclusions}

In the framework of the VLT-FLAMES Tarantula Survey, we have analysed
the optical spectra of 105 apparently single O-type dwarfs in the
30~Doradus region of the LMC. We used the {\sc iacob-gbat} to estimate
stellar and wind parameters for our sample, and used
both classical techniques and the Bayesian \textsc{bonnsai} tool to
estimate evolutionary masses. This study is the largest quantitative analysis 
of O dwarfs in the LMC, with the most complete coverage of spectral types to date. 
Our summary and conclusions are as follows:

\begin{itemize}

\item[$\bullet$] We were only able to obtain upper limits on \logq\ for $\sim$70\% of the stars. Analysis using HHeN diagnostics 
was necessary in 20 cases because we lack reliable \ion{He}{i} lines. In addition, we found indications of 
a possible binary nature in 14
stars through anomalously low helium abundances (Y(He)\,$<$\,0.08), and
10 stars with relatively high gravities. The O2-type stars in our
sample also show inconsistencies that could be related
to multiplicity (and/or nitrogen peculiarities).\\

\item[$\bullet$] We provide a new effective temperature\,--\,spectral type
  calibration from a linear fit to our data (excluding O2 stars). We
  find good agreement with previous results, but are unable at present
  to reach a firm conclusion on the need for a separate, steeper
  relation for spectral types earlier than O4.\\

\item[$\bullet$] From H--R and Kiel diagrams of our sample (which
  mostly spans 3.8\,$<$\,\grav\,$<$\,4.2) there appear to be
  relatively few stars younger than 1\,Myr. Possible explanations
  include underestimated gravities from \fastwind, missing
  stars still embedded in their natal clouds, or that the main episode
  of star formation in 30~Dor simply occurred 1 to 4\,Myr ago.\\

\item[$\bullet$] Most of the rapid rotators
  (\vsini\,$\geq$\,250\,\kms) in our sample have masses below
  $\sim$\,25~M$_{\odot}$ and are closer to the ZAMS than the
  rest of the sample. Most are also located outside the NGC\,2060 and
  NGC\,2070 clusters, suggesting a potential runaway nature.\\

\item[$\bullet$] The fastest rotators (\vsini\,$\geq$\,340\,\kms) are
  found mostly at intermediate gravities (3.9\,$<$\,\gravc\,$<$\,4.1),
  with none at lower gravities. We investigated if this was due to an
  observational bias, whereby lower-mass stars evolve out of the
  O-type sample as they age. We concluded that the estimated
  contribution of these additional stars is insufficient to explain
  the differences in the \vsini\,--\,\gravc\ distribution at
  intermediate and lower gravities, and note that this behaviour is
  not predicted by current evolutionary models. \\

\item[$\bullet$] We compared evolutionary and spectroscopic mass
  estimates for the sample, finding a non-negligible number of stars
  with differences of 25 to 50\%. We also found that the $M_{\rm
    ev}$/$M_{\rm sp}$ ratio decreases with increasing $M_{\rm sp}$,
  typically with $M_{\rm ev}$\,$>$\,$M_{\rm sp}$ for stars with
  $M_{\rm sp}$\,$\leq$\,20~M$_{\odot}$, and $M_{\rm ev}$\,$<$\,$M_{\rm
    sp}$ for $M_{\rm sp}$\,$\geq$\,30~M$_{\odot}$.  However, this
  trend is probably caused by a simple observational bias, whereby the
  O-type dwarfs all have $M_{\rm ev}$\,$>$\,15\,M$_{\odot}$; similar
  results for early B-dwarfs from the VFTS will need to be
  included for a complete view of the estimated masses. 
  We find no compelling evidence of a systematic mass discrepancy.\\

\item[$\bullet$] When we investigated the possible mass discrepancy, we
  noted that our gravities tended to be lower than those predicted by
  evolutionary models, which could be explained by the fact that
  \bonnsai\ takes into account that stars spend more time near the
  ZAMS than far away from it, which might favour slightly higher gravities.\\

\item[$\bullet$] At \logl\,$>$\,5.1 our stars have a high dispersion
  in the WLR, although the majority are consistent with theoretical
  predictions for the LMC within the estimated uncertainties.
  Unclumped models and the large uncertainty on the adopted terminal
  velocities could explain stars with winds apparently stronger than
  the theoretical WLR for Galactic metallicity, but there is no
  satisfactory explanation for those stars with winds estimated to be
  weaker than predicted for the SMC. We obtain a flatter fit to the
  WLR compared to that from \cite{mokiem2007WLR}, probably because
  their study also included O-type supergiants close to the Eddington
  limit. At \logl\,$<$\,5.1 we could not reliably estimate \logq\ for
  our sample because of the current lack of UV observations.

\end{itemize}

The physical properties of the O-type stars from the VFTS presented
here and by \citet{oscar17} will underpin studies to clarify the
physics underlying the evolution of massive stars and the mechanisms
that affect it.  Estimates of nitrogen abundances for the whole O-type
sample \citep[Sim\'on-D\'iaz et al. in prep.][]{grin16} -- one of the
original motivations of the VFTS -- will enable new studies of
rotational mixing and chemical evolution.  Moreover, together with the
ongoing {\em HST} study of proper motions (led by D.~J.~Lennon),
nitrogen abundances will help to constrain the binary nature and
possible runaway origin of the fastest rotators. Future UV
observations are required to better constrain the wind properties of
our sample, particularly at lower luminosities where we are only
able to give upper limits on their intensity from optical spectroscopy.

\begin{acknowledgements}
  CS-S acknowledges support from the Joint Committee ESO-Government of
  Chile and DIDULS programme from Universidad de La Serena under grant
  DIDULS Regular PR16145. SS-D and AH acknowledge support from the
  Spanish Ministry of Economy and Competitiveness (MINECO) under
  grants AYA2015-68012-C2-1 and SEV-2015-0548. This paper made use of
  the IAC Supercomputing facility HTCondor
  (http://research.cs.wisc.edu/htcondor/), recently expanded and
  improved thanks to FEDER funds granted by the Ministry of Economy
  and Competitiveness, project code IACA13-3E-2493. OHR-A acknowledges
  funding from the European Union's Horizon 2020 research and
  innovation programme under the Marie Sk{\l}odowska-Curie grant
  agreement No 665593 awarded to the Science and Technology Facilities
  Council. NJG is part of the International Max Planck Research School
  (IMPRS) for Astronomy and Astrophysics at the Universities of Bonn
  and Cologne. STScI is operated by the Association of Universities for Research in Astronomy, Inc., 
  under NASA contract NAS5-26555. 
\end{acknowledgements}



\begin{thebibliography}{119}
\expandafter\ifx\csname natexlab\endcsname\relax\def\natexlab#1{#1}\fi

\bibitem[{{Aldoretta} {et~al.}(2015){Aldoretta}, {Caballero-Nieves}, {Gies},
  {Nelan}, {Wallace}, {Hartkopf}, {Henry}, {Jao}, {Ma{\'{\i}}z Apell{\'a}niz},
  {Mason}, {Moffat}, {Norris}, {Richardson}, \& {Williams}}]{aldoretta}
{Aldoretta}, E.~J., {Caballero-Nieves}, S.~M., {Gies}, D.~R., {et~al.} 2015,
  \aj, 149, 26

\bibitem[{{Bestenlehner} {et~al.}(2014){Bestenlehner}, {Gr{\"a}fener}, {Vink},
  {Najarro}, {de Koter}, {Sana}, {Evans}, {Crowther}, {H{\'e}nault-Brunet},
  {Herrero}, {Langer}, {Schneider}, {Sim{\'o}n-D{\'{\i}}az}, {Taylor}, \&
  {Walborn}}]{besten14}
{Bestenlehner}, J.~M., {Gr{\"a}fener}, G., {Vink}, J.~S., {et~al.} 2014, \aap,
  570, A38

\bibitem[{{Bestenlehner} {et~al.}(2011){Bestenlehner}, {Vink}, {Gr{\"a}fener},
  {Najarro}, {Evans}, {Bastian}, {Bonanos}, {Bressert}, {Crowther}, {Doran},
  {Friedrich}, {H{\'e}nault-Brunet}, {Herrero}, {de Koter}, {Langer}, {Lennon},
  {Ma{\'{\i}}z Apell{\'a}niz}, {Sana}, {Soszynski}, \& {Taylor}}]{besten11}
{Bestenlehner}, J.~M., {Vink}, J.~S., {Gr{\"a}fener}, G., {et~al.} 2011, \aap,
  530, L14

\bibitem[{{Bouret} {et~al.}(2003){Bouret}, {Lanz}, {Hillier}, {Heap}, {Hubeny},
  {Lennon}, {Smith}, \& {Evans}}]{b03}
{Bouret}, J.-C., {Lanz}, T., {Hillier}, D.~J., {et~al.} 2003, \apj, 595, 1182

\bibitem[{{Bromm} {et~al.}(2009){Bromm}, {Yoshida}, {Hernquist}, \&
  {McKee}}]{bromm09}
{Bromm}, V., {Yoshida}, N., {Hernquist}, L., \& {McKee}, C.~F. 2009, \nat, 459,
  49

\bibitem[{{Brott} {et~al.}(2011){Brott}, {de Mink}, {Cantiello}, {Langer}, {de
  Koter}, {Evans}, {Hunter}, {Trundle}, \& {Vink}}]{brott}
{Brott}, I., {de Mink}, S.~E., {Cantiello}, M., {et~al.} 2011, \aap, 530, A115

\bibitem[{{Castro} {et~al.}(2014){Castro}, {Fossati}, {Langer},
  {Sim{\'o}n-D{\'{\i}}az}, {Schneider}, \& {Izzard}}]{castro2014}
{Castro}, N., {Fossati}, L., {Langer}, N., {et~al.} 2014, \aap, 570, L13

\bibitem[{{Castro} {et~al.}(2012){Castro}, {Urbaneja}, {Herrero}, {Garcia},
  {Sim{\'o}n-D{\'{\i}}az}, {Bresolin}, {Pietrzy{\'n}ski}, {Kudritzki}, \&
  {Gieren}}]{castro2012}
{Castro}, N., {Urbaneja}, M.~A., {Herrero}, A., {et~al.} 2012, \aap, 542, A79

\bibitem[{{Crowther} {et~al.}(2016){Crowther}, {Caballero-Nieves}, {Bostroem},
  {Ma{\'{\i}}z Apell{\'a}niz}, {Schneider}, {Walborn}, {Angus}, {Brott},
  {Bonanos}, {de Koter}, {de Mink}, {Evans}, {Gr{\"a}fener}, {Herrero},
  {Howarth}, {Langer}, {Lennon}, {Puls}, {Sana}, \& {Vink}}]{crowther2016}
{Crowther}, P.~A., {Caballero-Nieves}, S.~M., {Bostroem}, K.~A., {et~al.} 2016,
  \mnras, 458, 624

\bibitem[{{de Mink} {et~al.}(2013){de Mink}, {Langer}, {Izzard}, {Sana}, \& {de
  Koter}}]{demink13}
{de Mink}, S.~E., {Langer}, N., {Izzard}, R.~G., {Sana}, H., \& {de Koter}, A.
  2013, \apj, 764, 166

\bibitem[{{de Mink} {et~al.}(2014){de Mink}, {Sana}, {Langer}, {Izzard}, \&
  {Schneider}}]{demink14}
{de Mink}, S.~E., {Sana}, H., {Langer}, N., {Izzard}, R.~G., \& {Schneider},
  F.~R.~N. 2014, \apj, 782, 7

\bibitem[{{Denissenkov}(1994)}]{denissenkov94}
{Denissenkov}, P.~A. 1994, \aap, 287, 113

\bibitem[{{Doran} \& {Crowther}(2011)}]{doran11}
{Doran}, E.~I. \& {Crowther}, P.~A. 2011, Bulletin de la Societe Royale des
  Sciences de Liege, 80, 129

\bibitem[{{Doran} {et~al.}(2013){Doran}, {Crowther}, {de Koter}, {Evans},
  {McEvoy}, {Walborn}, {Bastian}, {Bestenlehner}, {Gr{\"a}fener}, {Herrero},
  {K{\"o}hler}, {Ma{\'{\i}}z Apell{\'a}niz}, {Najarro}, {Puls}, {Sana},
  {Schneider}, {Taylor}, {van Loon}, \& {Vink}}]{doran13}
{Doran}, E.~I., {Crowther}, P.~A., {de Koter}, A., {et~al.} 2013, \aap, 558,
  A134

\bibitem[{{Dufton} {et~al.}(2011){Dufton}, {Dunstall}, {Evans}, {Brott},
  {Cantiello}, {de Koter}, {de Mink}, {Fraser}, {H{\'e}nault-Brunet},
  {Howarth}, {Langer}, {Lennon}, {Markova}, {Sana}, \& {Taylor}}]{dufton11}
{Dufton}, P.~L., {Dunstall}, P.~R., {Evans}, C.~J., {et~al.} 2011, \apjl, 743,
  L22

\bibitem[{{Dufton} {et~al.}(2013){Dufton}, {Langer}, {Dunstall}, {Evans},
  {Brott}, {de Mink}, {Howarth}, {Kennedy}, {McEvoy}, {Potter},
  {Ram{\'{\i}}rez-Agudelo}, {Sana}, {Sim{\'o}n-D{\'{\i}}az}, {Taylor}, \&
  {Vink}}]{dufton13}
{Dufton}, P.~L., {Langer}, N., {Dunstall}, P.~R., {et~al.} 2013, \aap, 550,
  A109

\bibitem[{{Dunstall} {et~al.}(2015){Dunstall}, {Dufton}, {Sana}, {Evans},
  {Howarth}, {Sim{\'o}n-D{\'{\i}}az}, {de Mink}, {Langer}, {Ma{\'{\i}}z
  Apell{\'a}niz}, \& {Taylor}}]{dunstall15}
{Dunstall}, P.~R., {Dufton}, P.~L., {Sana}, H., {et~al.} 2015, \aap, 580, A93

\bibitem[{{Ekstr{\"o}m} {et~al.}(2012){Ekstr{\"o}m}, {Georgy}, {Eggenberger},
  {Meynet}, {Mowlavi}, {Wyttenbach}, {Granada}, {Decressin}, {Hirschi},
  {Frischknecht}, {Charbonnel}, \& {Maeder}}]{ekstrom2012}
{Ekstr{\"o}m}, S., {Georgy}, C., {Eggenberger}, P., {et~al.} 2012, \aap, 537,
  A146

\bibitem[{{Evans} {et~al.}(2012){Evans}, {Hainich}, {Oskinova}, {Gallagher},
  {Chu}, {Gruendl}, {Hamann}, {H{\'e}nault-Brunet}, \& {Todt}}]{evans12}
{Evans}, C.~J., {Hainich}, R., {Oskinova}, L.~M., {et~al.} 2012, \apj, 753, 173

\bibitem[{{Evans} {et~al.}(2015){Evans}, {Kennedy}, {Dufton}, {Howarth},
  {Walborn}, {Markova}, {Clark}, {de Mink}, {de Koter}, {Dunstall},
  {H{\'e}nault-Brunet}, {Ma{\'{\i}}z Apell{\'a}niz}, {McEvoy}, {Sana},
  {Sim{\'o}n-D{\'{\i}}az}, {Taylor}, \& {Vink}}]{evans15}
{Evans}, C.~J., {Kennedy}, M.~B., {Dufton}, P.~L., {et~al.} 2015, \aap, 574,
  A13

\bibitem[{{Evans} {et~al.}(2011){Evans}, {Taylor}, {H{\'e}nault-Brunet},
  {Sana}, {de Koter}, {Sim{\'o}n-D{\'{\i}}az}, {Carraro}, {Bagnoli}, {Bastian},
  {Bestenlehner}, {Bonanos}, {Bressert}, {Brott}, {Campbell}, {Cantiello},
  {Clark}, {Costa}, {Crowther}, {de Mink}, {Doran}, {Dufton}, {Dunstall},
  {Friedrich}, {Garcia}, {Gieles}, {Gr{\"a}fener}, {Herrero}, {Howarth},
  {Izzard}, {Langer}, {Lennon}, {Ma{\'{\i}}z Apell{\'a}niz}, {Markova},
  {Najarro}, {Puls}, {Ramirez}, {Sab{\'{\i}}n-Sanjuli{\'a}n}, {Smartt},
  {Stroud}, {van Loon}, {Vink}, \& {Walborn}}]{evans11}
{Evans}, C.~J., {Taylor}, W.~D., {H{\'e}nault-Brunet}, V., {et~al.} 2011, \aap,
  530, A108

\bibitem[{{Fullerton} {et~al.}(2006){Fullerton}, {Massa}, \&
  {Prinja}}]{fullerton06}
{Fullerton}, A.~W., {Massa}, D.~L., \& {Prinja}, R.~K. 2006, \apj, 637, 1025

\bibitem[{{Gibson}(2000)}]{gibson}
{Gibson}, B.~K. 2000, \memsai, 71, 693

\bibitem[{{Gr{\"a}fener} {et~al.}(2011){Gr{\"a}fener}, {Vink}, {de Koter}, \&
  {Langer}}]{grafener2011}
{Gr{\"a}fener}, G., {Vink}, J.~S., {de Koter}, A., \& {Langer}, N. 2011, \aap,
  535, A56

\bibitem[{{Gray}(2005)}]{gray}
{Gray}, D.~F. 2005, {The Observation and Analysis of Stellar Photospheres} (3rd
  Edition, by D.F.~Gray.~ISBN 0521851866. UK: Cambridge University Press,
  2005.)

\bibitem[{{Grin} {et~al.}(2016){Grin}, {Ramirez-Agudelo}, {de Koter}, {Sana},
  {Puls}, {Brott}, {Crowther}, {Dufton}, {Evans}, {Graefener}, {Herrero},
  {Langer}, {Lennon}, {van Loon}, {Markova}, {de Mink}, {Najarro}, {Schneider},
  {Taylor}, {Tramper}, {Vink}, \& {Walborn}}]{grin16}
{Grin}, N.~J., {Ramirez-Agudelo}, O.~H., {de Koter}, A., {et~al.} 2016, ArXiv
  e-prints

\bibitem[{{Groenewegen} {et~al.}(1989){Groenewegen}, {Lamers}, \&
  {Pauldrach}}]{groen89}
{Groenewegen}, M.~A.~T., {Lamers}, H.~J.~G.~L.~M., \& {Pauldrach}, A.~W.~A.
  1989, \aap, 221, 78

\bibitem[{{Hanson}(1998)}]{hanson}
{Hanson}, M.~M. 1998, in ASP Conference Series, Vol. 131, Properties of Hot
  Luminous Stars, ed. I.~{Howarth}, 1

\bibitem[{{Herrero} {et~al.}(1999){Herrero}, {Corral}, {Villamariz}, \&
  {Mart{\'{\i}}n}}]{h99}
{Herrero}, A., {Corral}, L.~J., {Villamariz}, M.~R., \& {Mart{\'{\i}}n}, E.~L.
  1999, \aap, 348, 542

\bibitem[{{Herrero} {et~al.}(2009){Herrero}, {Garcia}, \& {Najarro}}]{h09}
{Herrero}, A., {Garcia}, M., \& {Najarro}, F. 2009, \apss, 320, 149

\bibitem[{{Herrero} {et~al.}(1992){Herrero}, {Kudritzki}, {Vilchez}, {Kunze},
  {Butler}, \& {Haser}}]{h92}
{Herrero}, A., {Kudritzki}, R.~P., {Vilchez}, J.~M., {et~al.} 1992, \aap, 261,
  209

\bibitem[{{Herrero} {et~al.}(2002){Herrero}, {Puls}, \& {Najarro}}]{h02}
{Herrero}, A., {Puls}, J., \& {Najarro}, F. 2002, \aap, 396, 949

\bibitem[{{Hillier}(1991)}]{hillier91}
{Hillier}, D.~J. 1991, \aap, 247, 455

\bibitem[{{Hillier} \& {Miller}(1998)}]{hm98}
{Hillier}, D.~J. \& {Miller}, D.~L. 1998, \apj, 496, 407

\bibitem[{{Huenemoerder} {et~al.}(2012){Huenemoerder}, {Oskinova}, {Ignace},
  {Waldron}, {Todt}, {Hamaguchi}, \& {Kitamoto}}]{huenem12}
{Huenemoerder}, D.~P., {Oskinova}, L.~M., {Ignace}, R., {et~al.} 2012, \apjl,
  756, L34

\bibitem[{{Hunter} {et~al.}(2008){Hunter}, {Brott}, {Lennon}, {Langer},
  {Dufton}, {Trundle}, {Smartt}, {de Koter}, {Evans}, \& {Ryans}}]{hunter2008}
{Hunter}, I., {Brott}, I., {Lennon}, D.~J., {et~al.} 2008, \apjl, 676, L29

\bibitem[{{Ilee} {et~al.}(2016){Ilee}, {Cyganowski}, {Nazari}, {Hunter},
  {Brogan}, {Forgan}, \& {Zhang}}]{ilee16}
{Ilee}, J.~D., {Cyganowski}, C.~J., {Nazari}, P., {et~al.} 2016, \mnras, 462,
  4386

\bibitem[{{K{\"o}hler} {et~al.}(2015){K{\"o}hler}, {Langer}, {de Koter}, {de
  Mink}, {Crowther}, {Evans}, {Gr{\"a}fener}, {Sana}, {Sanyal}, {Schneider}, \&
  {Vink}}]{kohler2015}
{K{\"o}hler}, K., {Langer}, N., {de Koter}, A., {et~al.} 2015, \aap, 573, A71

\bibitem[{{Kudritzki}(1980)}]{kudri80}
{Kudritzki}, R.-P. 1980, \aap, 85, 174

\bibitem[{{Kudritzki} {et~al.}(1992){Kudritzki}, {Hummer}, {Pauldrach}, {Puls},
  {Najarro}, \& {Imhoff}}]{k92}
{Kudritzki}, R.-P., {Hummer}, D.~G., {Pauldrach}, A.~W.~A., {et~al.} 1992,
  \aap, 257, 655

\bibitem[{{Kudritzki} {et~al.}(1995){Kudritzki}, {Lennon}, \& {Puls}}]{k95}
{Kudritzki}, R.-P., {Lennon}, D.~J., \& {Puls}, J. 1995, in Science with the
  VLT, ed. J.~R. {Walsh} \& I.~J. {Danziger}, 246

\bibitem[{{Kudritzki} \& {Puls}(2000)}]{kp2000}
{Kudritzki}, R.-P. \& {Puls}, J. 2000, \araa, 38, 613

\bibitem[{{Kudritzki} {et~al.}(2008){Kudritzki}, {Urbaneja}, {Bresolin},
  {Przybilla}, {Gieren}, \& {Pietrzy{\'n}ski}}]{k08}
{Kudritzki}, R.-P., {Urbaneja}, M.~A., {Bresolin}, F., {et~al.} 2008, \apj,
  681, 269

\bibitem[{{Kudritzki} {et~al.}(2013){Kudritzki}, {Urbaneja}, {Gazak}, {Macri},
  {Hosek}, {Bresolin}, \& {Przybilla}}]{k13}
{Kudritzki}, R.-P., {Urbaneja}, M.~A., {Gazak}, Z., {et~al.} 2013, \apjl, 779,
  L20

\bibitem[{{Kuiper}(1960)}]{kuiper}
{Kuiper}, N.~H. 1960, Proceedings of the Koninklijke Nederlandse Akademie van
  Wetenschappen, 63, 38

\bibitem[{{Langer}(1992)}]{langer92}
{Langer}, N. 1992, \aap, 265, L17

\bibitem[{{Langer}(2012)}]{langer12}
{Langer}, N. 2012, \araa, 50, 107

\bibitem[{{Langer} {et~al.}(2008){Langer}, {Cantiello}, {Yoon}, {Hunter},
  {Brott}, {Lennon}, {de Mink}, \& {Verheijdt}}]{langer08}
{Langer}, N., {Cantiello}, M., {Yoon}, S.-C., {et~al.} 2008, in IAU Symposium,
  Vol. 250, IAU Symposium, ed. F.~{Bresolin}, P.~A. {Crowther}, \& J.~{Puls},
  167--178

\bibitem[{{Langer} \& {Kudritzki}(2014)}]{langerkudri2014}
{Langer}, N. \& {Kudritzki}, R.~P. 2014, \aap, 564, A52

\bibitem[{{Lefever} {et~al.}(2007){Lefever}, {Puls}, \& {Aerts}}]{lefever2007}
{Lefever}, K., {Puls}, J., \& {Aerts}, C. 2007, in Astronomical Society of the
  Pacific Conference Series, Vol. 364, The Future of Photometric,
  Spectrophotometric and Polarimetric Standardization, ed. C.~{Sterken}, 545

\bibitem[{{Leitherer} {et~al.}(1992){Leitherer}, {Robert}, \&
  {Drissen}}]{leitherer92}
{Leitherer}, C., {Robert}, C., \& {Drissen}, L. 1992, \apj, 401, 596

\bibitem[{{Lucy}(2012)}]{lucy12}
{Lucy}, L.~B. 2012, \aap, 544, A120

\bibitem[{{Maeder}(1987)}]{maeder87}
{Maeder}, A. 1987, \aap, 173, 247

\bibitem[{{Maeder} \& {Meynet}(2000)}]{mm2000}
{Maeder}, A. \& {Meynet}, G. 2000, \araa, 38, 143

\bibitem[{{Mahy} {et~al.}(2015){Mahy}, {Rauw}, {De Becker}, {Eenens}, \&
  {Flores}}]{mahy2015}
{Mahy}, L., {Rauw}, G., {De Becker}, M., {Eenens}, P., \& {Flores}, C.~A. 2015,
  \aap, 577, A23

\bibitem[{{Ma{\'{\i}}z Apell{\'a}niz}(2004)}]{jma2004}
{Ma{\'{\i}}z Apell{\'a}niz}, J. 2004, \pasp, 116, 859

\bibitem[{{Ma{\'{\i}}z Apell{\'a}niz} {et~al.}(2014){Ma{\'{\i}}z
  Apell{\'a}niz}, {Evans}, {Barb{\'a}}, {Gr{\"a}fener}, {Bestenlehner},
  {Crowther}, {Garc{\'{\i}}a}, {Herrero}, {Sana}, {Sim{\'o}n-D{\'{\i}}az},
  {Taylor}, {van Loon}, {Vink}, \& {Walborn}}]{jma2014}
{Ma{\'{\i}}z Apell{\'a}niz}, J., {Evans}, C.~J., {Barb{\'a}}, R.~H., {et~al.}
  2014, \aap, 564, A63

\bibitem[{{Markova} {et~al.}(2004){Markova}, {Puls}, {Repolust}, \&
  {Markov}}]{markova04}
{Markova}, N., {Puls}, J., {Repolust}, T., \& {Markov}, H. 2004, \aap, 413, 693

\bibitem[{{Martins} {et~al.}(2005){Martins}, {Schaerer}, {Hillier},
  {Meynadier}, {Heydari-Malayeri}, \& {Walborn}}]{martins05gal}
{Martins}, F., {Schaerer}, D., {Hillier}, D.~J., {et~al.} 2005, \aap, 441, 735

\bibitem[{{Mason} {et~al.}(1998){Mason}, {Gies}, {Hartkopf}, {Bagnuolo}, {ten
  Brummelaar}, \& {McAlister}}]{mason1998}
{Mason}, B.~D., {Gies}, D.~R., {Hartkopf}, W.~I., {et~al.} 1998, \aj, 115, 821

\bibitem[{{Mason} {et~al.}(2009){Mason}, {Hartkopf}, {Gies}, {Henry}, \&
  {Helsel}}]{mason2009}
{Mason}, B.~D., {Hartkopf}, W.~I., {Gies}, D.~R., {Henry}, T.~J., \& {Helsel},
  J.~W. 2009, \aj, 137, 3358

\bibitem[{{Massey} {et~al.}(2004){Massey}, {Bresolin}, {Kudritzki}, {Puls}, \&
  {Pauldrach}}]{massey04}
{Massey}, P., {Bresolin}, F., {Kudritzki}, R.~P., {Puls}, J., \& {Pauldrach},
  A.~W.~A. 2004, \apj, 608, 1001

\bibitem[{{Massey} {et~al.}(2013){Massey}, {Neugent}, {Hillier}, \&
  {Puls}}]{massey13}
{Massey}, P., {Neugent}, K.~F., {Hillier}, D.~J., \& {Puls}, J. 2013, \apj,
  768, 6

\bibitem[{{Massey} {et~al.}(2005){Massey}, {Puls}, {Pauldrach}, {Bresolin},
  {Kudritzki}, \& {Simon}}]{massey05}
{Massey}, P., {Puls}, J., {Pauldrach}, A.~W.~A., {et~al.} 2005, \apj, 627, 477

\bibitem[{{Massey} {et~al.}(2009){Massey}, {Zangari}, {Morrell}, {Puls},
  {DeGioia-Eastwood}, {Bresolin}, \& {Kudritzki}}]{massey09}
{Massey}, P., {Zangari}, A.~M., {Morrell}, N.~I., {et~al.} 2009, \apj, 692, 618

\bibitem[{{Mokiem} {et~al.}(2007{\natexlab{a}}){Mokiem}, {de Koter}, {Evans},
  {Puls}, {Smartt}, {Crowther}, {Herrero}, {Langer}, {Lennon}, {Najarro},
  {Villamariz}, \& {Vink}}]{mokiem2007LMC}
{Mokiem}, M.~R., {de Koter}, A., {Evans}, C.~J., {et~al.} 2007{\natexlab{a}},
  \aap, 465, 1003

\bibitem[{{Mokiem} {et~al.}(2007{\natexlab{b}}){Mokiem}, {de Koter}, {Vink},
  {Puls}, {Evans}, {Smartt}, {Crowther}, {Herrero}, {Langer}, {Lennon},
  {Najarro}, \& {Villamariz}}]{mokiem2007WLR}
{Mokiem}, M.~R., {de Koter}, A., {Vink}, J.~S., {et~al.} 2007{\natexlab{b}},
  \aap, 473, 603

\bibitem[{{Morel} {et~al.}(2015){Morel}, {Castro}, {Fossati}, {Hubrig},
  {Langer}, {Przybilla}, {Sch{\"o}ller}, {Carroll}, {Ilyin}, {Irrgang},
  {Oskinova}, {Schneider}, {D{\'{\i}}az}, {Briquet}, {Gonz{\'a}lez},
  {Kharchenko}, {Nieva}, {Scholz}, {de Koter}, {Hamann}, {Herrero},
  {Ma{\'{\i}}z Apell{\'a}niz}, {Sana}, {Arlt}, {Barb{\'a}}, {Dufton},
  {Kholtygin}, {Mathys}, {Piskunov}, {Reisenegger}, {Spruit}, \&
  {Yoon}}]{morel2015}
{Morel}, T., {Castro}, N., {Fossati}, L., {et~al.} 2015, in IAU Symposium, Vol.
  307, New Windows on Massive Stars, ed. G.~{Meynet}, C.~{Georgy}, J.~{Groh},
  \& P.~{Stee}, 342--347

\bibitem[{{Morrell} {et~al.}(2014){Morrell}, {Massey}, {Neugent}, {Penny}, \&
  {Gies}}]{morrell2014}
{Morrell}, N.~I., {Massey}, P., {Neugent}, K.~F., {Penny}, L.~R., \& {Gies},
  D.~R. 2014, \apj, 789, 139

\bibitem[{{Muijres} {et~al.}(2012){Muijres}, {Vink}, {de Koter}, {M{\"u}ller},
  \& {Langer}}]{muijres}
{Muijres}, L.~E., {Vink}, J.~S., {de Koter}, A., {M{\"u}ller}, P.~E., \&
  {Langer}, N. 2012, \aap, 537, A37

\bibitem[{{M{\"u}ller} \& {Vink}(2014)}]{muellervink}
{M{\"u}ller}, P.~E. \& {Vink}, J.~S. 2014, \aap, 564, A57

\bibitem[{{Packet}(1981)}]{packet81}
{Packet}, W. 1981, \aap, 102, 17

\bibitem[{{Pasquini} {et~al.}(2002){Pasquini}, {Avila}, {Blecha}, {Cacciari},
  {Cayatte}, {Colless}, {Damiani}, {de Propris}, {Dekker}, {di Marcantonio},
  {Farrell}, {Gillingham}, {Guinouard}, {Hammer}, {Kaufer}, {Hill}, {Marteaud},
  {Modigliani}, {Mulas}, {North}, {Popovic}, {Rossetti}, {Royer}, {Santin},
  {Schmutzer}, {Simond}, {Vola}, {Waller}, \& {Zoccali}}]{pasquini}
{Pasquini}, L., {Avila}, G., {Blecha}, A., {et~al.} 2002, The Messenger, 110, 1

\bibitem[{{Petit} {et~al.}(2017){Petit}, {Keszthelyi}, {MacInnis}, {Cohen},
  {Townsend}, {Wade}, {Thomas}, {Owocki}, {Puls}, \& {ud-Doula}}]{petit2016}
{Petit}, V., {Keszthelyi}, Z., {MacInnis}, R., {et~al.} 2017, \mnras, 466, 1052

\bibitem[{{Platais} {et~al.}(2015){Platais}, {van der Marel}, {Lennon},
  {Anderson}, {Bellini}, {Sabbi}, {Sana}, \& {Bedin}}]{platais2015}
{Platais}, I., {van der Marel}, R.~P., {Lennon}, D.~J., {et~al.} 2015, \aj,
  150, 89

\bibitem[{{Puls} {et~al.}(2006){Puls}, {Markova}, {Scuderi}, {Stanghellini},
  {Taranova}, {Burnley}, \& {Howarth}}]{puls06_clumping}
{Puls}, J., {Markova}, N., {Scuderi}, S., {et~al.} 2006, \aap, 454, 625

\bibitem[{{Puls} {et~al.}(2005){Puls}, {Urbaneja}, {Venero}, {Repolust},
  {Springmann}, {Jokuthy}, \& {Mokiem}}]{puls05}
{Puls}, J., {Urbaneja}, M.~A., {Venero}, R., {et~al.} 2005, \aap, 435, 669

\bibitem[{{Puls} {et~al.}(2008){Puls}, {Vink}, \& {Najarro}}]{puls08}
{Puls}, J., {Vink}, J.~S., \& {Najarro}, F. 2008, \aapr, 16, 209

\bibitem[{{Ram{\'{\i}}rez-Agudelo} {et~al.}(2015){Ram{\'{\i}}rez-Agudelo},
  {Sana}, {de Mink}, {H{\'e}nault-Brunet}, {de Koter}, {Langer}, {Tramper},
  {Gr{\"a}fener}, {Evans}, {Vink}, {Dufton}, \& {Taylor}}]{oscar15}
{Ram{\'{\i}}rez-Agudelo}, O.~H., {Sana}, H., {de Mink}, S.~E., {et~al.} 2015,
  \aap, 580, A92

\bibitem[{{Ram{\'{\i}}rez-Agudelo} {et~al.}(2013){Ram{\'{\i}}rez-Agudelo},
  {Sim{\'o}n-D{\'{\i}}az}, {Sana}, {de Koter}, {Sab{\'{\i}}n-Sanjul{\'{\i}}an},
  {de Mink}, {Dufton}, {Gr{\"a}fener}, {Evans}, {Herrero}, {Langer}, {Lennon},
  {Ma{\'{\i}}z Apell{\'a}niz}, {Markova}, {Najarro}, {Puls}, {Taylor}, \&
  {Vink}}]{oscar}
{Ram{\'{\i}}rez-Agudelo}, O.~H., {Sim{\'o}n-D{\'{\i}}az}, S., {Sana}, H.,
  {et~al.} 2013, \aap, 560, A29

\bibitem[{{Ram{\'{\i}}rez-Agudelo}(2017)}]{oscar17}
{Ram{\'{\i}}rez-Agudelo}, O.~H. e.~a. 2017, A\&A, in press

\bibitem[{{Repolust} {et~al.}(2004){Repolust}, {Puls}, \& {Herrero}}]{rph04}
{Repolust}, T., {Puls}, J., \& {Herrero}, A. 2004, \aap, 415, 349

\bibitem[{{Rivero Gonz{\'a}lez} {et~al.}(2012{\natexlab{a}}){Rivero
  Gonz{\'a}lez}, {Puls}, {Massey}, \& {Najarro}}]{rivero12iii}
{Rivero Gonz{\'a}lez}, J.~G., {Puls}, J., {Massey}, P., \& {Najarro}, F.
  2012{\natexlab{a}}, \aap, 543, A95

\bibitem[{{Rivero Gonz{\'a}lez} {et~al.}(2011){Rivero Gonz{\'a}lez}, {Puls}, \&
  {Najarro}}]{rivero12i}
{Rivero Gonz{\'a}lez}, J.~G., {Puls}, J., \& {Najarro}, F. 2011, \aap, 536, A58

\bibitem[{{Rivero Gonz{\'a}lez} {et~al.}(2012{\natexlab{b}}){Rivero
  Gonz{\'a}lez}, {Puls}, {Najarro}, \& {Brott}}]{rivero12ii}
{Rivero Gonz{\'a}lez}, J.~G., {Puls}, J., {Najarro}, F., \& {Brott}, I.
  2012{\natexlab{b}}, \aap, 537, A79

\bibitem[{{Sabbi} {et~al.}(2016){Sabbi}, {Lennon}, {Anderson}, {Cignoni}, {van
  der Marel}, {Zaritsky}, {De Marchi}, {Panagia}, {Gouliermis}, {Grebel},
  {Gallagher}, {Smith}, {Sana}, {Aloisi}, {Tosi}, {Evans}, {Arab}, {Boyer}, {de
  Mink}, {Gordon}, {Koekemoer}, {Larsen}, {Ryon}, \& {Zeidler}}]{sabbi2016}
{Sabbi}, E., {Lennon}, D.~J., {Anderson}, J., {et~al.} 2016, \apjs, 222, 11

\bibitem[{{Sabbi} {et~al.}(2012){Sabbi}, {Lennon}, {Gieles}, {de Mink},
  {Walborn}, {Anderson}, {Bellini}, {Panagia}, {van der Marel}, \& {Ma{\'{\i}}z
  Apell{\'a}niz}}]{sabbi2012}
{Sabbi}, E., {Lennon}, D.~J., {Gieles}, M., {et~al.} 2012, \apjl, 754, L37

\bibitem[{{Sab{\'{\i}}n-Sanjuli{\'a}n}
  {et~al.}(2014){Sab{\'{\i}}n-Sanjuli{\'a}n}, {Sim{\'o}n-D{\'{\i}}az},
  {Herrero}, {Walborn}, {Puls}, {Ma{\'{\i}}z Apell{\'a}niz}, {Evans}, {Brott},
  {de Koter}, {Garcia}, {Markova}, {Najarro}, {Ram{\'{\i}}rez-Agudelo}, {Sana},
  {Taylor}, \& {Vink}}]{cssj}
{Sab{\'{\i}}n-Sanjuli{\'a}n}, C., {Sim{\'o}n-D{\'{\i}}az}, S., {Herrero}, A.,
  {et~al.} 2014, \aap, 564, A39

\bibitem[{{Salpeter}(1955)}]{salpeter55}
{Salpeter}, E.~E. 1955, \apj, 121, 161

\bibitem[{{Sana} {et~al.}(2013){Sana}, {de Koter}, {de Mink}, {Dunstall},
  {Evans}, {H{\'e}nault-Brunet}, {Ma{\'{\i}}z Apell{\'a}niz},
  {Ram{\'{\i}}rez-Agudelo}, {Taylor}, {Walborn}, {Clark}, {Crowther},
  {Herrero}, {Gieles}, {Langer}, {Lennon}, \& {Vink}}]{sana13}
{Sana}, H., {de Koter}, A., {de Mink}, S.~E., {et~al.} 2013, \aap, 550, A107

\bibitem[{{Sana} {et~al.}(2012){Sana}, {de Mink}, {de Koter}, {Langer},
  {Evans}, {Gieles}, {Gosset}, {Izzard}, {Le Bouquin}, \& {Schneider}}]{sana12}
{Sana}, H., {de Mink}, S.~E., {de Koter}, A., {et~al.} 2012, Science, 337, 444

\bibitem[{{Sana} {et~al.}(2014){Sana}, {Le Bouquin}, {Lacour}, {Berger},
  {Duvert}, {Gauchet}, {Norris}, {Olofsson}, {Pickel}, {Zins}, {Absil}, {de
  Koter}, {Kratter}, {Schnurr}, \& {Zinnecker}}]{sana14}
{Sana}, H., {Le Bouquin}, J.-B., {Lacour}, S., {et~al.} 2014, \apjs, 215, 15

\bibitem[{{Santolaya-Rey} {et~al.}(1997){Santolaya-Rey}, {Puls}, \&
  {Herrero}}]{santolaya}
{Santolaya-Rey}, A.~E., {Puls}, J., \& {Herrero}, A. 1997, \aap, 323, 488

\bibitem[{{Schneider} {et~al.}(2014){Schneider}, {Langer}, {de Koter}, {Brott},
  {Izzard}, \& {Lau}}]{bonnsai}
{Schneider}, F.~R.~N., {Langer}, N., {de Koter}, A., {et~al.} 2014, \aap, 570,
  A66

\bibitem[{{Schneider} {et~al.}(2016){Schneider}, {Podsiadlowski}, {Langer},
  {Castro}, \& {Fossati}}]{fabian2016}
{Schneider}, F.~R.~N., {Podsiadlowski}, P., {Langer}, N., {Castro}, N., \&
  {Fossati}, L. 2016, \mnras, 457, 2355

\bibitem[{{Sim{\'o}n-D{\'{\i}}az} {et~al.}(2011){Sim{\'o}n-D{\'{\i}}az},
  {Castro}, {Herrero}, {Puls}, {Garcia}, \&
  {Sab{\'{\i}}n-Sanjuli{\'a}n}}]{iacob11}
{Sim{\'o}n-D{\'{\i}}az}, S., {Castro}, N., {Herrero}, A., {et~al.} 2011,
  Journal of Physics Conference Series, 328, 012021

\bibitem[{{Sim{\'o}n-D{\'{\i}}az} {et~al.}(2017){Sim{\'o}n-D{\'{\i}}az},
  {Godart}, {Castro}, {Herrero}, {Aerts}, {Puls}, {Telting}, \&
  {Grassitelli}}]{ssd16}
{Sim{\'o}n-D{\'{\i}}az}, S., {Godart}, M., {Castro}, N., {et~al.} 2017, \aap,
  597, A22

\bibitem[{{Sim{\'o}n-D{\'{\i}}az} \& {Herrero}(2014)}]{ssd13iacob}
{Sim{\'o}n-D{\'{\i}}az}, S. \& {Herrero}, A. 2014, \aap, 562, A135

\bibitem[{{Sim{\'o}n-D{\'{\i}}az} {et~al.}(2014){Sim{\'o}n-D{\'{\i}}az},
  {Herrero}, {Sab{\'{\i}}n-Sanjuli{\'a}n}, {Najarro}, {Garcia}, {Puls},
  {Castro}, \& {Evans}}]{ssd_letter14}
{Sim{\'o}n-D{\'{\i}}az}, S., {Herrero}, A., {Sab{\'{\i}}n-Sanjuli{\'a}n}, C.,
  {et~al.} 2014, \aap, 570, L6

\bibitem[{{Sota} {et~al.}(2014){Sota}, {Ma{\'{\i}}z Apell{\'a}niz}, {Morrell},
  {Barb{\'a}}, {Walborn}, {Gamen}, {Arias}, \& {Alfaro}}]{sota14}
{Sota}, A., {Ma{\'{\i}}z Apell{\'a}niz}, J., {Morrell}, N.~I., {et~al.} 2014,
  \apjs, 211, 10

\bibitem[{{Sota} {et~al.}(2011){Sota}, {Ma{\'{\i}}z Apell{\'a}niz}, {Walborn},
  {Alfaro}, {Barb{\'a}}, {Morrell}, {Gamen}, \& {Arias}}]{sota}
{Sota}, A., {Ma{\'{\i}}z Apell{\'a}niz}, J., {Walborn}, N.~R., {et~al.} 2011,
  \apjs, 193, 24

\bibitem[{{Sundqvist} {et~al.}(2014){Sundqvist}, {Puls}, \&
  {Owocki}}]{sundqvist14}
{Sundqvist}, J.~O., {Puls}, J., \& {Owocki}, S.~P. 2014, \aap, 568, A59

\bibitem[{{{\v S}urlan} {et~al.}(2013){{\v S}urlan}, {Hamann}, {Aret},
  {Kub{\'a}t}, {Oskinova}, \& {Torres}}]{surlan13}
{{\v S}urlan}, B., {Hamann}, W.-R., {Aret}, A., {et~al.} 2013, \aap, 559, A130

\bibitem[{{Vink}(2012)}]{vink2012}
{Vink}, J.~S. 2012, in IAU Symposium, Vol. 279, Death of Massive Stars:
  Supernovae and Gamma-Ray Bursts, ed. P.~{Roming}, N.~{Kawai}, \& E.~{Pian},
  29--33

\bibitem[{{Vink} {et~al.}(2001){Vink}, {de Koter}, \& {Lamers}}]{vink2001}
{Vink}, J.~S., {de Koter}, A., \& {Lamers}, H.~J.~G.~L.~M. 2001, \aap, 369, 574

\bibitem[{{Vink} {et~al.}(2011){Vink}, {Muijres}, {Anthonisse}, {de Koter},
  {Gr{\"a}fener}, \& {Langer}}]{vink2011}
{Vink}, J.~S., {Muijres}, L.~E., {Anthonisse}, B., {et~al.} 2011, \aap, 531,
  A132

\bibitem[{{Wade} \& {MiMeS Collaboration}(2015)}]{wade2015}
{Wade}, G.~A. \& {MiMeS Collaboration}. 2015, in Astronomical Society of the
  Pacific Conference Series, Vol. 494, Physics and Evolution of Magnetic and
  Related Stars, ed. Y.~Y. {Balega}, I.~I. {Romanyuk}, \& D.~O. {Kudryavtsev},
  30

\bibitem[{{Wade} {et~al.}(2016){Wade}, {Neiner}, {Alecian}, {Grunhut}, {Petit},
  {Batz}, {Bohlender}, {Cohen}, {Henrichs}, {Kochukhov}, {Landstreet},
  {Manset}, {Martins}, {Mathis}, {Oksala}, {Owocki}, {Rivinius}, {Shultz},
  {Sundqvist}, {Townsend}, {ud-Doula}, {Bouret}, {Braithwaite}, {Briquet},
  {Carciofi}, {David-Uraz}, {Folsom}, {Fullerton}, {Leroy}, {Marcolino},
  {Moffat}, {Naz{\'e}}, {Louis}, {Auri{\`e}re}, {Bagnulo}, {Bailey},
  {Barb{\'a}}, {Blaz{\`e}re}, {B{\"o}hm}, {Catala}, {Donati}, {Ferrario},
  {Harrington}, {Howarth}, {Ignace}, {Kaper}, {L{\"u}ftinger}, {Prinja},
  {Vink}, {Weiss}, \& {Yakunin}}]{wade2016}
{Wade}, G.~A., {Neiner}, C., {Alecian}, E., {et~al.} 2016, \mnras, 456, 2

\bibitem[{{Walborn}(2009)}]{w09}
{Walborn}, N.~R. 2009, {Optically observable zero-age main-sequence O stars:}
  (Massive Stars: From Pop III and GRBs to the Milky Way.~Space Telescope
  Science Institute Symposium Series No.~20.~Edited by Mario Livio and Eva
  Villaver.~ Cambridge University Press, 2009, ISSN 9780521762632, p.167-177),
  167--177

\bibitem[{{Walborn} {et~al.}(1999){Walborn}, {Drissen}, {Parker}, {Saha},
  {MacKenty}, \& {White}}]{w99}
{Walborn}, N.~R., {Drissen}, L., {Parker}, J.~W., {et~al.} 1999, \aj, 118, 1684

\bibitem[{{Walborn} {et~al.}(2002{\natexlab{a}}){Walborn}, {Howarth}, {Lennon},
  {Massey}, {Oey}, {Moffat}, {Skalkowski}, {Morrell}, {Drissen}, \&
  {Parker}}]{w02}
{Walborn}, N.~R., {Howarth}, I.~D., {Lennon}, D.~J., {et~al.}
  2002{\natexlab{a}}, \aj, 123, 2754

\bibitem[{{Walborn} {et~al.}(2002{\natexlab{b}}){Walborn},
  {Ma{\'{\i}}z-Apell{\'a}niz}, \& {Barb{\'a}}}]{walborn2002}
{Walborn}, N.~R., {Ma{\'{\i}}z-Apell{\'a}niz}, J., \& {Barb{\'a}}, R.~H.
  2002{\natexlab{b}}, \aj, 124, 1601

\bibitem[{{Walborn} {et~al.}(2014){Walborn}, {Sana}, {Sim{\'o}n-D{\'{\i}}az},
  {Ma{\'{\i}}z Apell{\'a}niz}, {Taylor}, {Evans}, {Markova}, {Lennon}, \& {de
  Koter}}]{walborn13}
{Walborn}, N.~R., {Sana}, H., {Sim{\'o}n-D{\'{\i}}az}, S., {et~al.} 2014, \aap,
  564, A40

\bibitem[{{Walborn} {et~al.}(2012){Walborn}, {Sana}, {Taylor},
  {Sim{\'o}n-D{\'{\i}}az}, \& {Evans}}]{w11}
{Walborn}, N.~R., {Sana}, H., {Taylor}, W.~D., {Sim{\'o}n-D{\'{\i}}az}, S., \&
  {Evans}, C.~J. 2012, in ASP Conference Series, Vol. 465, Proceedings of a
  Scientific Meeting in Honor of Anthony F. J. Moffat, ed. L.~{Drissen},
  C.~{Rubert}, N.~{St-Louis}, \& A.~F.~J. {Moffat}, 490

\bibitem[{{Weidner} \& {Vink}(2010)}]{weidner10}
{Weidner}, C. \& {Vink}, J.~S. 2010, \aap, 524, A98

\bibitem[{{Woosley} \& {Heger}(2006)}]{woosley06}
{Woosley}, S.~E. \& {Heger}, A. 2006, \apj, 637, 914

\bibitem[{{Woosley} {et~al.}(2002){Woosley}, {Heger}, \& {Weaver}}]{woosley02}
{Woosley}, S.~E., {Heger}, A., \& {Weaver}, T.~A. 2002, Reviews of Modern
  Physics, 74, 1015

\bibitem[{{Yorke}(1986)}]{yorke86}
{Yorke}, H.~W. 1986, \araa, 24, 49

\bibitem[{{Zinnecker} \& {Yorke}(2007)}]{zinnecker07}
{Zinnecker}, H. \& {Yorke}, H.~W. 2007, \araa, 45, 481

\end{thebibliography}


\newpage

\begin{appendix}

\onecolumn
\section{Stellar and wind parameters}

{\scriptsize
\setlength\LTcapwidth{\linewidth}
\begin{longtable}{llccccccccccl}
\caption{\label{tab1_HHe} Stellar and wind parameters obtained from quantitative analysis of our sample of O dwarfs.} \\
\hline\hline 
\\ [-1.5 ex]
VFTS & Spectral Type &\vsini&$T_{\rm eff}$&$\Delta T_{\rm eff}$&\grav$^{(1)}$&\gravc$^{(2)}$&$\Delta$\grav$^{(3)}$& 
Y(He)$^{(4)}$&\logq&$\Delta$\logq&log$D_{\rm mom}^{(5)}$&Comments$^{(6)}$\\
&&[\kms]&[K]&[K]&[cgs]&[cgs]&[cgs]&&&&&\\
\hline
\\ [-2. ex]

\endfirsthead
\caption{\it{continued}} \\
\hline\hline
\\ [-1.5 ex]
VFTS & Spectral Type &\vsini&$T_{\rm eff}$&$\Delta T_{\rm eff}$&\grav$^{(1)}$&\gravc$^{(2)}$&$\Delta$\grav$^{(3)}$& 
Y(He)$^{(4)}$&\logq&$\Delta$\logq&log$D_{\rm mom}^{(5)}$&Comments$^{(6)}$\\
&&[\kms]&[K]&[K]&[cgs]&[cgs]&[cgs]&&&&&\\
\hline
\\ [-2. ex]
\endhead
\hline
\multicolumn{13}{r}{\it{continued on next page}}\\
\endfoot
\endlastfoot
014 & O8.5 Vz           & \pp90 & 37100 & \pp600 &    3.91 &  --  & 0.10 & $<$0.06 & $<$$-$13.0 &  -- &      -- & SBs \\
021 & O9.5 IV           & \pp40 & 33800 & \pp900 &    3.90 & 3.90 & 0.10 & \pl0.10 & $<$$-$13.0 &  -- & $<$27.9 & SB? \\
065 & O8 V(n)           &   165 & 37100 &   1100 &    4.06 & 4.08 & 0.16 & \pl0.10 & $<$$-$13.0 &  -- & $<$27.9 & ... \\
067 & O9.5 Vz           & \pp40 & 35200 &   1100 &    4.12 & 4.12 & 0.19 & \pl0.08 & $<$$-$13.0 &  -- & $<$27.6 & SB? \\
074 & O9 Vn             &   265 & 35100 &   1300 &    4.18 & 4.23 & 0.21 & \pl0.10 & $<$$-$12.9 &  -- & $<$28.0 & ... \\
089 & O6.5 V((f))z      & \pp50 & 39700 & \pp700 &    4.02 & 4.02 & 0.12 & \pl0.13 & $<$$-$13.3 &  -- & $<$27.6 & ... \\
093 & O9.2 III-IV       & \pp60 & 34700 & \pp500 &    3.87 & 3.87 & 0.10 & \pl0.10 & \pl$-$13.1 & 0.4 & \pl28.1 & SBvs \\
096 & O6 V((n))((fc))z  &   125 & 40100 & \pp300 &    3.90 & 3.91 & 0.10 & \pl0.09 & \pl$-$13.0 & 0.4 & \pl28.7 & SBvs VM2 \\
110 & O6 V((n))z        &   175 & 39900 &   1000 &    3.86 & 3.88 & 0.10 & $<$0.06 & $<$$-$13.0 &  -- & $<$28.2 & VM2? \\
117 & O6: Vz            & \pp75 & 41300 &   1500 &    4.14 & 4.14 & 0.16 & \pl0.12 & $<$$-$13.0 &  -- & $<$28.0 & SB? \\
123 & O6.5 Vz           & \pp65 & 40400 & \pp700 &    4.10 & 4.10 & 0.12 & \pl0.13 & $<$$-$13.3 &  -- & $<$27.6 & SB? \\
130 & O8.5 V((n))       &   170 & 36500 &   1300 &    4.09 & 4.11 & 0.19 & \pl0.08 & $<$$-$13.0 &  -- & $<$28.4 & ... \\
132 & O9.5 Vz           & \pp40 & 35600 & \pp700 &    4.18 & 4.18 & 0.10 & \pl0.10 & $<$$-$13.0 &  -- & $<$27.9 & ... \\
138 & O9 Vn             &   350 & 34600 & \pp900 &    4.10 & 4.20 & 0.14 & \pl0.10 & $<$$-$13.0 &  -- & $<$27.9 & SB2? \\
149 & O9.5 V            &   125 & 35000 &   1400 &    4.12 & 4.13 & 0.24 & \pl0.12 & $<$$-$13.0 &  -- & $<$27.9 & ... \\
154 & O8.5 V            & \pp55 & 37400 & \pp700 &    4.12 & 4.12 & 0.13 & \pl0.09 & $<$$-$13.0 &  -- & $<$28.7 & SBs \\
168 & O8.5 Vz           & \pp40 & 37300 & \pp500 &    4.02 & 4.02 & 0.10 & \pl0.10 & $<$$-$13.5 &  -- & $<$27.5 & SB? \\
223 & O9.5 IV           & \pp40 & 34800 & \pp500 &    4.02 & 4.02 & 0.10 & \pl0.09 & $<$$-$13.1 &  -- & $<$28.2 & SBvs \\
249 & O8 Vn             &   300 & 36500 & \pp800 &    4.04 & 4.11 & 0.11 & \pl0.10 & $<$$-$13.3 &  -- & $<$27.6 & ... \\
250 & O9.2 V((n))       &   155 & 35400 & \pp800 &    4.12 & 4.14 & 0.15 & \pl0.09 & $<$$-$13.0 &  -- & $<$28.0 & ... \\
251 & O9.5 IV           & \pp40 & 33700 & \pp600 &    4.01 & 4.01 & 0.10 & \pl0.10 & $<$$-$13.1 &  -- & $<$27.8 & SB? \\
252 & O8.5 Vz           &   100 & 37000 & \pp500 &    4.21 & 4.22 & 0.10 & \pl0.11 & $<$$-$13.0 &  -- & $<$27.9 & ... \\
266 & O8 V((f))z        & \pp40 & 38000 & \pp200 &    4.01 & 4.01 & 0.10 & \pl0.10 & $<$$-$13.0 &  -- & $<$28.0 & ... \\
280 & O9 V((n))         &   150 & 34400 & \pp600 &    3.82 & 3.85 & 0.10 & \pl0.10 & $<$$-$13.4 &  -- & $<$27.5 & ... \\
285 & O7.5 Vnnn         &   600 & 35300 & \pp900 &    3.63 & 4.08 & 0.10 & \pl0.14 & $<$$-$13.0 &  -- & $<$27.9 & ... \\
290 & O9.5 IV           & \pp40 & 34000 & \pp500 &    3.99 & 3.99 & 0.10 & \pl0.10 & $<$$-$13.1 &  -- & $<$27.8 & SB? \\
303 & O9.5 IV           & \pp60 & 34700 & \pp500 & \ps4.36*& 4.36 & 0.10 & \pl0.09 & $<$$-$13.0 &  -- & $<$28.7 & VM2 \\
355 & O4 V((n))((fc))z  &   135 & 43400 & \pp600 &    3.84 & 3.86 & 0.10 & \pl0.09 & $<$$-$12.9 &  -- & $<$28.2 & SB2 NC \\
356 & O6: V(n)z         &   215 & 39300 &   1300 &    3.99 & 4.03 & 0.13 & \pl0.10 & $<$$-$13.0 &  -- & $<$28.1 & SB? \\
361 & O8.5 V            & \pp70 & 36900 & \pp700 &    4.07 & 4.07 & 0.10 & $<$0.06 & $<$$-$12.7 &  -- & $<$28.8 & ... \\
369 & O9.7 V            & \pp40 & 33400 &   1200 &    4.10 & 4.10 & 0.18 & $<$0.08 & $<$$-$13.0 &  -- & $<$27.9 & ... \\
380 & O6-7 Vz           & \pp65 & 39100 & \pp700 &    4.13 & 4.13 & 0.10 & $<$0.08 & $<$$-$12.9 &  -- & $<$28.0 & ... \\
385 & O4-5 V((n))((fc)) &   120 & 42900 &   1700 &    3.86 & 3.87 & 0.10 & \pl0.09 & \pl$-$12.5 & 0.2 & \pl28.5 & SBs \\
392 & O6-7 V((f))z      & \pp40 & 37600 & \pp800 &    3.87 & 3.87 & 0.10 & \pl0.10 & \pl$-$13.1 & 0.4 & \pl27.8 & ... \\
398 & O5.5 V((n))((f))z & \pp65 & 41200 &   1000 &    4.03 & 4.03 & 0.10 & $<$0.06 & \pl$-$12.6 & 0.2 & \pl28.9 & SBvs \\
418 & O5 V((n))((fc))z  &   135 & 43200 &   1700 &    4.09 & 4.10 & 0.13 & \pl0.09 & $<$$-$13.0 &  -- & $<$28.1 & SB? \\
419 & O9: V(n)          &   145 & 33100 & \pp900 &    3.61 & 3.64 & 0.10 & \pl0.11 & $<$$-$12.7 &  -- & $<$28.3 & ... \\
470 & O6: V((f))z       & \pp75 & 39300 & \pp600 &    3.93 & 3.94 & 0.10 & \pl0.10 & $<$$-$13.0 &  -- & $<$27.7 & ... \\
472 & O6 Vz             & \pp40 & 40400 & \pp900 &    4.12 & 4.12 & 0.12 & \pl0.11 & $<$$-$12.9 &  -- & $<$28.0 & ... \\
483 & O9 V              & \pp40 & 33700 & \pp900 &    4.09 &   -- & 0.11 & \pl0.11 & \pl$-$12.8 & 0.3 &      -- & SB? \\
484 & O6-7 V((n))       &   120 & 35700 & \pp700 &    3.67 & 3.68 & 0.10 & $<$0.06 & \pl$-$12.5 & 0.2 & \pl28.7 & ... \\
488 & O6 V((f))z        & \pp55 & 40700 & \pp700 &    3.87 & 3.87 & 0.10 & \pl0.09 & \pl$-$12.8 & 0.3 & \pl28.2 & ... \\
491 & O6 V((fc))        & \pp50 & 40400 & \pp800 &    3.84 & 3.84 & 0.10 & \pl0.09 & \pl$-$12.6 & 0.2 & \pl28.6 &   SB? \\
493 & O9 V              &   200 & 37100 &   1000 & \ps4.28*& 4.27 & 0.10 & $<$0.07 & $<$$-$13.0 &  -- & $<$28.5 & ... \\
494 & O8 V(n)           &   230 & 38900 &   1700 &    4.18 & 4.21 & 0.21 & \pl0.09 & $<$$-$13.0 &  -- & $<$28.2 & SB2? \\
498 & O9.5 V            & \pp40 & 33200 & \pp800 &    4.12 & 4.12 & 0.15 & \pl0.09 & $<$$-$13.0 &  -- & $<$28.3 & ... \\
505 & O9.5 V-III        &   100 & 34000 & \pp700 & \ps4.30*& 4.27 & 0.10 & $<$0.07 & $<$$-$13.0 &  -- & $<$28.1 & VM2? \\
511 & O5 V((n))((fc))z  &   105 & 43700 &   1700 & \ps4.27*& 4.25 & 0.11 & \pl0.10 & $<$$-$13.0 &  -- & $<$28.6 & SB1s \\
517 & O9.5 V-III((n))   &   120 & 33000 & \pp500 &    4.01 & 4.02 & 0.10 & \pl0.12 & $<$$-$13.0 &  -- & $<$28.5 & ... \\
521 & O9 V(n)           &   150 & 34800 & \pp600 &    4.12 & 4.13 & 0.10 & \pl0.09 & $<$$-$13.0 &  -- & $<$28.7 & VM2 \\
536 & O6 Vz             & \pp40 & 41500 &   1500 & \ps4.24*& 4.23 & 0.15 & \pl0.08 & $<$$-$13.0 &  -- & $<$28.3 & SB? \\
549 & O6.5 Vz           &   110 & 39800 &   1200 &    4.04 & 4.05 & 0.16 & \pl0.09 & $<$$-$13.0 &  -- & $<$28.0 & SB? \\
554 & O9.7 V            & \pp45 & 34100 & \pp800 & \ps4.30*&   -- & 0.10 & $<$0.06 & \pl$-$12.7 & 0.3 &      -- & ... \\
560 & O9.5 V            & \pp40 & 33600 &   1200 &    4.20 & 4.20 & 0.16 & \pl0.09 & $<$$-$13.0 &  -- & $<$27.7 & ... \\
582 & O9.5 V((n))       &   115 & 35000 & \pp800 & \ps4.29*&   -- & 0.10 & $<$0.07 & $<$$-$13.0 &  -- &      -- & ... \\
592 & O9.5 Vn           &   295 & 33600 &   1000 & \ps4.25*& 4.28 & 0.13 & $<$0.08 & $<$$-$13.0 &  -- & $<$28.2 & ... \\
597 & O8-9 V(n)         &   210 & 35400 & \pp700 &    3.90 & 3.94 & 0.11 & \pl0.11 & $<$$-$13.3 &  -- & $<$27.6 & ... \\
601 & O5-6 V((n))z      &   125 & 40300 & \pp500 &    3.93 & 3.94 & 0.10 & \pl0.09 & $<$$-$13.0 &  -- & $<$28.5 & ... \\
611 & O8 V(n)           &   210 & 37400 & \pp900 &    4.09 & 4.13 & 0.14 & \pl0.10 & $<$$-$13.3 &  -- & $<$27.6 & ... \\
627 & O9.7 V            & \pp50 & 33600 & \pp600 &    4.11 & 4.11 & 0.12 & \pl0.10 & $<$$-$13.0 &  -- & $<$27.9 & ... \\
635 & O9.5 IV           & \pp60 & 34100 & \pp500 &    4.00 & 4.00 & 0.10 & \pl0.09 & $<$$-$13.0 &  -- & $<$28.0 & SBvs \\
638 & O8.5 Vz           & \pp45 & 36900 & \pp500 &    4.20 & 4.20 & 0.10 & \pl0.10 & $<$$-$13.0 &  -- & $<$27.8 & ... \\
639 & O9.7 V            & \pp65 & 33700 & \pp500 &    4.18 & 4.18 & 0.10 & \pl0.09 & $<$$-$13.0 &  -- & $<$28.2 & SB? \\
649 & O9.5 V            &   105 & 34800 & \pp600 &    4.18 & 4.19 & 0.10 & \pl0.09 & $<$$-$13.0 &  -- & $<$28.0 & SB2 \\
660 & O9.5 Vnn          &   515 & 32300 &   1000 &    3.95 & 4.15 & 0.16 & \pl0.11 & $<$$-$13.3 &  -- & $<$27.9 & ... \\
663 & O8.5 V            & \pp90 & 36500 &   1700 &    4.02 & 4.03 & 0.29 & \pl0.09 & $<$$-$12.7 &  -- & $<$28.1 & SB? \\
677 & O9.5 V            & \pp40 & 35900 &   1100 & \ps4.24*& 4.20 & 0.13 & $<$0.06 & \pl$-$12.8 & 0.3 & \pl28.6 & VM3 \\
679 & O9.5 V            & \pp40 & 33200 & \pp900 &    4.10 & 4.10 & 0.15 & $<$0.06 & $<$$-$12.7 &  -- & $<$28.3 & SB? \\
704 & O9.2 V(n)         &   240 & 34200 &   1500 &    3.98 &  --  & 0.22 & \pl0.09 & $<$$-$12.7 &  -- &      -- & SB? \\
706 & O6-7 Vnnz         &   375 & 38000 &   1200 &    3.80 & 3.95 & 0.13 & \pl0.11 & $<$$-$12.9 &  -- & $<$28.0 & ... \\
710 & O9.5 IV           & \pp60 & 35000 & \pp800 & \ps4.24*& 4.24 & 0.12 & \pl0.09 & $<$$-$13.0 &  -- & $<$27.8 & SB? \\
716 & O9.5 IV           &   105 & 33200 & \pp600 &    3.95 & 3.96 & 0.10 & \pl0.09 & $<$$-$13.0 &  -- & $<$28.0 & SBs \\
717 & O9 IV             & \pp50 & 35000 & \pp500 &    3.89 & 3.89 & 0.10 & \pl0.09 & $<$$-$13.0 &  -- & $<$28.2 & SB?VM? \\
722 & O7 Vnnz           &   405 & 36600 & \pp800 &    3.84 & 4.01 & 0.10 & \pl0.11 & $<$$-$13.4 &  -- & $<$27.5 & SB? NC \\
724 & O7 Vnnz           &   370 & 37600 &   3300 &    3.78 & 3.93 & 0.41 & \pl0.19 & $<$$-$13.0 &  -- & $<$27.9 & NC \\
737 & O9 V              & \pp50 & 37500 & \pp700 & \ps4.30*& 4.30 & 0.10 & \pl0.08 & \pl$-$13.1 & 0.4 & \pl28.5 & ... \\
746 & O6 Vnn            &   275 & 39900 &   1200 &    3.86 & 3.92 & 0.10 & \pl0.08 & \pl$-$12.6 & 0.2 & \pl28.5 & ... \\
751 & O7-8 Vnnz         &   360 & 36000 &   1500 &    3.90 & 4.01 & 0.25 & \pl0.10 & $<$$-$13.0 &  -- & $<$28.1 & NC \\
761 & O6.5 V((n))((f))z &   110 & 40300 & \pp700 &    4.15 & 4.16 & 0.10 & \pl0.18 & $<$$-$13.5 &  -- & $<$27.5 & NC \\
768 & O8 Vn             &   290 & 35100 &   1200 &    3.88 & 3.95 & 0.18 & \pl0.10 & $<$$-$13.3 &  -- & $<$28.2 & SB2? NC \\
770 & O7 Vnn            &   350 & 37800 &   1100 &    3.95 & 4.06 & 0.15 & \pl0.10 & $<$$-$13.0 &  -- & $<$28.0 & ... \\
775 & O9.2 V            & \pp40 & 35900 &   1300 &    4.14 & 4.14 & 0.20 & \pl0.12 & $<$$-$13.0 &  -- & $<$27.9 & SB? NC \\
778 & O9.5 V            &   125 & 34200 &   1400 &    4.18 & 4.19 & 0.21 & \pl0.09 & $<$$-$12.9 &  -- & $<$28.3 & ... \\
849 & O7 Vz             & \pp95 & 39800 & \pp600 &    4.16 & 4.17 & 0.11 & \pl0.11 & $<$$-$13.4 &  -- & $<$27.6 & NC \\
892 & O9 V              & \pp40 & 35800 & \pp600 &    3.98 & 3.98 & 0.10 & \pl0.11 & $<$$-$13.2 &  -- & $<$27.6 & NC \\

\hline
\\[-2. ex]
\multicolumn{13}{l}{$^{(1)}$\, Cases with a particularly high gravity (\grav\,$>$\,4.2) 
are indicated with an asterisk.}\\
\multicolumn{13}{l}{$^{(2)}$\, $\rm{\log{g_c}=\log{[g+(v\sin{i})^2/R]}}$ \citep[see][]{h92,rph04}. }\\
\multicolumn{13}{l}{$^{(3)}$\, Formal errors adopting a minimum value of 0.1~dex.} \\
\multicolumn{13}{l}{$^{(4)}$\, $\rm{\Delta}$Y(He)\,=\,0.02.}\\
\multicolumn{13}{l}{$^{(5)}$\, We adopt $\rm{\Delta}$log$D_{\rm mom}$\,=\,0.4~dex.}\\
\multicolumn{13}{l}{$^{(6)}$\, Relevant comments from Table~1 of
  \cite{walborn13}: SB\,$=$\,spectroscopic binary
  (SB1\,$=$\,single lined, SB2\,$=$\,double lined); s\,$=$\,small-amplitude shift (10\,-\,20\,km\,s$^{-1}$);}\\
\multicolumn{13}{l}{\, vs\,$=$\,very small-amplitude shift
  ($<$10\,km\,s$^{-1}$ ); SB?\,$=$\,stellar absorption
  displaced from nebular emission lines but no radial-velocity variation measured; SB2?\,$=$\,confirmed SB}\\
\multicolumn{13}{l}{\, with possible second component or unconfirmed
  SB but with two components visible in the line of sight; VMn\,$=$\,
  visual multiple of n components within the 1\farcs2 Medusa fibre,}\\
\multicolumn{13}{l}{\, as determined from the {\em HST}/WFC3 images; NC\,$=$\,no coverage in {\em HST} imaging.}\\
\end{longtable}
}

{\tiny
\setlength\LTcapwidth{\linewidth}
\begin{longtable}{llccccccl}
  \caption{\label{tab1_HHeN} First estimates of stellar and wind parameters from HHeN analysis of stars for which \ion{He}{i}/\textsc{ii} analysis alone was considered unreliable.} \\
  \hline\hline
  \\ [-1.5 ex]
  VFTS& Spectral Type &v\,sin\,i&$T_{\rm eff}$&\grav&\grav$_c$&\logq&log$D_{\rm mom}$&Comments\\
  &&[\kms]&[K]&[cgs]&[cgs]&&&\\
  \hline
   \\ [-2. ex]
\endfirsthead
\caption{\it{continued}} \\
\hline\hline
VFTS& Spectral Type &v\,sin\,i&$T_{\rm eff}$&\grav&\grav$_c$&\logq&log$D_{\rm mom}$&Comments\\
 &&[\kms]&[K]&[cgs]&[cgs]&&&\\
\hline
\endhead
\hline
\multicolumn{9}{r}{\it{continued on next page}}\\
\endfoot
\hline
\endlastfoot
072 & O2 V-III(n)((f*))  &   205 & 54000 & 4.00 & 4.02 & \pl$-$12.5 & \pl28.9 & NC \\
169 & O2.5 V(n)((f*))    &   200 & 47000 & 3.90 & 3.92 & \pl$-$12.5 & \pl29.0 & SB? \\
216 & O4 V((fc))         &   100 & 43000 & 3.80 & 3.81 & \pl$-$12.7 & \pl28.8 & SB? \\
382 & O4-5 V((fc))z      & \pp75 & 40000 & 3.80 & 3.81 & \pl$-$13.0 & $<$28.3 & ... \\
410 & O7-8 V             & \pp40 & 34000 & 3.80 & 3.80 & \pl$-$12.7 & \pl28.5 & VM3 \\
435 & O7-8 V             & \pp80 & 37000 & 3.90 & 3.91 & $<$$-$13.0 & $<$28.3 & z? \\
436 & O7-8 V             & \pp60 & 35000 & 3.90 & 3.90 & \pl$-$13.0 & $<$27.8 & SB? z? \\
468 & O2 V((f*))+OB      & \pp80 & 52000 & 4.20 & 4.20 & \pl$-$12.3 & \pl29.8 & VM4 \\
506 & ON2 V((n))((f*))   &   100 & 55000 & 4.20 & 4.20 & \pl$-$12.5 & \pl29.8 & SB1s \\
537 & O5 V((fc))z        & \pp60 & 39000 & 3.80 & 3.80 & \pl$-$13.0 & \pl27.8 & ... \\
550 & O5 V((fc))z        & \pp50 & 39000 & 3.80 & 3.80 & \pl$-$13.0 & \pl27.9 & ... \\
564 & O6-8 V((f))        & \pp40 & 37000 & 4.10 & 4.10 & \pl$-$12.7 & \pl29.0 & z? \\
577 & O6 V((fc))z        & \pp40 & 38000 & 4.00 & 4.00 & $<$$-$13.0 & $<$28.2 & ... \\
581 & O4-5 V((fc))       & \pp70 & 41000 & 3.70 & 3.71 & \pl$-$12.7 & \pl28.2 & ... \\
586 & O4 V((n))((fc))z   &   100 & 45000 & 4.00 & 4.01 & \pl$-$13.0 & \pl28.1 & SB? \\
609 & O9-9.5 V-III       &   100 & 33000 & 3.80 & 3.82 & $<$$-$12.7 & $<$27.7 & SB? \\
621 & O2 V((f*))z        & \pp80 & 54000 & 4.20 & 4.20 & \pl$-$12.7 & \pl29.3 & VM3 \\
648 & O5.5 IV(f)         & \pp55 & 40000 & 3.80 & 3.80 & \pl$-$12.5 & \pl29.0 & SBvs \\
755 & O3 Vn((f*))        &   285 & 46000 & 3.90 & 3.96 & \pl$-$12.7 & \pl28.6 & ... \\
797 & O3.5 V((n))((fc))z &   140 & 45000 & 3.80 & 3.82 & \pl$-$13.0 & \pl28.0 & SB? \\

\hline
\end{longtable}
}


\section{Stellar radii, luminosities, and masses}

\begin{landscape}
\begin{center}
\scriptsize
\setlength\LTcapwidth{\linewidth}
\begin{longtable}{llcccccccccccccccl}
  \caption{\label{tab2_HHe} Radii, luminosities, and spectroscopic
    masses obtained from our analysis, and evolutionary masses
    estimated from the Kiel (GT) and H--R (LT) diagrams and from the
    \textsc{bonnsai} tool (B).}
\\
\hline\hline
\\[-1.5 ex]

VFTS&Spectral Type &$T_{\rm eff}$&\grav$_c$&$M_V$&$R/R_{\odot}$&$\Delta 
R/R_{\odot}$&\logl&$\Delta$\logl&M$_{\rm sp}/M_{\odot}$&$\Delta M_{\rm 
sp}$&$M_{\rm ev}$\,(GT)&$\Delta$$M_{\rm ev}$\,(GT)&$M_{\rm 
ev}$\,(LT)&$\Delta$$M_{\rm ev}$\,(LT)&$M_{\rm ev}$\,(B)&$\Delta$$M_{\rm 
ev}$\,(B) & Comments\\
 
&&[K]&[cgs]&&&&&&[$M_{\odot}$]&[$M_{\odot}$]&[$M_{\odot}$]&[$M_{\odot
}$]&[$M_{\odot}$]&[$M_{\odot}$]&[$M_{\odot}$]&[$M_{\odot}$] & \\
 \hline
 \\[-2. ex] 
\endfirsthead
\caption{\it{continued}} \\
\hline\hline
\\[-1.5 ex]

 VFTS&Spectral Type &$T_{\rm eff}$&\grav$_c$&$M_V$&$R/R_{\odot}$&$\Delta 
R/R_{\odot}$&\logl&$\Delta$\logl&M$_{\rm sp}/M_{\odot}$&$\Delta M_{\rm 
sp}$&$M_{\rm ev}$\,(GT)&$\Delta$$M_{\rm ev}$\,(GT)&$M_{\rm 
ev}$\,(LT)&$\Delta$$M_{\rm ev}$\,(LT)&$M_{\rm ev}$\,(B)&$\Delta$$M_{\rm 
ev}$\,(B) & Comments\\
 
&&[K]&[cgs]&&&&&&[$M_{\odot}$]&[$M_{\odot}$]&[$M_{\odot}$]&[$M_{\odot
}$]&[$M_{\odot}$]&[$M_{\odot}$] &\\
\hline
\\[-2. ex]
\endhead
\hline
\multicolumn{18}{r}{\it{continued on next page}}\\
\endfoot
\hline
\endlastfoot
014 & O8.5 Vz          & 37100 &   -- &   \pp-- &   \pp-- &   -- &   -- &   -- &   -- &  \pp-- &   -- &  \pp-- &  --  &  \pp-- &   -- &  -- & SBs \\
021 & O9.5 IV          & 33800 & 3.90 & $-$4.20 & \pp7.80 & 0.80 & 4.86 & 0.14 & 17.6 & \pp5.4 & 22.3 & \pp2.4 & 19.9 & \pp4.6 & 19.6 & 1.5 & SB? \\
065 & O8 V(n)          & 37100 & 4.08 & $-$3.78 & \pp6.10 & 0.60 & 4.80 & 0.15 & 16.3 & \pp6.8 & 24.0 & \pp1.3 & 19.9 & \pp5.0 & 22.2 & 1.6 & ... \\
067 & O9.5 Vz          & 35200 & 4.12 & $-$3.33 & \pp5.20 & 0.70 & 4.56 & 0.17 & 13.0 & \pp6.7 & 20.4 & \pp8.4 & 15.0 & \pp5.0 & 18.6 & 1.4 & SB? \\
074 & O9 Vn            & 35100 & 4.23 & $-$3.66 & \pp6.00 & 0.75 & 4.69 & 0.15 & 17.7 & \pp9.6 & 20.1 & \pp8.1 & 19.8 & \pp3.9 & 19.6 & 1.6 & ... \\
089 & O6.5 V((f))z     & 39700 & 4.02 & $-$4.30 & \pp7.30 & 0.70 & 5.09 & 0.18 & 20.4 & \pp6.8 & 31.0 & \pp2.5 & 28.6 & \pp5.3 & 28.0 & 2.1 & ... \\
093 & O9.2 III-IV      & 34700 & 3.87 & $-$4.70 & \pp9.70 & 0.75 & 5.10 & 0.10 & 25.5 & \pp4.9 & 23.9 & \pp2.0 & 24.3 & \pp6.4 & 23.4 & 1.5 & SBvs \\
096 & O6 V((n))((fc))z & 40100 & 3.91 & $-$5.72 &   14.30 & 1.45 & 5.67 & 0.26 & 60.7 &   13.0 & 36.7 & \pp4.4 & 45.1 &   11.7 & 35.2 & 4.4 & SBvs VM2 \\
110 & O6 V((n))z       & 39900 & 3.88 & $-$5.04 &   10.50 & 1.00 & 5.40 & 0.20 & 30.5 & \pp8.6 & 37.4 & \pp5.0 & 34.7 &   10.1 & 33.4 & 4.0 & VM2? \\
117 & O6: Vz           & 41300 & 4.14 & $-$4.06 & \pp6.50 & 0.70 & 5.02 & 0.26 & 21.3 & \pp9.1 & 32.0 &   22.0 & 24.8 & \pp9.3 & 29.0 & 3.4 & SB? \\
123 & O6.5 Vz          & 40400 & 4.10 & $-$4.05 & \pp6.50 & 0.40 & 4.99 & 0.13 & 19.4 & \pp5.9 & 30.5 &   10.5 & 24.8 & \pp7.8 & 28.0 & 1.6 & SB? \\
130 & O8.5 V((n))      & 36500 & 4.11 & $-$4.52 & \pp8.70 & 0.75 & 5.06 & 0.12 & 35.6 &   16.7 & 22.7 & \pp7.7 & 24.7 & \pp6.0 & 24.0 & 1.9 & ... \\
132 & O9.5 Vz          & 35600 & 4.18 & $-$3.65 & \pp5.90 & 0.60 & 4.71 & 0.13 & 19.2 & \pp5.6 & 20.3 &   14.3 & 20.0 & \pp4.1 & 19.8 & 1.0 & ... \\
138 & O9 Vn            & 34600 & 4.20 & $-$3.53 & \pp5.70 & 0.60 & 4.60 & 0.13 & 18.8 & \pp7.2 & 18.8 &   36.7 & 14.9 & \pp5.0 & 19.2 & 1.1 & SB2? \\
149 & O9.5 V           & 35000 & 4.13 & $-$3.65 & \pp6.00 & 0.80 & 4.68 & 0.15 & 16.5 &   10.1 & 20.5 & \pp1.2 & 19.7 & \pp3.8 & 19.0 & 1.6 & ... \\
154 & O8.5 V           & 37400 & 4.12 & $-$5.04 &   10.90 & 0.95 & 5.30 & 0.12 & 57.1 &   19.8 & 23.9 &   11.9 & 30.2 & \pp8.2 & 26.6 & 2.0 & SBs \\
168 & O8.5 Vz          & 37300 & 4.02 & $-$4.12 & \pp7.10 & 0.50 & 4.92 & 0.11 & 19.3 & \pp3.8 & 25.4 & \pp1.6 & 24.3 & \pp4.2 & 23.4 & 1.1 & SB? \\
223 & O9.5 IV          & 34800 & 4.02 & $-$4.57 & \pp9.10 & 1.15 & 5.05 & 0.13 & 31.6 & \pp9.9 & 21.7 & \pp1.9 & 24.0 & \pp5.1 & 21.4 & 1.4 & SBvs \\
249 & O8 Vn            & 36500 & 4.11 & $-$3.83 & \pp6.30 & 0.70 & 4.78 & 0.14 & 18.7 & \pp6.3 & 22.7 & \pp7.7 & 19.9 & \pp4.9 & 22.2 & 1.3 & ... \\
250 & O9.2 V((n))      & 35400 & 4.14 & $-$3.85 & \pp6.60 & 0.55 & 4.76 & 0.12 & 21.9 & \pp8.4 & 20.3 &   10.3 & 19.9 & \pp4.3 & 20.2 & 1.1 & ... \\
251 & O9.5 IV          & 33700 & 4.01 & $-$3.82 & \pp6.60 & 0.80 & 4.72 & 0.11 & 16.3 & \pp5.2 & 20.2 & \pp1.6 & 19.4 & \pp3.9 & 18.6 & 0.9 & SB? \\
252 & O8.5 Vz          & 37000 & 4.22 & $-$3.63 & \pp5.70 & 0.45 & 4.73 & 0.12 & 19.7 & \pp3.8 & 22.2 &   33.7 & 19.9 & \pp4.8 & 21.6 & 0.9 & ... \\
266 & O8 V((f))z       & 38000 & 4.01 & $-$4.34 & \pp7.80 & 0.55 & 5.05 & 0.12 & 22.7 & \pp3.6 & 27.3 & \pp2.2 & 24.8 & \pp6.9 & 25.4 & 1.4 & ... \\
280 & O9 V((n))        & 34400 & 3.85 & $-$4.24 & \pp8.00 & 0.85 & 4.88 & 0.12 & 16.5 & \pp4.4 & 23.7 & \pp2.0 & 22.4 & \pp3.2 & 21.0 & 1.4 & ... \\
285 & O7.5 Vnnn        & 35300 & 4.08 & $-$3.90 & \pp6.60 & 0.75 & 4.77 & 0.20 & 19.1 & \pp6.2 & 21.3 & \pp1.5 & 19.9 & \pp4.3 & 20.0 & 1.7 & ... \\
290 & O9.5 IV          & 34000 & 3.99 & $-$3.91 & \pp6.90 & 0.60 & 4.76 & 0.09 & 17.0 & \pp3.3 & 21.0 & \pp1.8 & 19.8 & \pp3.9 & 19.2 & 0.9 & SB? \\
303 & O9.5 IV          & 34700 & 4.36 & $-$4.42 & \pp8.60 & 0.85 & 5.00 & 0.11 & 61.8 &   12.9 & 10.0 &   45.6 & 23.6 & \pp4.7 & 19.8 & 0.9 & VM2 \\
355 & O4 V((n))((fc))z & 43400 & 3.86 & $-$5.11 &   10.40 & 0.95 & 5.52 & 0.19 & 28.6 & \pp6.5 & 57.3 &   12.0 & 42.3 &   13.5 & 43.4 & 5.4 & SB2 NC \\
356 & O6: V(n)z        & 39300 & 4.03 & $-$4.48 & \pp8.10 & 0.75 & 5.14 & 0.20 & 25.7 & \pp9.0 & 29.5 & \pp2.1 & 28.8 & \pp5.9 & 27.6 & 2.8 & SB? \\
361 & O8.5 V           & 36900 & 4.07 & $-$4.97 &   10.60 & 1.25 & 5.27 & 0.15 & 48.2 &   15.1 & 23.9 & \pp1.4 & 29.2 & \pp8.4 & 25.2 & 1.9 & ... \\
369 & O9.7 V           & 33400 & 4.10 & $-$3.69 & \pp6.30 & 0.75 & 4.67 & 0.14 & 18.2 & \pp8.7 & 18.4 & \pp1.1 & 19.1 & \pp3.6 & 17.6 & 1.3 & ... \\
380 & O6-7 Vz          & 39100 & 4.13 & $-$3.92 & \pp6.30 & 0.50 & 4.92 & 0.15 & 19.5 & \pp5.1 & 27.1 &   15.1 & 24.9 & \pp4.6 & 25.6 & 1.4 & ... \\
385 & O4-5 V((n))((fc))& 42900 & 3.87 & $-$5.18 &   10.70 & 1.60 & 5.55 & 0.29 & 24.0 & \pp9.8 & 35.4 &   13.5 & 41.0 &   14.3 & 41.0 & 7.8 & SBs \\
392 & O6-7 V((f))z     & 37600 & 3.87 & $-$4.49 & \pp8.30 & 1.00 & 5.11 & 0.23 & 18.6 & \pp6.2 & 30.2 & \pp3.2 & 26.8 & \pp6.0 & 26.2 & 2.8 & ... \\
398 & O5.5 V((n))((f))z& 41200 & 4.03 & $-$5.14 &   10.80 & 1.00 & 5.47 & 0.17 & 45.6 &   11.2 & 34.7 & \pp2.9 & 38.2 &   11.8 & 34.2 & 3.5 & SBvs \\
418 & O5 V((n))((fc))z & 43200 & 4.10 & $-$4.43 & \pp7.50 & 0.70 & 5.24 & 0.18 & 25.8 & \pp9.1 & 38.8 &   18.8 & 34.9 & \pp7.4 & 33.8 & 4.1 & SB? \\
419 & O9: V(n)         & 33100 & 3.64 & $-$4.80 &   10.60 & 1.30 & 5.07 & 0.24 & 17.9 & \pp6.3 & 27.3 & \pp4.0 & 23.6 & \pp5.2 & 22.4 & 2.8 & ... \\
470 & O6: V((f))z      & 39300 & 3.94 & $-$4.06 & \pp6.70 & 0.60 & 4.97 & 0.18 & 14.3 & \pp3.2 & 32.7 & \pp3.3 & 24.9 & \pp6.0 & 27.8 & 2.1 & ... \\
472 & O6 Vz            & 40400 & 4.12 & $-$4.07 & \pp6.60 & 0.55 & 5.01 & 0.15 & 21.0 & \pp6.8 & 29.9 &   17.9 & 24.8 & \pp8.2 & 27.8 & 1.9 & ... \\
483 & O9 V             & 33700 &   -- &   \pp-- &   \pp-- &   -- &   -- &   -- &   -- &  \pp-- &  --  &  \pp-- &  --  &  \pp-- &   -- & --  & SB? \\
484 & O6-7 V((n))      & 35700 & 3.68 & $-$5.38 &   13.00 & 1.05 & 5.41 & 0.14 & 29.5 & \pp7.2 & 32.9 & \pp5.5 & 32.6 & \pp9.7 & 31.0 & 3.2 & ... \\
488 & O6 V((f))z       & 40700 & 3.87 & $-$4.80 & \pp9.20 & 1.05 & 5.33 & 0.25 & 22.9 & \pp6.4 & 41.3 & \pp6.3 & 33.8 & \pp8.8 & 34.6 & 4.2 & ... \\
491 & O6 V((fc))       & 40400 & 3.84 & $-$5.11 &   10.80 & 0.85 & 5.43 & 0.16 & 29.4 & \pp7.1 & 41.8 & \pp6.8 & 36.3 &   10.4 & 35.4 & 3.9 &     SB? \\
493 & O9 V             & 37100 & 4.27 & $-$4.42 & \pp8.20 & 1.20 & 5.06 & 0.16 & 45.7 &   17.0 & 21.8 &   34.1 & 24.7 & \pp6.5 & 23.4 & 1.6 & ... \\
494 & O8 V(n)          & 38900 & 4.21 & $-$4.20 & \pp7.20 & 0.85 & 5.03 & 0.20 & 28.0 &   15.1 & 26.1 &   19.1 & 24.8 & \pp7.2 & 25.6 & 2.9 & SB2? \\
498 & O9.5 V           & 33200 & 4.12 & $-$4.29 & \pp8.50 & 1.25 & 4.88 & 0.14 & 34.8 &   15.8 & 17.9 & \pp5.9 & 21.8 & \pp3.1 & 18.6 & 1.3 & ... \\
505 & O9.5 V-III       & 34000 & 4.27 & $-$3.64 & \pp6.20 & 0.75 & 4.66 & 0.13 & 26.1 & \pp7.6 & 17.7 &   37.7 & 19.3 & \pp3.6 & 17.8 & 0.9 & VM2? \\
511 & O5 V((n))((fc))z & 43700 & 4.25 & $-$4.91 & \pp9.30 & 0.75 & 5.46 & 0.15 & 56.1 &   16.9 & 36.1 &   28.1 & 40.6 &   12.2 & 38.2 & 4.3 & SB1s \\
517 & O9.5 V-III((n))  & 33000 & 4.02 & $-$4.81 &   10.80 & 1.10 & 5.09 & 0.10 & 44.6 &   10.0 & 19.1 & \pp1.5 & 23.6 & \pp5.7 & 20.8 & 1.3 & ... \\
521 & O9 V(n)          & 34800 & 4.13 & $-$4.97 &   11.00 & 1.30 & 5.21 & 0.12 & 59.6 &   18.7 & 19.7 & \pp7.7 & 27.4 & \pp6.3 & 22.4 & 1.5 & VM2 \\
536 & O6 Vz            & 41500 & 4.23 & $-$4.39 & \pp7.50 & 0.70 & 5.19 & 0.17 & 34.9 &   13.7 & 30.7 &   26.4 & 29.7 & \pp8.8 & 30.2 & 3.1 & SB? \\
549 & O6.5 Vz          & 39800 & 4.05 & $-$4.31 & \pp7.30 & 0.60 & 5.09 & 0.15 & 21.8 & \pp8.8 & 30.2 & \pp2.1 & 28.6 & \pp5.3 & 27.6 & 2.3 & SB? \\
554 & O9.7 V           & 34100 &   -- &   \pp-- &   \pp-- &   -- &   -- &   -- &   -- &  \pp-- &   -- &  \pp-- &   -- &  \pp-- &   -- &  -- & ... \\
560 & O9.5 V           & 33600 & 4.20 & $-$3.32 & \pp5.30 & 0.90 & 4.52 & 0.18 & 13.2 & \pp6.6 & 18.5 & \pp3.5 & 15.0 & \pp4.9 & 16.8 & 1.3 & ... \\
582 & O9.5 V((n))      & 35000 &   -- &   \pp-- &   \pp-- &   -- &   -- &   -- &   -- &  \pp-- &   -- &  \pp-- &   -- &  \pp-- &   -- & --  & ... \\
592 & O9.5 Vn          & 33600 & 4.28 & $-$3.72 & \pp6.40 & 0.95 & 4.69 & 0.13 & 28.5 &   12.0 & 17.0 &   38.3 & 19.3 & \pp3.7 & 18.4 & 1.2 & ... \\
597 & O8-9 V(n)        & 35400 & 3.94 & $-$4.14 & \pp7.40 & 0.80 & 4.87 & 0.14 & 17.4 & \pp5.8 & 23.8 & \pp1.9 & 22.8 & \pp3.3 & 21.6 & 1.5 & ... \\
601 & O5-6 V((n))z     & 40300 & 3.94 & $-$5.41 &   12.40 & 1.00 & 5.55 & 0.18 & 48.9 & \pp9.7 & 35.8 & \pp3.8 & 40.1 &   14.5 & 34.6 & 3.6 & ... \\
611 & O8 V(n)          & 37400 & 4.13 & $-$3.77 & \pp6.00 & 0.60 & 4.79 & 0.14 & 17.7 & \pp6.7 & 23.8 &   11.8 & 19.9 & \pp5.0 & 22.8 & 1.5 & ... \\
627 & O9.7 V           & 33600 & 4.11 & $-$3.67 & \pp6.20 & 0.80 & 4.67 & 0.13 & 18.1 & \pp6.8 & 18.5 & \pp3.5 & 19.2 & \pp3.6 & 17.8 & 0.9 & ... \\
635 & O9.5 IV          & 34100 & 4.00 & $-$4.10 & \pp7.50 & 0.75 & 4.83 & 0.12 & 20.5 & \pp5.0 & 21.0 & \pp1.7 & 19.9 & \pp4.4 & 19.4 & 1.0 & SBvs \\
638 & O8.5 Vz          & 36900 & 4.20 & $-$3.48 & \pp5.30 & 0.45 & 4.68 & 0.13 & 16.2 & \pp3.3 & 22.3 &   33.6 & 20.0 & \pp4.5 & 21.4 & 0.9 & ... \\
639 & O9.7 V           & 33700 & 4.18 & $-$3.95 & \pp7.10 & 0.85 & 4.78 & 0.12 & 27.8 & \pp7.4 & 17.9 &   11.9 & 19.9 & \pp3.8 & 18.2 & 0.8 & SB? \\
649 & O9.5 V           & 34800 & 4.19 & $-$3.72 & \pp6.20 & 0.50 & 4.71 & 0.12 & 21.7 & \pp5.0 & 19.1 &   13.1 & 19.7 & \pp3.9 & 19.0 & 0.9 & SB2 \\
660 & O9.5 Vnn         & 32300 & 4.15 & $-$4.00 & \pp7.50 & 1.35 & 4.73 & 0.20 & 29.0 &   14.9 & 16.2 & \pp6.2 & 18.9 & \pp3.9 & 17.2 & 1.1 & ... \\
663 & O8.5 Vz?         & 36500 & 4.03 & $-$3.78 & \pp6.20 & 0.55 & 4.77 & 0.12 & 15.0 &   10.4 & 23.8 & \pp1.6 & 19.9 & \pp4.8 & 20.6 & 1.8 & SB? \\
677 & O9.5 V           & 35900 & 4.20 & $-$4.42 & \pp8.40 & 1.20 & 5.03 & 0.18 & 40.8 &   16.9 & 20.5 &   35.2 & 24.3 & \pp4.9 & 21.4 & 1.7 & VM3 \\
679 & O9.5 V           & 33200 & 4.10 & $-$3.88 & \pp7.00 & 1.30 & 4.72 & 0.20 & 22.5 &   11.4 & 18.2 & \pp1.1 & 19.3 & \pp3.8 & 17.4 & 1.3 & SB? \\
704 & O9.2 V(n)        & 34200 &   -- &   \pp-- &   \pp-- &   -- &   -- &   -- &  --  &  \pp-- &   -- &  \pp-- &   -- &  \pp-- &   -- &  -- & SB? \\
706 & O6-7 Vnnz        & 38000 & 3.95 & $-$4.24 & \pp7.50 & 1.35 & 5.02 & 0.26 & 18.3 & \pp8.6 & 28.8 & \pp2.4 & 24.8 & \pp6.0 & 25.0 & 2.8 & ... \\
710 & O9.5 IV          & 35000 & 4.24 & $-$3.36 & \pp5.20 & 0.55 & 4.57 & 0.10 & 17.1 & \pp6.0 & 19.1 &   36.5 & 15.0 & \pp5.0 & 18.4 & 0.9 & SB? \\
716 & O9.5 IV          & 33200 & 3.96 & $-$4.15 & \pp7.90 & 0.80 & 4.82 & 0.11 & 20.8 & \pp6.4 & 20.3 & \pp1.9 & 19.9 & \pp4.0 & 18.8 & 1.0 & SBs \\
717 & O9 IV            & 35000 & 3.89 & $-$4.66 & \pp9.50 & 0.80 & 5.09 & 0.12 & 25.6 & \pp5.6 & 24.0 & \pp1.9 & 24.4 & \pp6.3 & 23.6 & 1.6 & SB?VM? \\
722 & O7 Vnnz          & 36600 & 4.01 & $-$4.08 & \pp7.00 & 0.60 & 4.91 & 0.13 & 18.3 & \pp5.3 & 24.4 & \pp1.5 & 24.0 & \pp4.0 & 22.6 & 1.7 & SB? NC \\
724 & O7 Vnnz          & 37600 & 3.93 & $-$4.35 & \pp7.80 & 2.65 & 5.01 & 0.47 & 18.9&    12.0 & 28.4 & \pp2.5 & 24.8 & \pp5.2 & 21.6 & 4.4 & NC \\
737 & O9 Vz?           & 37500 & 4.30 & $-$4.47 & \pp8.30 & 0.70 & 5.11 & 0.12 & 50.2&    14.3 & 21.9 &   36.1 & 26.5 & \pp6.0 & 24.0 & 1.3 & ... \\
746 & O6 Vnn           & 39900 & 3.92 & $-$4.77 & \pp9.30 & 2.20 & 5.29 & 0.24 & 26.2&    13.6 & 35.4 & \pp4.0 & 32.3 & \pp8.1 & 31.2 & 3.9 & ... \\
751 & O7-8 Vnnz        & 36000 & 4.01 & $-$4.38 & \pp8.20 & 1.35 & 5.01 & 0.31 & 25.1&    16.6 & 23.6 & \pp1.6 & 24.2 & \pp4.8 & 22.0 & 2.6 & NC \\
761 & O6.5 V((n))((f))z& 40300 & 4.16 & $-$4.06 & \pp6.60 & 0.45 & 4.99 & 0.13 & 23.0&  \pp6.1 & 29.1 &   21.1 & 24.8 & \pp7.6 & 27.8 & 1.5 & NC \\
768 & O8 Vn            & 35100 & 3.95 & $-$4.66 & \pp9.50 & 1.10 & 5.09 & 0.22 & 29.3&    13.9 & 23.3 & \pp2.0 & 24.5 & \pp6.3 & 22.2 & 2.5 & SB2? NC \\
770 & O7 Vnn           & 37800 & 4.06 & $-$4.16 & \pp7.10 & 1.15 & 4.98 & 0.27 & 21.1&    10.0 & 25.6 & \pp1.5 & 24.8 & \pp4.6 & 24.6 & 2.4 & ... \\
775 & O9.2 V           & 35900 & 4.14 & $-$3.68 & \pp5.90 & 0.50 & 4.72 & 0.12 & 17.5&  \pp8.6 & 21.2 &   11.2 & 19.9 & \pp4.3 & 20.0 & 1.5 & SB? NC \\
778 & O9.5 V           & 34200 & 4.19 & $-$4.03 & \pp7.30 & 1.30 & 4.81 & 0.18 & 28.8&    17.3 & 18.6 &   11.6 & 19.9 & \pp4.3 & 19.0 & 1.7 & ... \\
849 & O7 Vz            & 39800 & 4.17 & $-$4.13 & \pp6.80 & 0.55 & 5.02 & 0.13 & 25.0&  \pp7.5 & 27.9 &   20.9 & 24.8 & \pp7.8 & 27.0 & 1.4 & NC \\
892 & O9 V             & 35800 & 3.98 & $-$4.01 & \pp6.80 & 0.55 & 4.85 & 0.12 & 16.1&  \pp3.4 & 23.7 & \pp1.6 & 19.9 & \pp5.8 & 21.2 & 1.2 & NC \\

\end{longtable}
\end{center}
\end{landscape}

\begin{landscape}
\begin{center}
\scriptsize
\setlength\LTcapwidth{\linewidth}
\begin{longtable}{llccccccccccl}
  \caption{\label{tab2_HHeN} First estimates of radii, luminosities,
    and spectroscopic masses obtained from HHeN analysis, and
    evolutionary masses estimated from the Kiel (GT) and H--R (LT)
    diagrams and from the \textsc{bonnsai} tool (B), for the subsample of
    stars for which \ion{He}{i}/\textsc{ii} analysis alone was
    considered
    unreliable. } \\
  \hline\hline
  \\[-1.5 ex]
     VFTS&Spectral Type&$T_{\rm
    eff}$&\grav$_c$&$M_V$&$R/R_{\odot}$&\logl&$M_{\rm sp}$&$M_{\rm
    ev}$\,(GT)&$M_{\rm
    ev}$\,(LT)&$M_{\rm
    ev}$\,(B)&$\Delta$$M_{\rm
    ev}$\,(B)& Comments\\
 
&&[K]&[cgs]&&&&[$M_{\odot}$]&[$M_{\odot}$]&[$M_{\odot}$]&[$M_{\odot}$]&[
$M_{\odot}$]&\\
 \hline
 \\[-2. ex] 
\endfirsthead
\caption{continued.} \\
\hline\hline
 VFTS&Spectral Type &$T_{\rm eff}$&\grav$_c$&$M_V$&$R/R_{\odot}$&\logl&$M_{\rm 
sp}$&$M_{\rm ev}$\,(GT)&$M_{\rm 
ev}$\,(LT)&$M_{\rm ev}$\,(B)&$\Delta$$M_{\rm 
ev}$\,(B)&Comments\\
 
&&[K]&[cgs]&&&&[$M_{\odot}$]&[$M_{\odot}$]&[$M_{\odot}$]&[$M_{\odot}$]&[
$M_{\odot}$]&\\
\hline
\\[-2. ex]
\endhead
\hline
\endfoot
072 & O2 V-III(n)((f*))  & 54000 & 4.02 & $-$5.75 &   12.39 & 6.07 &   \pp58.6 &$\gtrsim$175&  \pp95 &  \pp94.0 &   25.2 & NC \\
169 & O2.5 V(n)((f*))    & 47000 & 3.92 & $-$5.79 &   13.58 & 5.91 &   \pp56.0 & \pp58 &  \pp56 &  \pp62.0 & \pp9.2 & SB? \\
216 & O4 V((fc))         & 43000 & 3.81 & $-$5.88 &   14.93 & 5.83 &   \pp52.5 & \pp57 &  \pp57 &  \pp52.8 & \pp7.4 & SB? \\
382 & O4-5 V((fc))z      & 40000 & 3.81 & $-$4.81 & \pp9.46 & 5.31 &   \pp21.1 & \pp42 &  \pp33 &  \pp32.2 & \pp3.4 & ... \\
410 & O7-8 V             & 34000 & 3.80 & $-$4.84 &   10.72 & 5.14 &   \pp26.5 & \pp24 &  \pp25 &  \pp23.2 & \pp2.3 & VM3 \\
435 & O7-8 V             & 37000 & 3.91 & $-$4.64 & \pp9.15 & 5.15 &   \pp24.3 & \pp28 &  \pp28 &  \pp24.8 & \pp2.0 & z? \\
436 & O7-8 V             & 35000 & 3.90 & $-$4.10 & \pp7.43 & 4.87 &   \pp16.0 & \pp24 &  \pp23 &  \pp20.4 & \pp1.8 & SB? z? \\
468 & O2 V((f*))+OB      & 52000 & 4.20 & $-$6.12 &   14.96 & 6.17 &     129.4 & \pp85 & 110 &  \pp92.4 &14.2 & VM4 \\
506 & ON2 V((n))((f*))   & 55000 & 4.20 & $-$6.61 &   18.17 & 6.43 &     190.9 &125 & 160 & 136.8 &24.2 & SB1s \\
537 & O5 V((fc))z        & 39000 & 3.80 & $-$4.59 & \pp8.67 & 5.19 &   \pp17.3 & \pp39 & 293 &  \pp29.0 & \pp2.9 & ... \\
550 & O5 V((fc))z        & 39000 & 3.80 & $-$4.60 & \pp8.74 & 5.20 &   \pp17.6 & \pp39 &  \pp29 &  \pp29.0 & \pp2.9 & ... \\
564 & O6-8 V((f))        & 37000 & 4.10 & $-$5.09 &   11.31 & 5.33 &   \pp58.8 & \pp24 &  \pp31 &  \pp27.6 & \pp2.6 & z? \\
577 & O6 V((fc))z        & 38000 & 4.00 & $-$4.40 & \pp8.06 & 5.08 &   \pp23.7 & \pp28 &  \pp27 &  \pp25.6 & \pp3.0 & ... \\
581 & O4-5 V((fc))       & 41000 & 3.71 & $-$5.00 &   10.22 & 5.42 &   \pp19.1 & \pp57 &  \pp37 &  \pp37.8 & \pp3.9 & ... \\
586 & O4 V((n))((fc))z   & 45000 & 4.01 & $-$4.73 & \pp8.48 & 5.42 &   \pp26.9 & \pp52 &  \pp40 & \pp40.8 & \pp4.3 & SB? \\
609 & O9-9.5 V-III       & 33000 & 3.82 & $-$3.37 & \pp5.55 & 4.52 & \pp\pp7.4 & \pp23 &  \pp15 &  \pp16.4 & \pp1.4 & SB? \\
621 & O2 V((f*))z        & 54000 & 4.20 & $-$6.14 &   14.77 & 6.22 &     126.2 &110 & 120 & 106.6 &17.1& VM3 \\
648 & O5.5 IV(f)         & 40000 & 3.80 & $-$5.67 &   14.12 & 5.66 &   \pp45.9 & \pp43 &  \pp44 &  \pp41.2 & \pp5.2 & SBvs \\
755 & O3 Vn((f*))        & 46000 & 3.96 & $-$5.21 &   10.51 & 5.65 &   \pp36.8 & \pp58 &  \pp49 &  \pp49.8 & \pp7.3 & ... \\
797 & O3.5 V((n))((fc))z & 45000 & 3.82 & $-$5.15 &   10.36 & 5.60 &   \pp25.9 & \pp57 &  \pp47 &  \pp47.4 & \pp6.2 & SB? \\

\end{longtable}
\end{center}
\end{landscape}

\twocolumn{
}

\end{appendix}

\end{document}